\documentclass{jfm}

\usepackage{graphicx}
\usepackage{newtxtext}
\usepackage{newtxmath}
\usepackage{natbib}
\usepackage{hyperref}
\hypersetup{
    colorlinks = true,
    urlcolor   = blue,
    citecolor  = blue,
}

\newcommand{\RomanNumeralCaps}[1]
\linenumbers
\usepackage{tcolorbox}

\usepackage{color}
\usepackage{xcolor}
\definecolor{AliceBlue}{rgb}{0.94,0.97,1.00}
\definecolor{AntiqueWhite1}{rgb}{1.00,0.94,0.86}
\definecolor{AntiqueWhite2}{rgb}{0.93,0.87,0.80}
\definecolor{AntiqueWhite3}{rgb}{0.80,0.75,0.69}
\definecolor{AntiqueWhite4}{rgb}{0.55,0.51,0.47}
\definecolor{AntiqueWhite}{rgb}{0.98,0.92,0.84}
\definecolor{BlanchedAlmond}{rgb}{1.00,0.92,0.80}
\definecolor{BlueViolet}{rgb}{0.54,0.17,0.89}
\definecolor{CadetBlue1}{rgb}{0.60,0.96,1.00}
\definecolor{CadetBlue2}{rgb}{0.56,0.90,0.93}
\definecolor{CadetBlue3}{rgb}{0.48,0.77,0.80}
\definecolor{CadetBlue4}{rgb}{0.33,0.53,0.55}
\definecolor{CadetBlue}{rgb}{0.37,0.62,0.63}
\definecolor{CornflowerBlue}{rgb}{0.39,0.58,0.93}
\definecolor{DarkBlue}{rgb}{0.00,0.00,0.55}
\definecolor{DarkCyan}{rgb}{0.00,0.55,0.55}
\definecolor{DarkGoldenrod1}{rgb}{1.00,0.73,0.06}
\definecolor{DarkGoldenrod2}{rgb}{0.93,0.68,0.05}
\definecolor{DarkGoldenrod3}{rgb}{0.80,0.58,0.05}
\definecolor{DarkGoldenrod4}{rgb}{0.55,0.40,0.03}
\definecolor{DarkGoldenrod}{rgb}{0.72,0.53,0.04}
\definecolor{DarkGray}{rgb}{0.66,0.66,0.66}
\definecolor{DarkGreen}{rgb}{0.00,0.39,0.00}
\definecolor{DarkGrey}{rgb}{0.66,0.66,0.66}
\definecolor{DarkKhaki}{rgb}{0.74,0.72,0.42}
\definecolor{DarkMagenta}{rgb}{0.55,0.00,0.55}
\definecolor{DarkOliveGreen1}{rgb}{0.79,1.00,0.44}
\definecolor{DarkOliveGreen2}{rgb}{0.74,0.93,0.41}
\definecolor{DarkOliveGreen3}{rgb}{0.64,0.80,0.35}
\definecolor{DarkOliveGreen4}{rgb}{0.43,0.55,0.24}
\definecolor{DarkOliveGreen}{rgb}{0.33,0.42,0.18}
\definecolor{DarkOrange1}{rgb}{1.00,0.50,0.00}
\definecolor{DarkOrange2}{rgb}{0.93,0.46,0.00}
\definecolor{DarkOrange3}{rgb}{0.80,0.40,0.00}
\definecolor{DarkOrange4}{rgb}{0.55,0.27,0.00}
\definecolor{DarkOrange}{rgb}{1.00,0.55,0.00}
\definecolor{DarkOrchid1}{rgb}{0.75,0.24,1.00}
\definecolor{DarkOrchid2}{rgb}{0.70,0.23,0.93}
\definecolor{DarkOrchid3}{rgb}{0.60,0.20,0.80}
\definecolor{DarkOrchid4}{rgb}{0.41,0.13,0.55}
\definecolor{DarkOrchid}{rgb}{0.60,0.20,0.80}
\definecolor{DarkRed}{rgb}{0.55,0.00,0.00}
\definecolor{DarkSalmon}{rgb}{0.91,0.59,0.48}
\definecolor{DarkSeaGreen1}{rgb}{0.76,1.00,0.76}
\definecolor{DarkSeaGreen2}{rgb}{0.71,0.93,0.71}
\definecolor{DarkSeaGreen3}{rgb}{0.61,0.80,0.61}
\definecolor{DarkSeaGreen4}{rgb}{0.41,0.55,0.41}
\definecolor{DarkSeaGreen}{rgb}{0.56,0.74,0.56}
\definecolor{DarkSlateBlue}{rgb}{0.28,0.24,0.55}
\definecolor{DarkSlateGray1}{rgb}{0.59,1.00,1.00}
\definecolor{DarkSlateGray2}{rgb}{0.55,0.93,0.93}
\definecolor{DarkSlateGray3}{rgb}{0.47,0.80,0.80}
\definecolor{DarkSlateGray4}{rgb}{0.32,0.55,0.55}
\definecolor{DarkSlateGray}{rgb}{0.18,0.31,0.31}
\definecolor{DarkSlateGrey}{rgb}{0.18,0.31,0.31}
\definecolor{DarkTurquoise}{rgb}{0.00,0.81,0.82}
\definecolor{DarkViolet}{rgb}{0.58,0.00,0.83}
\definecolor{DeepPink1}{rgb}{1.00,0.08,0.58}
\definecolor{DeepPink2}{rgb}{0.93,0.07,0.54}
\definecolor{DeepPink3}{rgb}{0.80,0.06,0.46}
\definecolor{DeepPink4}{rgb}{0.55,0.04,0.31}
\definecolor{DeepPink}{rgb}{1.00,0.08,0.58}
\definecolor{DeepSkyBlue1}{rgb}{0.00,0.75,1.00}
\definecolor{DeepSkyBlue2}{rgb}{0.00,0.70,0.93}
\definecolor{DeepSkyBlue3}{rgb}{0.00,0.60,0.80}
\definecolor{DeepSkyBlue4}{rgb}{0.00,0.41,0.55}
\definecolor{DeepSkyBlue}{rgb}{0.00,0.75,1.00}
\definecolor{DimGray}{rgb}{0.41,0.41,0.41}
\definecolor{DimGrey}{rgb}{0.41,0.41,0.41}
\definecolor{DodgerBlue1}{rgb}{0.12,0.56,1.00}
\definecolor{DodgerBlue2}{rgb}{0.11,0.53,0.93}
\definecolor{DodgerBlue3}{rgb}{0.09,0.45,0.80}
\definecolor{DodgerBlue4}{rgb}{0.06,0.31,0.55}
\definecolor{DodgerBlue}{rgb}{0.12,0.56,1.00}
\definecolor{FloralWhite}{rgb}{1.00,0.98,0.94}
\definecolor{ForestGreen}{rgb}{0.13,0.55,0.13}
\definecolor{GhostWhite}{rgb}{0.97,0.97,1.00}
\definecolor{GreenYellow}{rgb}{0.68,1.00,0.18}
\definecolor{HotPink1}{rgb}{1.00,0.43,0.71}
\definecolor{HotPink2}{rgb}{0.93,0.42,0.65}
\definecolor{HotPink3}{rgb}{0.80,0.38,0.56}
\definecolor{HotPink4}{rgb}{0.55,0.23,0.38}
\definecolor{HotPink}{rgb}{1.00,0.41,0.71}
\definecolor{IndianRed1}{rgb}{1.00,0.42,0.42}
\definecolor{IndianRed2}{rgb}{0.93,0.39,0.39}
\definecolor{IndianRed3}{rgb}{0.80,0.33,0.33}
\definecolor{IndianRed4}{rgb}{0.55,0.23,0.23}
\definecolor{IndianRed}{rgb}{0.80,0.36,0.36}
\definecolor{LavenderBlush1}{rgb}{1.00,0.94,0.96}
\definecolor{LavenderBlush2}{rgb}{0.93,0.88,0.90}
\definecolor{LavenderBlush3}{rgb}{0.80,0.76,0.77}
\definecolor{LavenderBlush4}{rgb}{0.55,0.51,0.53}
\definecolor{LavenderBlush}{rgb}{1.00,0.94,0.96}
\definecolor{LawnGreen}{rgb}{0.49,0.99,0.00}
\definecolor{LemonChiffon1}{rgb}{1.00,0.98,0.80}
\definecolor{LemonChiffon2}{rgb}{0.93,0.91,0.75}
\definecolor{LemonChiffon3}{rgb}{0.80,0.79,0.65}
\definecolor{LemonChiffon4}{rgb}{0.55,0.54,0.44}
\definecolor{LemonChiffon}{rgb}{1.00,0.98,0.80}
\definecolor{LightBlue1}{rgb}{0.75,0.94,1.00}
\definecolor{LightBlue2}{rgb}{0.70,0.87,0.93}
\definecolor{LightBlue3}{rgb}{0.60,0.75,0.80}
\definecolor{LightBlue4}{rgb}{0.41,0.51,0.55}
\definecolor{LightBlue}{rgb}{0.68,0.85,0.90}
\definecolor{LightCoral}{rgb}{0.94,0.50,0.50}
\definecolor{LightCyan1}{rgb}{0.88,1.00,1.00}
\definecolor{LightCyan2}{rgb}{0.82,0.93,0.93}
\definecolor{LightCyan3}{rgb}{0.71,0.80,0.80}
\definecolor{LightCyan4}{rgb}{0.48,0.55,0.55}
\definecolor{LightCyan}{rgb}{0.88,1.00,1.00}
\definecolor{LightGoldenrod1}{rgb}{1.00,0.93,0.55}
\definecolor{LightGoldenrod2}{rgb}{0.93,0.86,0.51}
\definecolor{LightGoldenrod3}{rgb}{0.80,0.75,0.44}
\definecolor{LightGoldenrod4}{rgb}{0.55,0.51,0.30}
\definecolor{LightGoldenrodYellow}{rgb}{0.98,0.98,0.82}
\definecolor{LightGoldenrod}{rgb}{0.93,0.87,0.51}
\definecolor{LightGray}{rgb}{0.83,0.83,0.83}
\definecolor{LightGreen}{rgb}{0.56,0.93,0.56}
\definecolor{LightGrey}{rgb}{0.83,0.83,0.83}
\definecolor{LightPink1}{rgb}{1.00,0.68,0.73}
\definecolor{LightPink2}{rgb}{0.93,0.64,0.68}
\definecolor{LightPink3}{rgb}{0.80,0.55,0.58}
\definecolor{LightPink4}{rgb}{0.55,0.37,0.40}
\definecolor{LightPink}{rgb}{1.00,0.71,0.76}
\definecolor{LightSalmon1}{rgb}{1.00,0.63,0.48}
\definecolor{LightSalmon2}{rgb}{0.93,0.58,0.45}
\definecolor{LightSalmon3}{rgb}{0.80,0.51,0.38}
\definecolor{LightSalmon4}{rgb}{0.55,0.34,0.26}
\definecolor{LightSalmon}{rgb}{1.00,0.63,0.48}
\definecolor{LightSeaGreen}{rgb}{0.13,0.70,0.67}
\definecolor{LightSkyBlue1}{rgb}{0.69,0.89,1.00}
\definecolor{LightSkyBlue2}{rgb}{0.64,0.83,0.93}
\definecolor{LightSkyBlue3}{rgb}{0.55,0.71,0.80}
\definecolor{LightSkyBlue4}{rgb}{0.38,0.48,0.55}
\definecolor{LightSkyBlue}{rgb}{0.53,0.81,0.98}
\definecolor{LightSlateBlue}{rgb}{0.52,0.44,1.00}
\definecolor{LightSlateGray}{rgb}{0.47,0.53,0.60}
\definecolor{LightSlateGrey}{rgb}{0.47,0.53,0.60}
\definecolor{LightSteelBlue1}{rgb}{0.79,0.88,1.00}
\definecolor{LightSteelBlue2}{rgb}{0.74,0.82,0.93}
\definecolor{LightSteelBlue3}{rgb}{0.64,0.71,0.80}
\definecolor{LightSteelBlue4}{rgb}{0.43,0.48,0.55}
\definecolor{LightSteelBlue}{rgb}{0.69,0.77,0.87}
\definecolor{LightYellow1}{rgb}{1.00,1.00,0.88}
\definecolor{LightYellow2}{rgb}{0.93,0.93,0.82}
\definecolor{LightYellow3}{rgb}{0.80,0.80,0.71}
\definecolor{LightYellow4}{rgb}{0.55,0.55,0.48}
\definecolor{LightYellow}{rgb}{1.00,1.00,0.88}
\definecolor{LimeGreen}{rgb}{0.20,0.80,0.20}
\definecolor{MediumAquamarine}{rgb}{0.40,0.80,0.67}
\definecolor{MediumBlue}{rgb}{0.00,0.00,0.80}
\definecolor{MediumOrchid1}{rgb}{0.88,0.40,1.00}
\definecolor{MediumOrchid2}{rgb}{0.82,0.37,0.93}
\definecolor{MediumOrchid3}{rgb}{0.71,0.32,0.80}
\definecolor{MediumOrchid4}{rgb}{0.48,0.22,0.55}
\definecolor{MediumOrchid}{rgb}{0.73,0.33,0.83}
\definecolor{MediumPurple1}{rgb}{0.67,0.51,1.00}
\definecolor{MediumPurple2}{rgb}{0.62,0.47,0.93}
\definecolor{MediumPurple3}{rgb}{0.54,0.41,0.80}
\definecolor{MediumPurple4}{rgb}{0.36,0.28,0.55}
\definecolor{MediumPurple}{rgb}{0.58,0.44,0.86}
\definecolor{MediumSeaGreen}{rgb}{0.24,0.70,0.44}
\definecolor{MediumSlateBlue}{rgb}{0.48,0.41,0.93}
\definecolor{MediumSpringGreen}{rgb}{0.00,0.98,0.60}
\definecolor{MediumTurquoise}{rgb}{0.28,0.82,0.80}
\definecolor{MediumVioletRed}{rgb}{0.78,0.08,0.52}
\definecolor{MidnightBlue}{rgb}{0.10,0.10,0.44}
\definecolor{MintCream}{rgb}{0.96,1.00,0.98}
\definecolor{MistyRose1}{rgb}{1.00,0.89,0.88}
\definecolor{MistyRose2}{rgb}{0.93,0.84,0.82}
\definecolor{MistyRose3}{rgb}{0.80,0.72,0.71}
\definecolor{MistyRose4}{rgb}{0.55,0.49,0.48}
\definecolor{MistyRose}{rgb}{1.00,0.89,0.88}
\definecolor{NavajoWhite1}{rgb}{1.00,0.87,0.68}
\definecolor{NavajoWhite2}{rgb}{0.93,0.81,0.63}
\definecolor{NavajoWhite3}{rgb}{0.80,0.70,0.55}
\definecolor{NavajoWhite4}{rgb}{0.55,0.47,0.37}
\definecolor{NavajoWhite}{rgb}{1.00,0.87,0.68}
\definecolor{NavyBlue}{rgb}{0.00,0.00,0.50}
\definecolor{OldLace}{rgb}{0.99,0.96,0.90}
\definecolor{OliveDrab1}{rgb}{0.75,1.00,0.24}
\definecolor{OliveDrab2}{rgb}{0.70,0.93,0.23}
\definecolor{OliveDrab3}{rgb}{0.60,0.80,0.20}
\definecolor{OliveDrab4}{rgb}{0.41,0.55,0.13}
\definecolor{OliveDrab}{rgb}{0.42,0.56,0.14}
\definecolor{OrangeRed1}{rgb}{1.00,0.27,0.00}
\definecolor{OrangeRed2}{rgb}{0.93,0.25,0.00}
\definecolor{OrangeRed3}{rgb}{0.80,0.22,0.00}
\definecolor{OrangeRed4}{rgb}{0.55,0.15,0.00}
\definecolor{OrangeRed}{rgb}{1.00,0.27,0.00}
\definecolor{PaleGoldenrod}{rgb}{0.93,0.91,0.67}
\definecolor{PaleGreen1}{rgb}{0.60,1.00,0.60}
\definecolor{PaleGreen2}{rgb}{0.56,0.93,0.56}
\definecolor{PaleGreen3}{rgb}{0.49,0.80,0.49}
\definecolor{PaleGreen4}{rgb}{0.33,0.55,0.33}
\definecolor{PaleGreen}{rgb}{0.60,0.98,0.60}
\definecolor{PaleTurquoise1}{rgb}{0.73,1.00,1.00}
\definecolor{PaleTurquoise2}{rgb}{0.68,0.93,0.93}
\definecolor{PaleTurquoise3}{rgb}{0.59,0.80,0.80}
\definecolor{PaleTurquoise4}{rgb}{0.40,0.55,0.55}
\definecolor{PaleTurquoise}{rgb}{0.69,0.93,0.93}
\definecolor{PaleVioletRed1}{rgb}{1.00,0.51,0.67}
\definecolor{PaleVioletRed2}{rgb}{0.93,0.47,0.62}
\definecolor{PaleVioletRed3}{rgb}{0.80,0.41,0.54}
\definecolor{PaleVioletRed4}{rgb}{0.55,0.28,0.36}
\definecolor{PaleVioletRed}{rgb}{0.86,0.44,0.58}
\definecolor{PapayaWhip}{rgb}{1.00,0.94,0.84}
\definecolor{PeachPuff1}{rgb}{1.00,0.85,0.73}
\definecolor{PeachPuff2}{rgb}{0.93,0.80,0.68}
\definecolor{PeachPuff3}{rgb}{0.80,0.69,0.58}
\definecolor{PeachPuff4}{rgb}{0.55,0.47,0.40}
\definecolor{PeachPuff}{rgb}{1.00,0.85,0.73}
\definecolor{PowderBlue}{rgb}{0.69,0.88,0.90}
\definecolor{RosyBrown1}{rgb}{1.00,0.76,0.76}
\definecolor{RosyBrown2}{rgb}{0.93,0.71,0.71}
\definecolor{RosyBrown3}{rgb}{0.80,0.61,0.61}
\definecolor{RosyBrown4}{rgb}{0.55,0.41,0.41}
\definecolor{RosyBrown}{rgb}{0.74,0.56,0.56}
\definecolor{RoyalBlue1}{rgb}{0.28,0.46,1.00}
\definecolor{RoyalBlue2}{rgb}{0.26,0.43,0.93}
\definecolor{RoyalBlue3}{rgb}{0.23,0.37,0.80}
\definecolor{RoyalBlue4}{rgb}{0.15,0.25,0.55}
\definecolor{RoyalBlue}{rgb}{0.25,0.41,0.88}
\definecolor{SaddleBrown}{rgb}{0.55,0.27,0.07}
\definecolor{SandyBrown}{rgb}{0.96,0.64,0.38}
\definecolor{SeaGreen1}{rgb}{0.33,1.00,0.62}
\definecolor{SeaGreen2}{rgb}{0.31,0.93,0.58}
\definecolor{SeaGreen3}{rgb}{0.26,0.80,0.50}
\definecolor{SeaGreen4}{rgb}{0.18,0.55,0.34}
\definecolor{SeaGreen}{rgb}{0.18,0.55,0.34}
\definecolor{SkyBlue1}{rgb}{0.53,0.81,1.00}
\definecolor{SkyBlue2}{rgb}{0.49,0.75,0.93}
\definecolor{SkyBlue3}{rgb}{0.42,0.65,0.80}
\definecolor{SkyBlue4}{rgb}{0.29,0.44,0.55}
\definecolor{SkyBlue}{rgb}{0.53,0.81,0.92}
\definecolor{SlateBlue1}{rgb}{0.51,0.44,1.00}
\definecolor{SlateBlue2}{rgb}{0.48,0.40,0.93}
\definecolor{SlateBlue3}{rgb}{0.41,0.35,0.80}
\definecolor{SlateBlue4}{rgb}{0.28,0.24,0.55}
\definecolor{SlateBlue}{rgb}{0.42,0.35,0.80}
\definecolor{SlateGray1}{rgb}{0.78,0.89,1.00}
\definecolor{SlateGray2}{rgb}{0.73,0.83,0.93}
\definecolor{SlateGray3}{rgb}{0.62,0.71,0.80}
\definecolor{SlateGray4}{rgb}{0.42,0.48,0.55}
\definecolor{SlateGray}{rgb}{0.44,0.50,0.56}
\definecolor{SlateGrey}{rgb}{0.44,0.50,0.56}
\definecolor{SpringGreen1}{rgb}{0.00,1.00,0.50}
\definecolor{SpringGreen2}{rgb}{0.00,0.93,0.46}
\definecolor{SpringGreen3}{rgb}{0.00,0.80,0.40}
\definecolor{SpringGreen4}{rgb}{0.00,0.55,0.27}
\definecolor{SpringGreen}{rgb}{0.00,1.00,0.50}
\definecolor{SteelBlue1}{rgb}{0.39,0.72,1.00}
\definecolor{SteelBlue2}{rgb}{0.36,0.67,0.93}
\definecolor{SteelBlue3}{rgb}{0.31,0.58,0.80}
\definecolor{SteelBlue4}{rgb}{0.21,0.39,0.55}
\definecolor{SteelBlue}{rgb}{0.27,0.51,0.71}
\definecolor{VioletRed1}{rgb}{1.00,0.24,0.59}
\definecolor{VioletRed2}{rgb}{0.93,0.23,0.55}
\definecolor{VioletRed3}{rgb}{0.80,0.20,0.47}
\definecolor{VioletRed4}{rgb}{0.55,0.13,0.32}
\definecolor{VioletRed}{rgb}{0.82,0.13,0.56}
\definecolor{WhiteSmoke}{rgb}{0.96,0.96,0.96}
\definecolor{YellowGreen}{rgb}{0.60,0.80,0.20}
\definecolor{aliceblue}{rgb}{0.94,0.97,1.00}
\definecolor{antiquewhite}{rgb}{0.98,0.92,0.84}
\definecolor{aquamarine1}{rgb}{0.50,1.00,0.83}
\definecolor{aquamarine2}{rgb}{0.46,0.93,0.78}
\definecolor{aquamarine3}{rgb}{0.40,0.80,0.67}
\definecolor{aquamarine4}{rgb}{0.27,0.55,0.45}
\definecolor{aquamarine}{rgb}{0.50,1.00,0.83}
\definecolor{azure1}{rgb}{0.94,1.00,1.00}
\definecolor{azure2}{rgb}{0.88,0.93,0.93}
\definecolor{azure3}{rgb}{0.76,0.80,0.80}
\definecolor{azure4}{rgb}{0.51,0.55,0.55}
\definecolor{azure}{rgb}{0.94,1.00,1.00}
\definecolor{beige}{rgb}{0.96,0.96,0.86}
\definecolor{bisque1}{rgb}{1.00,0.89,0.77}
\definecolor{bisque2}{rgb}{0.93,0.84,0.72}
\definecolor{bisque3}{rgb}{0.80,0.72,0.62}
\definecolor{bisque4}{rgb}{0.55,0.49,0.42}
\definecolor{bisque}{rgb}{1.00,0.89,0.77}
\definecolor{black}{rgb}{0.00,0.00,0.00}
\definecolor{blanchedalmond}{rgb}{1.00,0.92,0.80}
\definecolor{blue1}{rgb}{0.00,0.00,1.00}
\definecolor{blue2}{rgb}{0.00,0.00,0.93}
\definecolor{blue3}{rgb}{0.00,0.00,0.80}
\definecolor{blue4}{rgb}{0.00,0.00,0.55}
\definecolor{blueviolet}{rgb}{0.54,0.17,0.89}
\definecolor{blue}{rgb}{0.00,0.00,1.00}
\definecolor{brown1}{rgb}{1.00,0.25,0.25}
\definecolor{brown2}{rgb}{0.93,0.23,0.23}
\definecolor{brown3}{rgb}{0.80,0.20,0.20}
\definecolor{brown4}{rgb}{0.55,0.14,0.14}
\definecolor{brown}{rgb}{0.65,0.16,0.16}
\definecolor{burlywood1}{rgb}{1.00,0.83,0.61}
\definecolor{burlywood2}{rgb}{0.93,0.77,0.57}
\definecolor{burlywood3}{rgb}{0.80,0.67,0.49}
\definecolor{burlywood4}{rgb}{0.55,0.45,0.33}
\definecolor{burlywood}{rgb}{0.87,0.72,0.53}
\definecolor{cadetblue}{rgb}{0.37,0.62,0.63}
\definecolor{chartreuse1}{rgb}{0.50,1.00,0.00}
\definecolor{chartreuse2}{rgb}{0.46,0.93,0.00}
\definecolor{chartreuse3}{rgb}{0.40,0.80,0.00}
\definecolor{chartreuse4}{rgb}{0.27,0.55,0.00}
\definecolor{chartreuse}{rgb}{0.50,1.00,0.00}
\definecolor{chocolate1}{rgb}{1.00,0.50,0.14}
\definecolor{chocolate2}{rgb}{0.93,0.46,0.13}
\definecolor{chocolate3}{rgb}{0.80,0.40,0.11}
\definecolor{chocolate4}{rgb}{0.55,0.27,0.07}
\definecolor{chocolate}{rgb}{0.82,0.41,0.12}
\definecolor{coral1}{rgb}{1.00,0.45,0.34}
\definecolor{coral2}{rgb}{0.93,0.42,0.31}
\definecolor{coral3}{rgb}{0.80,0.36,0.27}
\definecolor{coral4}{rgb}{0.55,0.24,0.18}
\definecolor{coral}{rgb}{1.00,0.50,0.31}
\definecolor{cornflowerblue}{rgb}{0.39,0.58,0.93}
\definecolor{cornsilk1}{rgb}{1.00,0.97,0.86}
\definecolor{cornsilk2}{rgb}{0.93,0.91,0.80}
\definecolor{cornsilk3}{rgb}{0.80,0.78,0.69}
\definecolor{cornsilk4}{rgb}{0.55,0.53,0.47}
\definecolor{cornsilk}{rgb}{1.00,0.97,0.86}
\definecolor{cyan1}{rgb}{0.00,1.00,1.00}
\definecolor{cyan2}{rgb}{0.00,0.93,0.93}
\definecolor{cyan3}{rgb}{0.00,0.80,0.80}
\definecolor{cyan4}{rgb}{0.00,0.55,0.55}
\definecolor{cyan}{rgb}{0.00,1.00,1.00}
\definecolor{darkblue}{rgb}{0.00,0.00,0.55}
\definecolor{darkcyan}{rgb}{0.00,0.55,0.55}
\definecolor{darkgoldenrod}{rgb}{0.72,0.53,0.04}
\definecolor{darkgray}{rgb}{0.66,0.66,0.66}
\definecolor{darkgreen}{rgb}{0.00,0.39,0.00}
\definecolor{darkgrey}{rgb}{0.66,0.66,0.66}
\definecolor{darkkhaki}{rgb}{0.74,0.72,0.42}
\definecolor{darkmagenta}{rgb}{0.55,0.00,0.55}
\definecolor{darkolive}{rgb}{0.33,0.42,0.18}
\definecolor{darkorange}{rgb}{1.00,0.55,0.00}
\definecolor{darkorchid}{rgb}{0.60,0.20,0.80}
\definecolor{darkred}{rgb}{0.55,0.00,0.00}
\definecolor{darksalmon}{rgb}{0.91,0.59,0.48}
\definecolor{darksea}{rgb}{0.56,0.74,0.56}
\definecolor{darkslate}{rgb}{0.18,0.31,0.31}
\definecolor{darkslate}{rgb}{0.18,0.31,0.31}
\definecolor{darkslate}{rgb}{0.28,0.24,0.55}
\definecolor{darkturquoise}{rgb}{0.00,0.81,0.82}
\definecolor{darkviolet}{rgb}{0.58,0.00,0.83}
\definecolor{deeppink}{rgb}{1.00,0.08,0.58}
\definecolor{deepsky}{rgb}{0.00,0.75,1.00}
\definecolor{dimgray}{rgb}{0.41,0.41,0.41}
\definecolor{dimgrey}{rgb}{0.41,0.41,0.41}
\definecolor{dodgerblue}{rgb}{0.12,0.56,1.00}
\definecolor{firebrick1}{rgb}{1.00,0.19,0.19}
\definecolor{firebrick2}{rgb}{0.93,0.17,0.17}
\definecolor{firebrick3}{rgb}{0.80,0.15,0.15}
\definecolor{firebrick4}{rgb}{0.55,0.10,0.10}
\definecolor{firebrick}{rgb}{0.70,0.13,0.13}
\definecolor{floralwhite}{rgb}{1.00,0.98,0.94}
\definecolor{forestgreen}{rgb}{0.13,0.55,0.13}
\definecolor{gainsboro}{rgb}{0.86,0.86,0.86}
\definecolor{ghostwhite}{rgb}{0.97,0.97,1.00}
\definecolor{gold1}{rgb}{1.00,0.84,0.00}
\definecolor{gold2}{rgb}{0.93,0.79,0.00}
\definecolor{gold3}{rgb}{0.80,0.68,0.00}
\definecolor{gold4}{rgb}{0.55,0.46,0.00}
\definecolor{goldenrod1}{rgb}{1.00,0.76,0.15}
\definecolor{goldenrod2}{rgb}{0.93,0.71,0.13}
\definecolor{goldenrod3}{rgb}{0.80,0.61,0.11}
\definecolor{goldenrod4}{rgb}{0.55,0.41,0.08}
\definecolor{goldenrod}{rgb}{0.85,0.65,0.13}
\definecolor{gold}{rgb}{1.00,0.84,0.00}
\definecolor{gray0}{rgb}{0.00,0.00,0.00}
\definecolor{gray100}{rgb}{1.00,1.00,1.00}
\definecolor{gray10}{rgb}{0.10,0.10,0.10}
\definecolor{gray11}{rgb}{0.11,0.11,0.11}
\definecolor{gray12}{rgb}{0.12,0.12,0.12}
\definecolor{gray13}{rgb}{0.13,0.13,0.13}
\definecolor{gray14}{rgb}{0.14,0.14,0.14}
\definecolor{gray15}{rgb}{0.15,0.15,0.15}
\definecolor{gray16}{rgb}{0.16,0.16,0.16}
\definecolor{gray17}{rgb}{0.17,0.17,0.17}
\definecolor{gray18}{rgb}{0.18,0.18,0.18}
\definecolor{gray19}{rgb}{0.19,0.19,0.19}
\definecolor{gray1}{rgb}{0.01,0.01,0.01}
\definecolor{gray20}{rgb}{0.20,0.20,0.20}
\definecolor{gray21}{rgb}{0.21,0.21,0.21}
\definecolor{gray22}{rgb}{0.22,0.22,0.22}
\definecolor{gray23}{rgb}{0.23,0.23,0.23}
\definecolor{gray24}{rgb}{0.24,0.24,0.24}
\definecolor{gray25}{rgb}{0.25,0.25,0.25}
\definecolor{gray26}{rgb}{0.26,0.26,0.26}
\definecolor{gray27}{rgb}{0.27,0.27,0.27}
\definecolor{gray28}{rgb}{0.28,0.28,0.28}
\definecolor{gray29}{rgb}{0.29,0.29,0.29}
\definecolor{gray2}{rgb}{0.02,0.02,0.02}
\definecolor{gray30}{rgb}{0.30,0.30,0.30}
\definecolor{gray31}{rgb}{0.31,0.31,0.31}
\definecolor{gray32}{rgb}{0.32,0.32,0.32}
\definecolor{gray33}{rgb}{0.33,0.33,0.33}
\definecolor{gray34}{rgb}{0.34,0.34,0.34}
\definecolor{gray35}{rgb}{0.35,0.35,0.35}
\definecolor{gray36}{rgb}{0.36,0.36,0.36}
\definecolor{gray37}{rgb}{0.37,0.37,0.37}
\definecolor{gray38}{rgb}{0.38,0.38,0.38}
\definecolor{gray39}{rgb}{0.39,0.39,0.39}
\definecolor{gray3}{rgb}{0.03,0.03,0.03}
\definecolor{gray40}{rgb}{0.40,0.40,0.40}
\definecolor{gray41}{rgb}{0.41,0.41,0.41}
\definecolor{gray42}{rgb}{0.42,0.42,0.42}
\definecolor{gray43}{rgb}{0.43,0.43,0.43}
\definecolor{gray44}{rgb}{0.44,0.44,0.44}
\definecolor{gray45}{rgb}{0.45,0.45,0.45}
\definecolor{gray46}{rgb}{0.46,0.46,0.46}
\definecolor{gray47}{rgb}{0.47,0.47,0.47}
\definecolor{gray48}{rgb}{0.48,0.48,0.48}
\definecolor{gray49}{rgb}{0.49,0.49,0.49}
\definecolor{gray4}{rgb}{0.04,0.04,0.04}
\definecolor{gray50}{rgb}{0.50,0.50,0.50}
\definecolor{gray51}{rgb}{0.51,0.51,0.51}
\definecolor{gray52}{rgb}{0.52,0.52,0.52}
\definecolor{gray53}{rgb}{0.53,0.53,0.53}
\definecolor{gray54}{rgb}{0.54,0.54,0.54}
\definecolor{gray55}{rgb}{0.55,0.55,0.55}
\definecolor{gray56}{rgb}{0.56,0.56,0.56}
\definecolor{gray57}{rgb}{0.57,0.57,0.57}
\definecolor{gray58}{rgb}{0.58,0.58,0.58}
\definecolor{gray59}{rgb}{0.59,0.59,0.59}
\definecolor{gray5}{rgb}{0.05,0.05,0.05}
\definecolor{gray60}{rgb}{0.60,0.60,0.60}
\definecolor{gray61}{rgb}{0.61,0.61,0.61}
\definecolor{gray62}{rgb}{0.62,0.62,0.62}
\definecolor{gray63}{rgb}{0.63,0.63,0.63}
\definecolor{gray64}{rgb}{0.64,0.64,0.64}
\definecolor{gray65}{rgb}{0.65,0.65,0.65}
\definecolor{gray66}{rgb}{0.66,0.66,0.66}
\definecolor{gray67}{rgb}{0.67,0.67,0.67}
\definecolor{gray68}{rgb}{0.68,0.68,0.68}
\definecolor{gray69}{rgb}{0.69,0.69,0.69}
\definecolor{gray6}{rgb}{0.06,0.06,0.06}
\definecolor{gray70}{rgb}{0.70,0.70,0.70}
\definecolor{gray71}{rgb}{0.71,0.71,0.71}
\definecolor{gray72}{rgb}{0.72,0.72,0.72}
\definecolor{gray73}{rgb}{0.73,0.73,0.73}
\definecolor{gray74}{rgb}{0.74,0.74,0.74}
\definecolor{gray75}{rgb}{0.75,0.75,0.75}
\definecolor{gray76}{rgb}{0.76,0.76,0.76}
\definecolor{gray77}{rgb}{0.77,0.77,0.77}
\definecolor{gray78}{rgb}{0.78,0.78,0.78}
\definecolor{gray79}{rgb}{0.79,0.79,0.79}
\definecolor{gray7}{rgb}{0.07,0.07,0.07}
\definecolor{gray80}{rgb}{0.80,0.80,0.80}
\definecolor{gray81}{rgb}{0.81,0.81,0.81}
\definecolor{gray82}{rgb}{0.82,0.82,0.82}
\definecolor{gray83}{rgb}{0.83,0.83,0.83}
\definecolor{gray84}{rgb}{0.84,0.84,0.84}
\definecolor{gray85}{rgb}{0.85,0.85,0.85}
\definecolor{gray86}{rgb}{0.86,0.86,0.86}
\definecolor{gray87}{rgb}{0.87,0.87,0.87}
\definecolor{gray88}{rgb}{0.88,0.88,0.88}
\definecolor{gray89}{rgb}{0.89,0.89,0.89}
\definecolor{gray8}{rgb}{0.08,0.08,0.08}
\definecolor{gray90}{rgb}{0.90,0.90,0.90}
\definecolor{gray91}{rgb}{0.91,0.91,0.91}
\definecolor{gray92}{rgb}{0.92,0.92,0.92}
\definecolor{gray93}{rgb}{0.93,0.93,0.93}
\definecolor{gray94}{rgb}{0.94,0.94,0.94}
\definecolor{gray95}{rgb}{0.95,0.95,0.95}
\definecolor{gray96}{rgb}{0.96,0.96,0.96}
\definecolor{gray97}{rgb}{0.97,0.97,0.97}
\definecolor{gray98}{rgb}{0.98,0.98,0.98}
\definecolor{gray99}{rgb}{0.99,0.99,0.99}
\definecolor{gray9}{rgb}{0.09,0.09,0.09}
\definecolor{gray}{rgb}{0.75,0.75,0.75}
\definecolor{green1}{rgb}{0.00,1.00,0.00}
\definecolor{green2}{rgb}{0.00,0.93,0.00}
\definecolor{green3}{rgb}{0.00,0.80,0.00}
\definecolor{green4}{rgb}{0.00,0.55,0.00}
\definecolor{greenyellow}{rgb}{0.68,1.00,0.18}
\definecolor{green}{rgb}{0.00,1.00,0.00}
\definecolor{grey0}{rgb}{0.00,0.00,0.00}
\definecolor{grey100}{rgb}{1.00,1.00,1.00}
\definecolor{grey10}{rgb}{0.10,0.10,0.10}
\definecolor{grey11}{rgb}{0.11,0.11,0.11}
\definecolor{grey12}{rgb}{0.12,0.12,0.12}
\definecolor{grey13}{rgb}{0.13,0.13,0.13}
\definecolor{grey14}{rgb}{0.14,0.14,0.14}
\definecolor{grey15}{rgb}{0.15,0.15,0.15}
\definecolor{grey16}{rgb}{0.16,0.16,0.16}
\definecolor{grey17}{rgb}{0.17,0.17,0.17}
\definecolor{grey18}{rgb}{0.18,0.18,0.18}
\definecolor{grey19}{rgb}{0.19,0.19,0.19}
\definecolor{grey1}{rgb}{0.01,0.01,0.01}
\definecolor{grey20}{rgb}{0.20,0.20,0.20}
\definecolor{grey21}{rgb}{0.21,0.21,0.21}
\definecolor{grey22}{rgb}{0.22,0.22,0.22}
\definecolor{grey23}{rgb}{0.23,0.23,0.23}
\definecolor{grey24}{rgb}{0.24,0.24,0.24}
\definecolor{grey25}{rgb}{0.25,0.25,0.25}
\definecolor{grey26}{rgb}{0.26,0.26,0.26}
\definecolor{grey27}{rgb}{0.27,0.27,0.27}
\definecolor{grey28}{rgb}{0.28,0.28,0.28}
\definecolor{grey29}{rgb}{0.29,0.29,0.29}
\definecolor{grey2}{rgb}{0.02,0.02,0.02}
\definecolor{grey30}{rgb}{0.30,0.30,0.30}
\definecolor{grey31}{rgb}{0.31,0.31,0.31}
\definecolor{grey32}{rgb}{0.32,0.32,0.32}
\definecolor{grey33}{rgb}{0.33,0.33,0.33}
\definecolor{grey34}{rgb}{0.34,0.34,0.34}
\definecolor{grey35}{rgb}{0.35,0.35,0.35}
\definecolor{grey36}{rgb}{0.36,0.36,0.36}
\definecolor{grey37}{rgb}{0.37,0.37,0.37}
\definecolor{grey38}{rgb}{0.38,0.38,0.38}
\definecolor{grey39}{rgb}{0.39,0.39,0.39}
\definecolor{grey3}{rgb}{0.03,0.03,0.03}
\definecolor{grey40}{rgb}{0.40,0.40,0.40}
\definecolor{grey41}{rgb}{0.41,0.41,0.41}
\definecolor{grey42}{rgb}{0.42,0.42,0.42}
\definecolor{grey43}{rgb}{0.43,0.43,0.43}
\definecolor{grey44}{rgb}{0.44,0.44,0.44}
\definecolor{grey45}{rgb}{0.45,0.45,0.45}
\definecolor{grey46}{rgb}{0.46,0.46,0.46}
\definecolor{grey47}{rgb}{0.47,0.47,0.47}
\definecolor{grey48}{rgb}{0.48,0.48,0.48}
\definecolor{grey49}{rgb}{0.49,0.49,0.49}
\definecolor{grey4}{rgb}{0.04,0.04,0.04}
\definecolor{grey50}{rgb}{0.50,0.50,0.50}
\definecolor{grey51}{rgb}{0.51,0.51,0.51}
\definecolor{grey52}{rgb}{0.52,0.52,0.52}
\definecolor{grey53}{rgb}{0.53,0.53,0.53}
\definecolor{grey54}{rgb}{0.54,0.54,0.54}
\definecolor{grey55}{rgb}{0.55,0.55,0.55}
\definecolor{grey56}{rgb}{0.56,0.56,0.56}
\definecolor{grey57}{rgb}{0.57,0.57,0.57}
\definecolor{grey58}{rgb}{0.58,0.58,0.58}
\definecolor{grey59}{rgb}{0.59,0.59,0.59}
\definecolor{grey5}{rgb}{0.05,0.05,0.05}
\definecolor{grey60}{rgb}{0.60,0.60,0.60}
\definecolor{grey61}{rgb}{0.61,0.61,0.61}
\definecolor{grey62}{rgb}{0.62,0.62,0.62}
\definecolor{grey63}{rgb}{0.63,0.63,0.63}
\definecolor{grey64}{rgb}{0.64,0.64,0.64}
\definecolor{grey65}{rgb}{0.65,0.65,0.65}
\definecolor{grey66}{rgb}{0.66,0.66,0.66}
\definecolor{grey67}{rgb}{0.67,0.67,0.67}
\definecolor{grey68}{rgb}{0.68,0.68,0.68}
\definecolor{grey69}{rgb}{0.69,0.69,0.69}
\definecolor{grey6}{rgb}{0.06,0.06,0.06}
\definecolor{grey70}{rgb}{0.70,0.70,0.70}
\definecolor{grey71}{rgb}{0.71,0.71,0.71}
\definecolor{grey72}{rgb}{0.72,0.72,0.72}
\definecolor{grey73}{rgb}{0.73,0.73,0.73}
\definecolor{grey74}{rgb}{0.74,0.74,0.74}
\definecolor{grey75}{rgb}{0.75,0.75,0.75}
\definecolor{grey76}{rgb}{0.76,0.76,0.76}
\definecolor{grey77}{rgb}{0.77,0.77,0.77}
\definecolor{grey78}{rgb}{0.78,0.78,0.78}
\definecolor{grey79}{rgb}{0.79,0.79,0.79}
\definecolor{grey7}{rgb}{0.07,0.07,0.07}
\definecolor{grey80}{rgb}{0.80,0.80,0.80}
\definecolor{grey81}{rgb}{0.81,0.81,0.81}
\definecolor{grey82}{rgb}{0.82,0.82,0.82}
\definecolor{grey83}{rgb}{0.83,0.83,0.83}
\definecolor{grey84}{rgb}{0.84,0.84,0.84}
\definecolor{grey85}{rgb}{0.85,0.85,0.85}
\definecolor{grey86}{rgb}{0.86,0.86,0.86}
\definecolor{grey87}{rgb}{0.87,0.87,0.87}
\definecolor{grey88}{rgb}{0.88,0.88,0.88}
\definecolor{grey89}{rgb}{0.89,0.89,0.89}
\definecolor{grey8}{rgb}{0.08,0.08,0.08}
\definecolor{grey90}{rgb}{0.90,0.90,0.90}
\definecolor{grey91}{rgb}{0.91,0.91,0.91}
\definecolor{grey92}{rgb}{0.92,0.92,0.92}
\definecolor{grey93}{rgb}{0.93,0.93,0.93}
\definecolor{grey94}{rgb}{0.94,0.94,0.94}
\definecolor{grey95}{rgb}{0.95,0.95,0.95}
\definecolor{grey96}{rgb}{0.96,0.96,0.96}
\definecolor{grey97}{rgb}{0.97,0.97,0.97}
\definecolor{grey98}{rgb}{0.98,0.98,0.98}
\definecolor{grey99}{rgb}{0.99,0.99,0.99}
\definecolor{grey9}{rgb}{0.09,0.09,0.09}
\definecolor{grey}{rgb}{0.75,0.75,0.75}
\definecolor{honeydew1}{rgb}{0.94,1.00,0.94}
\definecolor{honeydew2}{rgb}{0.88,0.93,0.88}
\definecolor{honeydew3}{rgb}{0.76,0.80,0.76}
\definecolor{honeydew4}{rgb}{0.51,0.55,0.51}
\definecolor{honeydew}{rgb}{0.94,1.00,0.94}
\definecolor{hotpink}{rgb}{1.00,0.41,0.71}
\definecolor{indianred}{rgb}{0.80,0.36,0.36}
\definecolor{ivory1}{rgb}{1.00,1.00,0.94}
\definecolor{ivory2}{rgb}{0.93,0.93,0.88}
\definecolor{ivory3}{rgb}{0.80,0.80,0.76}
\definecolor{ivory4}{rgb}{0.55,0.55,0.51}
\definecolor{ivory}{rgb}{1.00,1.00,0.94}
\definecolor{khaki1}{rgb}{1.00,0.96,0.56}
\definecolor{khaki2}{rgb}{0.93,0.90,0.52}
\definecolor{khaki3}{rgb}{0.80,0.78,0.45}
\definecolor{khaki4}{rgb}{0.55,0.53,0.31}
\definecolor{khaki}{rgb}{0.94,0.90,0.55}
\definecolor{lavenderblush}{rgb}{1.00,0.94,0.96}
\definecolor{lavender}{rgb}{0.90,0.90,0.98}
\definecolor{lawngreen}{rgb}{0.49,0.99,0.00}
\definecolor{lemonchiffon}{rgb}{1.00,0.98,0.80}
\definecolor{lightblue}{rgb}{0.68,0.85,0.90}
\definecolor{lightcoral}{rgb}{0.94,0.50,0.50}
\definecolor{lightcyan}{rgb}{0.88,1.00,1.00}
\definecolor{lightgoldenrod}{rgb}{0.93,0.87,0.51}
\definecolor{lightgoldenrod}{rgb}{0.98,0.98,0.82}
\definecolor{lightgray}{rgb}{0.83,0.83,0.83}
\definecolor{lightgreen}{rgb}{0.56,0.93,0.56}
\definecolor{lightgrey}{rgb}{0.83,0.83,0.83}
\definecolor{lightpink}{rgb}{1.00,0.71,0.76}
\definecolor{lightsalmon}{rgb}{1.00,0.63,0.48}
\definecolor{lightsea}{rgb}{0.13,0.70,0.67}
\definecolor{lightsky}{rgb}{0.53,0.81,0.98}
\definecolor{lightslate}{rgb}{0.47,0.53,0.60}
\definecolor{lightslate}{rgb}{0.47,0.53,0.60}
\definecolor{lightslate}{rgb}{0.52,0.44,1.00}
\definecolor{lightsteel}{rgb}{0.69,0.77,0.87}
\definecolor{lightyellow}{rgb}{1.00,1.00,0.88}
\definecolor{limegreen}{rgb}{0.20,0.80,0.20}
\definecolor{linen}{rgb}{0.98,0.94,0.90}
\definecolor{magenta1}{rgb}{1.00,0.00,1.00}
\definecolor{magenta2}{rgb}{0.93,0.00,0.93}
\definecolor{magenta3}{rgb}{0.80,0.00,0.80}
\definecolor{magenta4}{rgb}{0.55,0.00,0.55}
\definecolor{magenta}{rgb}{1.00,0.00,1.00}
\definecolor{maroon1}{rgb}{1.00,0.20,0.70}
\definecolor{maroon2}{rgb}{0.93,0.19,0.65}
\definecolor{maroon3}{rgb}{0.80,0.16,0.56}
\definecolor{maroon4}{rgb}{0.55,0.11,0.38}
\definecolor{maroon}{rgb}{0.69,0.19,0.38}
\definecolor{mediumaquamarine}{rgb}{0.40,0.80,0.67}
\definecolor{mediumblue}{rgb}{0.00,0.00,0.80}
\definecolor{mediumorchid}{rgb}{0.73,0.33,0.83}
\definecolor{mediumpurple}{rgb}{0.58,0.44,0.86}
\definecolor{mediumsea}{rgb}{0.24,0.70,0.44}
\definecolor{mediumslate}{rgb}{0.48,0.41,0.93}
\definecolor{mediumspring}{rgb}{0.00,0.98,0.60}
\definecolor{mediumturquoise}{rgb}{0.28,0.82,0.80}
\definecolor{mediumviolet}{rgb}{0.78,0.08,0.52}
\definecolor{midnightblue}{rgb}{0.10,0.10,0.44}
\definecolor{mintcream}{rgb}{0.96,1.00,0.98}
\definecolor{mistyrose}{rgb}{1.00,0.89,0.88}
\definecolor{moccasin}{rgb}{1.00,0.89,0.71}
\definecolor{navajowhite}{rgb}{1.00,0.87,0.68}
\definecolor{navyblue}{rgb}{0.00,0.00,0.50}
\definecolor{navy}{rgb}{0.00,0.00,0.50}
\definecolor{oldlace}{rgb}{0.99,0.96,0.90}
\definecolor{olivedrab}{rgb}{0.42,0.56,0.14}
\definecolor{orange1}{rgb}{1.00,0.65,0.00}
\definecolor{orange2}{rgb}{0.93,0.60,0.00}
\definecolor{orange3}{rgb}{0.80,0.52,0.00}
\definecolor{orange4}{rgb}{0.55,0.35,0.00}
\definecolor{orangered}{rgb}{1.00,0.27,0.00}
\definecolor{orange}{rgb}{1.00,0.65,0.00}
\definecolor{orchid1}{rgb}{1.00,0.51,0.98}
\definecolor{orchid2}{rgb}{0.93,0.48,0.91}
\definecolor{orchid3}{rgb}{0.80,0.41,0.79}
\definecolor{orchid4}{rgb}{0.55,0.28,0.54}
\definecolor{orchid}{rgb}{0.85,0.44,0.84}
\definecolor{palegoldenrod}{rgb}{0.93,0.91,0.67}
\definecolor{palegreen}{rgb}{0.60,0.98,0.60}
\definecolor{paleturquoise}{rgb}{0.69,0.93,0.93}
\definecolor{paleviolet}{rgb}{0.86,0.44,0.58}
\definecolor{papayawhip}{rgb}{1.00,0.94,0.84}
\definecolor{peachpuff}{rgb}{1.00,0.85,0.73}
\definecolor{peru}{rgb}{0.80,0.52,0.25}
\definecolor{pink1}{rgb}{1.00,0.71,0.77}
\definecolor{pink2}{rgb}{0.93,0.66,0.72}
\definecolor{pink3}{rgb}{0.80,0.57,0.62}
\definecolor{pink4}{rgb}{0.55,0.39,0.42}
\definecolor{pink}{rgb}{1.00,0.75,0.80}
\definecolor{plum1}{rgb}{1.00,0.73,1.00}
\definecolor{plum2}{rgb}{0.93,0.68,0.93}
\definecolor{plum3}{rgb}{0.80,0.59,0.80}
\definecolor{plum4}{rgb}{0.55,0.40,0.55}
\definecolor{plum}{rgb}{0.87,0.63,0.87}
\definecolor{powderblue}{rgb}{0.69,0.88,0.90}
\definecolor{purple1}{rgb}{0.61,0.19,1.00}
\definecolor{purple2}{rgb}{0.57,0.17,0.93}
\definecolor{purple3}{rgb}{0.49,0.15,0.80}
\definecolor{purple4}{rgb}{0.33,0.10,0.55}
\definecolor{purple}{rgb}{0.63,0.13,0.94}
\definecolor{red1}{rgb}{1.00,0.00,0.00}
\definecolor{red2}{rgb}{0.93,0.00,0.00}
\definecolor{red3}{rgb}{0.80,0.00,0.00}
\definecolor{red4}{rgb}{0.55,0.00,0.00}
\definecolor{red}{rgb}{1.00,0.00,0.00}
\definecolor{rosybrown}{rgb}{0.74,0.56,0.56}
\definecolor{royalblue}{rgb}{0.25,0.41,0.88}
\definecolor{saddlebrown}{rgb}{0.55,0.27,0.07}
\definecolor{salmon1}{rgb}{1.00,0.55,0.41}
\definecolor{salmon2}{rgb}{0.93,0.51,0.38}
\definecolor{salmon3}{rgb}{0.80,0.44,0.33}
\definecolor{salmon4}{rgb}{0.55,0.30,0.22}
\definecolor{salmon}{rgb}{0.98,0.50,0.45}
\definecolor{sandybrown}{rgb}{0.96,0.64,0.38}
\definecolor{seagreen}{rgb}{0.18,0.55,0.34}
\definecolor{seashell1}{rgb}{1.00,0.96,0.93}
\definecolor{seashell2}{rgb}{0.93,0.90,0.87}
\definecolor{seashell3}{rgb}{0.80,0.77,0.75}
\definecolor{seashell4}{rgb}{0.55,0.53,0.51}
\definecolor{seashell}{rgb}{1.00,0.96,0.93}
\definecolor{sienna1}{rgb}{1.00,0.51,0.28}
\definecolor{sienna2}{rgb}{0.93,0.47,0.26}
\definecolor{sienna3}{rgb}{0.80,0.41,0.22}
\definecolor{sienna4}{rgb}{0.55,0.28,0.15}
\definecolor{sienna}{rgb}{0.63,0.32,0.18}
\definecolor{skyblue}{rgb}{0.53,0.81,0.92}
\definecolor{slateblue}{rgb}{0.42,0.35,0.80}
\definecolor{slategray}{rgb}{0.44,0.50,0.56}
\definecolor{slategrey}{rgb}{0.44,0.50,0.56}
\definecolor{snow1}{rgb}{1.00,0.98,0.98}
\definecolor{snow2}{rgb}{0.93,0.91,0.91}
\definecolor{snow3}{rgb}{0.80,0.79,0.79}
\definecolor{snow4}{rgb}{0.55,0.54,0.54}
\definecolor{snow}{rgb}{1.00,0.98,0.98}
\definecolor{springgreen}{rgb}{0.00,1.00,0.50}
\definecolor{steelblue}{rgb}{0.27,0.51,0.71}
\definecolor{tan1}{rgb}{1.00,0.65,0.31}
\definecolor{tan2}{rgb}{0.93,0.60,0.29}
\definecolor{tan3}{rgb}{0.80,0.52,0.25}
\definecolor{tan4}{rgb}{0.55,0.35,0.17}
\definecolor{tan}{rgb}{0.82,0.71,0.55}
\definecolor{thistle1}{rgb}{1.00,0.88,1.00}
\definecolor{thistle2}{rgb}{0.93,0.82,0.93}
\definecolor{thistle3}{rgb}{0.80,0.71,0.80}
\definecolor{thistle4}{rgb}{0.55,0.48,0.55}
\definecolor{thistle}{rgb}{0.85,0.75,0.85}
\definecolor{tomato1}{rgb}{1.00,0.39,0.28}
\definecolor{tomato2}{rgb}{0.93,0.36,0.26}
\definecolor{tomato3}{rgb}{0.80,0.31,0.22}
\definecolor{tomato4}{rgb}{0.55,0.21,0.15}
\definecolor{tomato}{rgb}{1.00,0.39,0.28}
\definecolor{turquoise1}{rgb}{0.00,0.96,1.00}
\definecolor{turquoise2}{rgb}{0.00,0.90,0.93}
\definecolor{turquoise3}{rgb}{0.00,0.77,0.80}
\definecolor{turquoise4}{rgb}{0.00,0.53,0.55}
\definecolor{turquoise}{rgb}{0.25,0.88,0.82}
\definecolor{violetred}{rgb}{0.82,0.13,0.56}
\definecolor{violet}{rgb}{0.93,0.51,0.93}
\definecolor{wheat1}{rgb}{1.00,0.91,0.73}
\definecolor{wheat2}{rgb}{0.93,0.85,0.68}
\definecolor{wheat3}{rgb}{0.80,0.73,0.59}
\definecolor{wheat4}{rgb}{0.55,0.49,0.40}
\definecolor{wheat}{rgb}{0.96,0.87,0.70}
\definecolor{whitesmoke}{rgb}{0.96,0.96,0.96}
\definecolor{white}{rgb}{1.00,1.00,1.00}
\definecolor{yellow1}{rgb}{1.00,1.00,0.00}
\definecolor{yellow2}{rgb}{0.93,0.93,0.00}
\definecolor{yellow3}{rgb}{0.80,0.80,0.00}
\definecolor{yellow4}{rgb}{0.55,0.55,0.00}
\definecolor{yellowgreen}{rgb}{0.60,0.80,0.20}
\definecolor{yellow}{rgb}{1.00,1.00,0.00}

\usepackage{ulem}
\definecolor{DarkGreen}{rgb}{0.00,0.39,0.00}
\definecolor{TurquoiseLight}{RGB}{20,180,130}

\definecolor{rev01}{RGB}{0,0,0}
\definecolor{rev02}{RGB}{0,0,0}
\definecolor{rev03}{RGB}{0,0,0}
\definecolor{rev}{RGB}{0,0,0}

\usepackage[color=TurquoiseLight,colorinlistoftodos]{todonotes}

\usepackage{soul}

\usepackage{amsmath,amsfonts}
\usepackage{amssymb}

\newcommand{\IBMmath}[0]{\textit{I\kern-1pt B\kern-1pt M}}
 
\newcommand\vf{\ensuremath{{\fontfamily{lmr}\selectfont\textit{v}}_f}}

\title{The effect of spanwise heterogeneous \textcolor{rev01}{surfaces} on mixed convection in turbulent channels}
\author{Kay Sch\"afer\aff{1}, Bettina Frohnapfel\aff{1} \and Juan Pedro Mellado\aff{2}}
\affiliation{\aff{1} Institute of Fluid Mechanics, Karlsruhe Institute of Technology, Karlsruhe, Germany
\aff{2} Meteorological Institute, University of Hamburg, Hamburg, Germany
}
\begin{document}

\maketitle

\begin{abstract}
Turbulent mixed convection in channel flows with heterogeneous \textcolor{rev01}{surfaces} is studied using direct numerical simulations.
The relative importance between buoyancy and shear effects, characterized by the bulk Richardson number $Ri_b$, is varied in order to cover the flow regimes of forced, mixed and natural convection, \textcolor{rev}{which are associated with different large-scale flow organization}.
The heterogeneous \textcolor{rev01}{surface} consists of streamwise-aligned ridges, which are known to induce secondary motion in case of forced convection. 
\textcolor{rev}{The large-scale streamwise rolls emerging under smooth-wall mixed convection conditions are significantly affected by the heterogeneous surfaces and their appearance is considerably reduced for dense ridge spacings. 
It is found that the formation of these rolls requires larger buoyancy forces than over smooth walls due to the additional drag induced by the ridges. 
Therefore, the transition from forced convection structures to rolls is delayed towards larger $Ri_b$ for spanwise heterogeneous surfaces.
}
\textcolor{rev}{The influence of the heterogeneous surface on the flow organization of mixed convection is particularly pronounced in the roll-to-cell transition range, where ridges favor the transition to convective cells at significantly lower $Ri_b$.
In addition, the convective cells are observed to align perpendicular to the ridges with decreasing ridge spacing.
We attribute this reorganization to the fact that
flow parallel to the ridges experience less drag than flow across the ridges, which is energetically more beneficial.
}
\textcolor{rev}{Furthermore, we find} that streamwise rolls exhibit very slow dynamics for $Ri_b=1$ and $Ri_b=3.2$ \textcolor{rev}{when the ridge spacing is in the order of the rolls' width}. 
For these cases the up- and downdrafts of the rolls move slowly across
the entire channel instead of being fixed in space \textcolor{rev}{as observed for the smooth-wall cases.}
\end{abstract}

\begin{keywords}
mixed convection, turbulent boundary layers, \textcolor{rev01}{heterogeneous surfaces}, secondary motions
\end{keywords}

\section{Introduction}
In many technical and environmental flows, shear and buoyancy effects occur simultaneously and both may significantly contribute to the heat and mass transfer of the flow. 
This transport process, also known as mixed convection, represents the combination of the two extreme cases of forced and natural convection, where the former is driven purely by a pressure gradient and the latter by vertical temperature differences. 
While in forced convection turbulence is produced by shear and in natural convection by buoyancy, both effects contribute to turbulence production in mixed convection flows.
The aforementioned flows often occur over rough surfaces, which can also strongly influence the flow properties. 
In the forced convection community, heterogeneous rough surfaces have received special attention in recent years, since in turbulent wall-bounded flows they can
induce secondary flows of Prandtl's second kind 
that can significantly change the transport properties of the flow. 
It is therefore the objective of the present study to investigate the effects of heterogeneous \textcolor{rev01}{surfaces} on the interplay of forced and natural convection in turbulent flows. 

\begin{figure}
    \centering
    \includegraphics[scale=0.95]{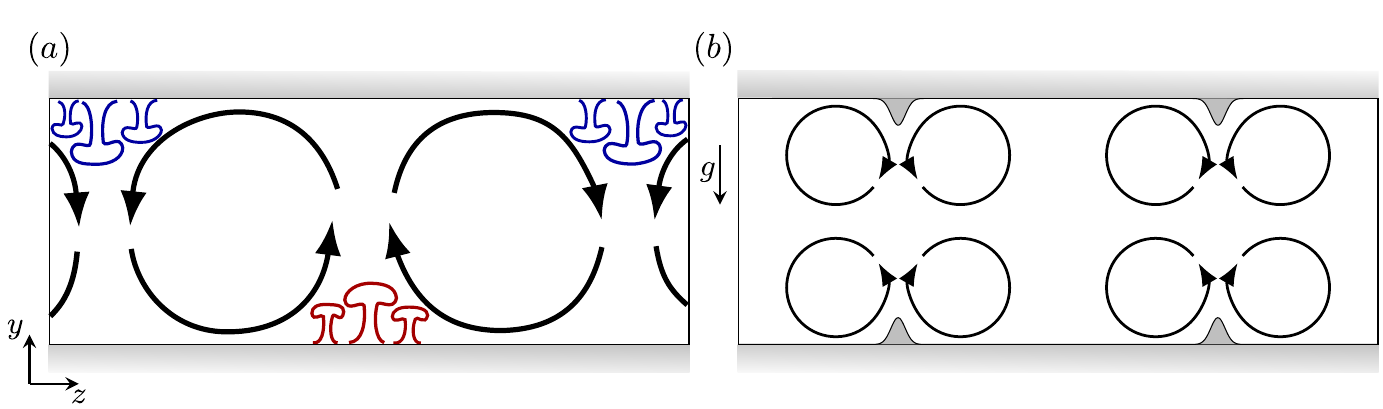}
    \caption{Schematic of different large-scale structures in the cross-sectional plane.
    In $(a)$ streamwise rolls can emerge in flows with buoyancy effects over smooth walls, while in $(b)$ secondary motions appear over rough walls in form of streamwise-aligned ridges. The direction of gravitational acceleration $g$ is given by the downward arrow.}
    \label{fig:sketch_structures}
\end{figure}

In mixed convection flows with unstable thermal stratification, where denser fluid is bedded over less dense fluid, different turbulent large-scale flow structures have been identified and the organisation depends on the relative strength between shear and buoyancy effects, such as in turbulent channel flows \citep{Pirozzoli_mixed_2017}, turbulent Couette flows \citep{blass_flow_2020} and in the atmospheric boundary layer (ABL) \citep{deardorff_numerical_1972,lemone_structure_1973,moeng_comparison_1994,khanna_ABL_1998}. 
The transition between different flow organisation is accompanied by an alteration of the effective heat and momentum transfer in the flow.
In case of strong buoyancy effects and weak shear, open cells form in the flow, which resemble Rayleigh-Bénard convection, while in the case of
weak to moderate buoyancy effects and strong shear, the flow organizes into horizontal rolls aligned with the main flow direction \citep{khanna_ABL_1998,Salesky_nature_2017,Pirozzoli_mixed_2017}.
These streamwise rolls are illustrated in the cross-sectional plane of the channel in figure \ref{fig:sketch_structures} $(a)$, with the up- and downdraft region of the roll occurring where localized buoyancy forces accumulate. 
At very strong shear, with negligible buoyancy effects, the flow organisation in turbulent channels resembles the one of classical Poiseuille flows \citep{Pirozzoli_mixed_2017} and in case of neutral ABL that of flat-plate boundary layers \citep{khanna_ABL_1998}.

The transition between the different flow topologies of mixed convection can be characterised by various stability parameters. 
In the atmospheric science community, the stability parameter $-z_i/L$ is used to classify the transitions between rolls and cellular structures, which expresses the ratio of the boundary layer depth $z_i$ and the Obukhov length $L$. 
The Monin-Obukhov similarity theory introduces the Obukhov length $L$ as a length scale that compares the effects of friction and buoyancy on the flow.  
$L$ is used for non-dimensionalisation to generalise Prandtl's mixing length theory for buoyancy effects of non-neutral stratifications \citep{obukhov_turbulence_1946,monin_basic_laws_1954,wyngaard_turbulence_2010}.
In this context, $|L|$ can be physically interpreted as the height at which the buoyant production of turbulent kinetic energy equals the one due to shear production, such that below $|L|$ mechanical production predominates, while above $|L|$ buoyant production is dominant.
In the case of convective boundary layers, large ratios of $-z_i/L$ are indicative for the formation of convective cells, while small values are typical for roll formation \citep{khanna_ABL_1998,Salesky_nature_2017}.
An alternative stability parameter to characterize the relative importance of buoyancy effects and shear is the Richardson number $Ri$. 
For mixed convection in turbulent channel flows, low $Ri$ correspond to pure forced convection, intermediate values of $Ri$ to roll formation and large $Ri$ values to natural convection with cell-like structures \citep{Pirozzoli_mixed_2017}.
Independent of the chosen stability parameter the exact range at which the transition between the different flow regimes occur is still under debate with recent studies focusing on the transition between rolls and cells \citep{Salesky_nature_2017} and the transition between neutral to moderately convective conditions \citep{jayaraman_transition_2021}.

Effects of surface heterogeneity, occurring for instance between urban-rural areas, on the transition between the different flow organisation in mixed convection has gained little attention, even though their impact for the quality of weather and climate prediction is important \citep{Bou-zeid_persistent_2020}.
Likewise, there is the chance that the flow topology may have an influence on the deformation of the surface topography, which in turn affects the flow and possibly results in a positive feedback.
For example, from satellite observations of deserts in the 1960s, it was proposed that quasi-streamwise rolls in the atmospheric boundary layer were responsible for the formation of sand dunes aligned with the mean flow direction \citep{hanna_formation_1969,shao_physics_2008}.
This is similar to observations from aqueous open-channel flows, where sand dunes can also form due to a positive feedback with large-scale flow structures. 
The formation of these large-scale flow structures is associated with irregularities in the surface topography, which leads to the formation of so-called secondary motions \citep{colombini_turbulence_driven_1993,scherer_role_2022}, which, unlike the streamwise rolls, can occur without buoyancy effects. 

These secondary motions, also named as secondary motions of Prandtl's second kind \citep{hinze_experimental_1973,anderson_numerical_2015}, 
can occur in turbulent flows over surfaces with spanwise variations in the wall properties, which can significantly alter the momentum and heat transport of turbulent flows \citep{stroh_secondary_2020}. 
Unlike streamwise rolls, secondary motions cannot be observed in instantaneous velocity fields, instead they occur only in time-averaged velocity fields, where they appear as counter-rotating vortices aligned with the main flow direction.
The secondary motions introduce a spanwise variation in the mean velocity field, which for instance in a turbulent boundary layer leads to spanwise modulation of the boundary layer thickness \citep{willingham_turbulent_2014}.  
These differences in the mean profile are associated with the formation of low- and high-momentum pathways in the mean velocity field, usually occurring at the updraft and downdraft regions of the secondary motions, respectively \citep{barros_observations_2014, willingham_turbulent_2014}.

The spanwise heterogeneous surfaces, triggering secondary motions, are roughly distinguished in two main surface types, namely in ridge-type and strip-type roughness \citep{Wang2006}.
Strip-type roughness is characterized by significant spanwise differences in drag, for example alternating streamwise strips with different wall-shear stress conditions \citep{willingham_turbulent_2014,anderson_numerical_2015,Chung_similartiy_2018} or alternating strips of smooth and rough walls, which do not feature large differences in the wall elevation \citep{hinze_experimental_1973,Wangsawijaya_effect_2020,stroh_rearrangement_2020,schafer_modelling_2022}.
Ridge-type roughness comprise notable spanwise wall elevation differences, such as streamwise-aligned ridges, which are studied numerically and experimentally for a wide range of different ridge shapes in turbulent boundary layers \citep{vanderwel_effects_2015,hwang_secondary_2018,medjnoun_characteristics_2018,Vanderwel2019,medjnoun_effects_2020}, turbulent channel flows \citep{Vanderwel2019,stroh_secondary_2020} and turbulent open-channel flows \citep{awasthi_numerical_2018,zampiron_secondary_2020}.

The ridge-type behaviour of secondary motions is characterized by updraft regions occurring above the center ridge position, while downdrafts descend in the valley region between adjacent ridges, which is illustrated in figure \ref{fig:sketch_structures} $(b)$. 
The spanwise distance between the ridges influences the spatial extent of the secondary motion  \citep{vanderwel_effects_2015,Vanderwel2019}. 
For large ridge spacings, which are of the order of several boundary layer thickness the flow is divided into regions disturbed by the secondary motion in the vicinity of the ridge and a homogeneous and unaffected region far from the ridges. 
In case of ridge spacings of the order of the boundary layer thickness, the secondary motion affects the entire flow region and introduces a spanwise heterogeneity of the mean flow \citep{hwang_secondary_2018,medjnoun_effects_2020}.
Further reduction of the ridge spacing leads to a decreasing spatial extent of the secondary motions, which scales with the ridge spacing.


\textcolor{rev03}{
The objective of the present study is to investigate the influence of spanwise heterogeneous surfaces on turbulent mixed convection flows as well as their influence on the transition between the different large-scale flow organization. 
Secondary motions feature in a mean sense some similarities with the roll motion of mixed convection, e.g. large-scale counter-rotating vortices as depicted in figure \ref{fig:sketch_structures}. Since secondary motions can be induced by streamwise-aligned ridges, the study investigates to which extend such surface structures can influence the formation and dynamics of streamwise rolls.
These questions are investigated using a simplified setting of a turbulent channel flow, similar to \cite{Pirozzoli_mixed_2017}, augmented with streamwise-aligned ridges on the walls. 
The different flow regimes of mixed convection are covered by systematically varying the relative strength of shear and buoyancy effects.
}

\section{Methodology}
\subsection{\textcolor{rev}{Flow configuration and numerical procedure}}
\begin{figure}
    \centering
    \includegraphics[]{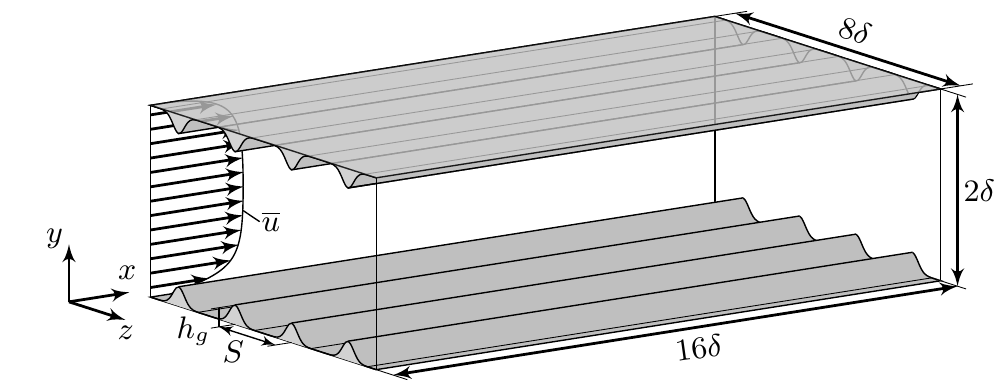}
    \caption{Sketch of the numerical channel domain  with streamwise-aligned Gaussian ridges at the walls.}
    \label{fig:sketch_Gaussian_ridges}
\end{figure}

The balance equations for mass, momentum and energy are considered in the Boussinesq approximation:
\begin{align}
    \frac{\partial u_i}{\partial x_i} &= 0, \\
    \frac{\partial u_i}{\partial t} + \frac{\partial u_i u_j}{\partial x_j}
        &= - \frac{1}{\rho_0}\frac{\partial p}{\partial x_i}
          + \nu \frac{\partial^2 u_i}{\partial x_j \partial x_j }
          + \beta g T \delta_{i2} 
          + \Pi \delta_{i1} 
          + F_{\IBMmath,i}, \\
    \frac{\partial T}{\partial t} + \frac{\partial T  u_j}{\partial x_j}
    &= \alpha \frac{\partial^2 T}{\partial x_j \partial x_j} + Q_\IBMmath,
\end{align}           
where $u_i$ and $x_i$ are the components of the velocity and position vector, with indices $1,2,3$ corresponding to the streamwise, wall-normal and spanwise direction.
As an index-free notation, the velocity components are also written as $(u,v,w) = (u_1,u_2,u_3)$.
The reference density is given by $\rho_0$, $p$ is the pressure, $\nu$ the kinematic viscosity, $\beta$ is the thermal expansion coefficient, $g$ the gravitational acceleration, $T$ the  temperature and $\alpha$ the temperature diffusivity. 
A constant volume flow rate is maintained by the forcing term $\Pi = u^2_\tau/\delta$, where $u_\tau$ is the friction 
velocity and $\delta$ the half-channel height.
In order to represent structured surfaces an external volume forcing term $F_{\IBMmath,i}$ is introduced by an immersed boundary method (IBM).
In the temperature equation the source term $Q_\IBMmath$ represents the forcing term to maintain a constant temperature at the surface.
\textcolor{rev}{
The structured surface of the present flow configuration consists of streamwise-aligned ridges on both walls shown in figure \ref{fig:sketch_Gaussian_ridges}.
}
Due to the presence of these 
ridges, the half-channel height $\delta$ reduces to the effective half-channel height $\delta_\textit{eff} = \delta - \langle \overline{h}_\textit{s} \rangle$, where $\langle \overline{h}_\textit{s} \rangle$ is the horizontally averaged mean height of the surface elevation of one wall.

In case of spanwise heterogeneous surfaces, where the spanwise direction is no longer statistically homogeneous, it is a common choice to decompose the velocity field into its 
temporal and streamwise mean $\overline{u}_i$ and related random fluctuation $u_i''$ given by 
\begin{equation}
    u_i (x,y,z,t) = \overline{u}_i (y,z) + u_i''(x,y,z,t).
\end{equation}
The mean velocity can be further decomposed into its global mean $\langle\overline{u}_i\rangle$, obtained through additional spanwise averaging 
indicated by angular brackets $\langle\cdot\rangle$, 
and its coherent or dispersive component $\tilde{u}_i$, which is the spatial fluctuation about $\langle\overline{u}_i\rangle$, and is defined as 
\textcolor{rev}{
\begin{equation}
    \overline{u}_i (y,z) = \langle\overline{u}_i\rangle(y) + \tilde{u}_i (y,z).
\end{equation}
}
The sum of coherent and random fluctuations is defined as fluctuation $u_i'$ such that
\begin{align}
    u_i (x,y,z,t) &= \langle\overline{u}_i\rangle(y) + \tilde{u}_i (y,z) + u_i''(x,y,z,t) \label{equ:decomp01}\\
                  &= \langle\overline{u}_i\rangle(y) + u_i'(x,y,z,t). \label{equ:decomp02}
\end{align}
In this study the spatial averages are based on intrinsic averaging. \textcolor{rev01}{This procedure excludes the values at the grid points inside the immersed (solid) body, while the values on the surface are included in the integration. 
In consequence, the average is computed through normalization with the fluid area only.
This affects the evaluation of}
global quantities which are integrated in time and all three spatial directions \textcolor{rev}{and are defined as} 
\begin{equation}
    \Phi= \frac{1}{2\delta_\textit{eff} \,\, L_z} \int_0^{L_z} \int_{y_b(z)}^{y_t(z)} \bar{\phi} \,\,\mathrm{d}y\, \mathrm{d}z,
    \label{equ:global_average}
\end{equation}
where $\phi$ represents an arbitrary quantity \textcolor{rev}{and $\Phi$ is its volume- and time-average, while $y_b(z)$ and $y_t(z)$ are the wall-normal surface elevation at the bottom and top wall, respectively.} 
Note, that this definition applies intrinsic averaging through its spanwise dependent wall-normal integration borders.

The considered flow is characterized by three non-dimensional numbers, namely the Prandtl number $Pr$ \textcolor{rev02}{which is the ratio of momentum and thermal diffusivity}, the bulk Reynolds number $Re_b$ \textcolor{rev02}{describing the ratio of inertial and viscous effects}, and the Rayleigh number $Ra$ \textcolor{rev02}{characterizing the ratio of buoyant and viscous effects}, and their 
definitions are given by 
\begin{gather} \stepcounter{equation}
    Pr = \frac{\nu}{\alpha},
    \qquad
    Re_b = \frac{u_{b}\,\, \delta_\textit{eff}}{\nu},
    \qquad
    Ra = \frac{(2\,\delta_\textit{eff})^3 \beta g \Delta T }{\alpha \nu}.
    \tag{\theequation\,a,b,c}
\end{gather}
The bulk velocity is defined as $u_b= 1/(2\delta_\textit{eff} \,\, L_z) \int_0^{L_z} \int_{y_b(z)}^{y_t(z)} \bar{u} \,\mathrm{d}y \mathrm{d}z$. 
$\Delta T$ is the imposed and constant temperature difference between 
the bottom and top wall surface $\Delta T = T_b - T_t$.
The Prandtl number is set to $Pr = 1$ for all considered cases. 
\textcolor{rev}{Following the work by \cite{Pirozzoli_mixed_2017} the bulk Richardson number is defined as}
\begin{equation}
Ri_b = \frac{2\delta_\textit{eff}\; \beta g \Delta T}{u_{b}^2} 
     = \frac{Ra}{4Re_b^2 Pr},
\end{equation}
\textcolor{rev02}{
to characterize the relative importance of buoyant and inertial effects. 
Please note, that positive values of $Ri_b$ indicate convectively unstable conditions, while in atmospheric flows positive values commonly indicate convectively stable conditions \citep{wyngaard_turbulence_2010}.
Another quantity widely used in the atmospheric boundary layer community to categorize the flow is the ratio of the boundary layer depth $z_i$ and the Obukhov length $L$, which is known as stability parameter $-z_i/L$ \citep{wyngaard_turbulence_2010}.
For the present channel flow configuration with ridges this translates into the ratio of $\delta_\textit{eff}$ and $L$, with $L = - u_{\tau}^3 / (\kappa\, Q \, \beta g )$, where $\kappa$ is the von K\'arm\'an constant with $\kappa=0.4$ \citep{wyngaard_turbulence_2010}, while $u_\tau$ and $Q$ are the friction velocity and the vertical heat flux, respectively (both defined in the next paragraph).
}

\textcolor{rev02}{The drag exerted on the flow is quantified by the skin friction coefficient $C_f$ and friction Reynolds number $Re_\tau$, which are defined as $C_f =2 u_{\tau}^2/u_{b}^2$ and $Re_\tau = u_{\tau}\, \delta_\textit{eff} / \nu$, with 
the friction velocity $u_{\tau} = (\tau_{w}/\rho)^{0.5}$ and the wall shear stress $\tau_{w}$.
The wall shear stress is determined by extrapolating the total shear stress from the bulk region ($0.5 \le y/\delta \le 1.5$) to the virtual wall location $y_0 = \delta - \delta_\textit{eff}$ \citep{chan-braun_force_2011}.
The heat transfer of the flow is characterized by the Nusselt number $\mathit{Nu} = 2\delta_\textit{eff}\, Q / (\alpha \Delta T)$, with the vertical heat flux $Q$ determined by evaluating the time and horizontally averaged temperature transport equation at the half-channel height $Q = \langle\overline{v'T'}\rangle|_\delta - \alpha\, \partial{\langle\bar{T}\rangle}/\partial{y}|_\delta$ \citep{stroh_secondary_2020}.
}

\textcolor{rev02}{
The turbulence level is quantified in this study  by the Reynolds number $Re_k = \sqrt{K}\delta_\textit{eff}/\nu$, 
where $K$ corresponds to the time- and volume-averaged turbulent kinetic energy. This quantity is computed by applying the averaging procedure given in equation \ref{equ:global_average} $ k = 0.5 \cdot \overline{u_i' u_i'}$.
}
A characteristic velocity scale for natural convection is the free-fall velocity $\vf = (2 \delta_\textit{eff}\; \beta g \Delta T)^{1/2}$
and together with the effective channel height the free fall time $t_f$ can be 
defined $t_f = 2\delta_\textit{eff} / \vf = ( 2\delta_\textit{eff} / (\beta g \Delta T) )^{1/2}$.
The time scale characterizing the forced convection processes is the bulk time unit $t_b = \delta_\textit{eff} / u_b$. 
From the given definitions, the ratio of $t_b$ and $t_f$ results in the following relationship $t_b/t_f = \sqrt{Ri_b}/2$.

The spanwise height distribution of \textcolor{rev}{the streamwise-aligned ridges follows a Gaussian distribution for each individual ridge, which}
is defined by 
\begin{equation}
    h_\textit{Gauss}(z) = \sum_{i=1}^{n_g} h_{g} \, \exp(- (z-z_{c,i})^2/(2 \sigma^2)),
\end{equation}
where $n_g$ is the total number of Gaussian ridges at one wall, $h_g$ is the maximum height of a single Gaussian ridge, $z_{c,i}$ is the spanwise 
center position of each individual ridge, given by $z_{c,i} = S \, (i + 0.5)$ with $S$ as the spanwise spacing between two Gaussian ridges. 
The parameter $\sigma$ represents the spanwise extent of an individual Gaussian ridge.
In this study the parameters of the Gaussian ridges is set to $h_g = 0.1\delta$ and $\sigma=0.05\delta$. 
The cross-sectional area occupied by a single ridge is given by $A_\textit{Gauss} = \sqrt{2\pi}\, h_{g} \sigma$.
Thus, the effective half-channel height is given by 
$\delta_\textit{eff} = \delta - \delta_\textit{melt}$ with the melt-down height $\delta_\textit{melt} = n_g A_\textit{Gauss} / L_z$.

The governing equations are numerically solved using direct numerical simulations (DNS) by means of the open-source code \text{Xcompact3d} \citep{Laizet_high-order_2009,Bartholomew_xcompact3d_2020} based on compact finite differences of 6th order and a 3rd order Runge-Kutta time integration scheme.
The representation of structured surfaces is achieved using an immersed boundary method based on polynomial reconstruction of the 
velocity and temperature field inside the solid region of the ridges \citep{Gautier_dns_2014}. 
The existing code was extended for the simulation of buoyancy effects and the code was validated with the 
data base for mixed convection and Rayleigh-B\'enard flows of \cite{Pirozzoli_mixed_2017} as documented in appendix \ref{app:validation_code}.
The simulations were performed on a domain size $L_x \times L_y \times L_z =  16\delta \times 2\delta \times 8\delta$, which is in agreement with \cite{Pirozzoli_mixed_2017}, 
who reported for this domain size insensitivity of the mean velocity and temperature profiles.
\subsection{Cases}

\begin{figure}
    \centering
    \includegraphics[]{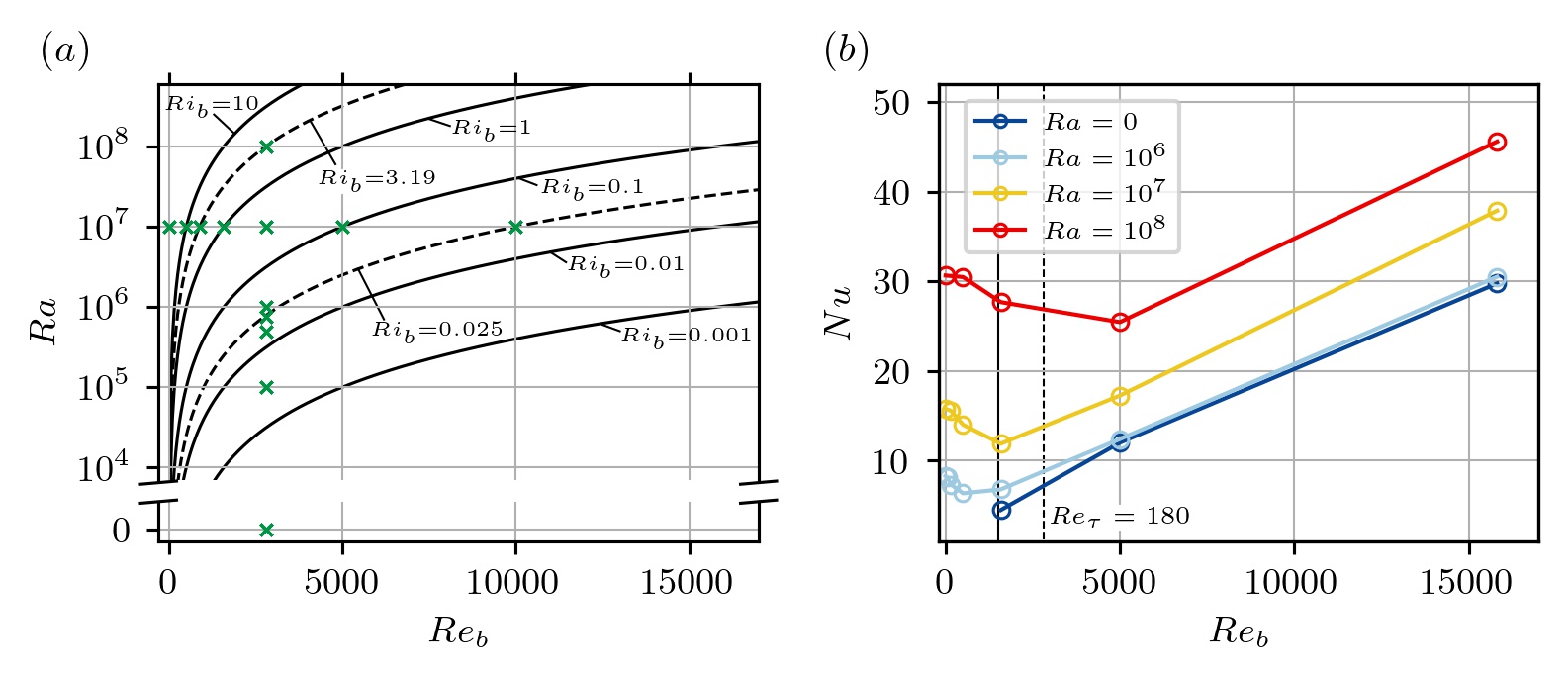}
    \caption{
    Parameter space of Rayleigh $Ra$ and bulk Reynolds number $Re_b$ in $(a)$.
    The green marks in $(a)$ indicate the flow parameters of the present simulations and solid and dashed black lines represent isolevels with constant $Ri_b$. The dashed lines highlight $Ri_b$ values, for which $Re$-effects are investigated.
    The Nusselt number $\mathit{Nu}$ over bulk Reynolds number $Re_b$ of the turbulent mixed convection channel flow from \cite{Pirozzoli_mixed_2017} for various Rayleigh numbers is shown in $(b)$.
    The vertical solid black line separates the transitional and turbulent range for pure forced convection flows.}
    \label{fig:Cf_Nu_Pirozzoli}
\end{figure}

The transition between forced convection structures and streamwise rolls as well as the transition between streamwise rolls and convective cells in mixed convection flows is controlled by the mean shear and buoyancy forcing, which are determined by the 
imposition of the Reynolds $Re_b$ and Rayleigh number $Ra$.
There are several possibilities to vary these two dimensionless numbers to achieve the same Richardson number $Ri_b$, as depicted by the black solid and dashed lines in Figure \ref{fig:Cf_Nu_Pirozzoli}\,(a).
The simplest approach is to fix one of the dimensionless numbers, while varying the other one and vice versa.
In the study of \cite{Pirozzoli_mixed_2017} a smooth wall channel flow is explored for a large parameter space of $Re_b$ and $Ra$, covering all flow regimes, which is shown in terms of the resulting $Nu$ in figure \ref{fig:Cf_Nu_Pirozzoli}\,(b).
For fixed $Ra$ the initial reduction of $Nu$ with increasing the $Re_b$ is associated with the emergence of streamwise rolls, which reduce the effective heat exchange of the convective plumes from the natural convection case. 
For larger values of $Re_b$ the flow transitions to the forced convection regime where $Nu$ increases with increasing $Re_b$.

\textcolor{rev02}{
In the current study the spanwise spacing of the Gaussian ridges $S$ is varied in the range of $S/\delta = 0.5,1,2,4,\infty$, where $S=\infty$ corresponds to the smooth wall case.
Furthermore, the bulk Richardson number $Ri_b$ is varied in such a way to cover the different flow regimes of mixed convection and their transition ranges. 
The variation of $Ri_b$ is achieved by two parameter sweeps, one at constant $Ra=10^7$ and varying $Re_b$ and the other sweep for changing $Ra$ at constant $Re_b = 2800$.
}
These \textcolor{rev02}{two} parameter sweeps for the simulations of the present study are represented in Figure \ref{fig:Cf_Nu_Pirozzoli}\,(a) with crossed green marks, which results in a total number of 65 direct numerical simulations.
These two parameter sweeps intersect in the vicinity of the minimum of $Nu$ for $Ra=10^7$ shown in figure \ref{fig:Cf_Nu_Pirozzoli} $(b)$, which allows us to study the parameter sensitivity of the transition processes from two sides. 
The two black dashed lines in figure \ref{fig:Cf_Nu_Pirozzoli} $(a)$ indicate isolines of constant $Ri_b$ at which the two transition ranges of mixed convection occur.
As will be shown later the lower isoline $Ri_b = 0.025$ lies within the transition range of forced convection structures to streamwise rolls, while the upper isoline $Ri_b = 3.19$ is in the transition range between streamwise rolls to convective cells.
For the present simulations two parameter points with similar $Ri_b$ exist within the two transition ranges, allowing us to study the effect of $Re$ on the flow organization.

From forced convection flows it is known, that the strongest secondary motion occur for $S/\delta \approx \mathcal{O}(1)$ with a spanwise extent of $\approx \delta$ \citep{Vanderwel2019}.
Since secondary motions of Prandtl's second kind occur only in turbulent flows, the parameter sweep with fixed $Re_b = 2800$ will accommodate these secondary motion for all $Ra$, which will allow us to study the effect of buoyancy on the secondary motions as well. 
In case of the $Ra$-sweep, this will partly be the case.
The spanwise extent of the convection cells and streamwise rolls found in \cite{Pirozzoli_mixed_2017} are roughly $4\delta$, such that the chosen values of $S$ cover the width of the different aforementioned flow structures. 


The grid resolutions for the simulations are chosen according to the ones used by \cite{Pirozzoli_mixed_2017} for a second order finite difference code. 
The grid requirements for mixed convection simulations in conjunction with Gaussian ridges represented by an immersed boundary method were investigated 
in a resolution study presented in Appendix \ref{app:grid_study}.
It is found, that the present grid resolution for smooth wall mixed convection cases at $Ra=10^7$ is sufficient for the representation of streamwise Gaussian ridges \textcolor{rev01}{and a further increase of the resolution results in no significant differences of the mean quantities and profiles}. 
Only for lower $Ra$ the spanwise grid resolution \textcolor{rev01}{needs to be} slightly increased \textcolor{rev01}{to achieve grid-independent statistical results for the streamwise-aligned ridge cases}.
The statistical time integration is carried out over at least $1500\,t_b$ for cases $Ri_b\le0.32$, except for the high $Re_b$ cases with $Ri_b = 0.025$ with time integration of at least $300\,t_b$. 
For cases with $Ri_b \ge 1.0$ the time integration comprises at least $400\,t_f$, while the high $Ra$ cases with $Ri_b = 3.2$ and $Ra = 10^8$ were averaged over at least $120\,t_f$.   

\begin{table}
    \begin{center}
    \begin{tabular}[b]{cccccccccccccccc}
    $\mathrm{Ra}$ & $\mathrm{Re}_{b}$ & $\mathrm{Ri}_{b}$ & $S/\delta$ & $\delta_\textit{eff}/\delta$& {$N_x\times N_y\times N_z$} 
         & $Re_\tau$ & $Re_k$ & $C_f (\cdot 10^{-3})$ & $\mathit{Nu}$ &  $-\delta_\textit{eff}/L$ \\[3pt] 
       \hline
      $0$    & 2800  & 0 & $\infty$ & 1     & $512 \times 193\times 384$ & 178.5 & 228.1 &  8.1 & 7.4 & - \\ 
      $0$    & 2800  & 0 & 4        & 0.997 & $512 \times 193\times 384$ & 180.0 & 236.9 & 8.3 & 7.5 & - \\ 
      $0$    & 2800  & 0 & 2        & 0.994 & $512 \times 193\times 384$ & 181.7 & 244.1 & 8.5 & 7.7 & - \\ 
      $0$    & 2800  & 0 & 1        & 0.987 & $512 \times 193\times 384$ & 184.2 & 250.8 & 8.8 & 8.0 & - \\ 
      $0$    & 2800  & 0 & 0.5      & 0.975 & $512 \times 193\times 384$ & 196.3 & 258.5 & 10.2 & 8.6 & - \\[5pt] 
      $10^5$ & 2800  & 0.003 & $\infty$ & 1     & $512 \times 193\times 256$ & 179.3 & 229.1 & 8.2 & 7.5 & 0.003 \\ 
      $10^5$ & 2800  & 0.003 & 4        & 0.997 & $512 \times 193\times 384$ & 180.4 & 238.3 & 8.3 & 7.7 & 0.003 \\ 
      $10^5$ & 2800  & 0.003 & 2        & 0.994 & $512 \times 193\times 384$ & 182.0 & 244.7 & 8.5 & 7.8 & 0.003 \\ 
      $10^5$ & 2800  & 0.003 & 1        & 0.987 & $512 \times 193\times 384$ & 184.7 & 251.2 & 8.9 & 8.1 & 0.003 \\ 
      $10^5$ & 2800  & 0.003 & 0.5      & 0.975 & $512 \times 193\times 384$ & 196.7 & 259.1 & 10.2 & 8.6 & 0.003 \\[5pt] 
      %
      %
      $5.0\cdot10^5$ & 2800  & 0.016  & $\infty$ & 1     & $512 \times 193\times 256$ & 180.5 & 234.3 & 8.3 & 8.1 & 0.017 \\ 
      $5.0\cdot10^5$ & 2800  & 0.016  & 4        & 0.997 & $512 \times 193\times 384$ & 181.8 & 240.9 & 8.5 & 8.1 & 0.017 \\ 
      $5.0\cdot10^5$ & 2800  & 0.016  & 2        & 0.994 & $512 \times 193\times 384$ & 183.2 & 247.3 & 8.6 & 8.2 & 0.017 \\ 
      $5.0\cdot10^5$ & 2800  & 0.016  & 1        & 0.987 & $512 \times 193\times 384$ & 186.1 & 253.0 & 9.0 & 8.4 & 0.016 \\ 
      $5.0\cdot10^5$ & 2800  & 0.016  & 0.5      & 0.975 & $512 \times 193\times 384$ & 198.1 & 261.1 & 10.4 & 9.0 & 0.014 \\[5pt] 
      $7.5\cdot10^5$ & 2800  & 0.024  & $\infty$ & 1     & $512 \times 193\times 256$ & 180.4 & 249.1 & 8.3 & 9.4 & 0.030 \\ 
      $7.5\cdot10^5$ & 2800  & 0.024  & 4        & 0.997 & $512 \times 193\times 384$ & 181.8 & 249.5 & 8.5 & 9.2 & 0.029 \\ 
      $7.5\cdot10^5$ & 2800  & 0.024  & 2        & 0.994 & $512 \times 193\times 384$ & 183.6 & 252.2 & 8.7 & 9.1 & 0.028 \\ 
      $7.5\cdot10^5$ & 2800  & 0.024  & 1        & 0.987 & $512 \times 193\times 384$ & 187.0 & 254.1 & 9.1 & 8.6 & 0.025 \\ 
      $7.5\cdot10^5$ & 2800  & 0.024  & 0.5      & 0.975 & $512 \times 193\times 384$ & 199.4 & 262.7 & 10.5 & 9.2 & 0.022 \\[5pt] 

      %
      $10^6$ & 2800  & 0.032 & $\infty$ & 1     & $512 \times 193\times 256$ & 178.9 & 257.1 & 8.2 &  9.7 & 0.042 \\ 
      $10^6$ & 2800  & 0.032 & 4        & 0.997 & $512 \times 193\times 384$ & 180.9 & 256.7 & 8.4 &  9.7 & 0.041 \\ 
      $10^6$ & 2800  & 0.032 & 2        & 0.994 & $512 \times 193\times 384$ & 183.2 & 261.1 & 8.6 &  9.8 & 0.040 \\ 
      $10^6$ & 2800  & 0.032 & 1        & 0.987 & $512 \times 193\times 384$ & 186.8 & 261.4 & 9.1 &  9.6 & 0.037 \\ 
      $10^6$ & 2800  & 0.032 & 0.5      & 0.975 & $512 \times 193\times 384$ & 198.9 & 266.2 & 10.5 & 10.0 & 0.032 \\[5pt] 
      %
      %
      $10^7$ & 0     & $\infty$ & $\infty$ & 1     & $ 1024 \times 257 \times 512 $ & -     & 237.2 & -     & 15.7 & $\infty$ \\ 
      $10^7$ & 0     & $\infty$ & 4        & 0.997 & $ 1024 \times 257 \times 512 $ & -     & 236.9 & -     & 16.0 & $\infty$ \\ 
      $10^7$ & 0     & $\infty$ & 2        & 0.994 & $ 1024 \times 257 \times 512 $ & -     & 236.7 & -     & 16.3 & $\infty$ \\ 
      $10^7$ & 0     & $\infty$ & 1        & 0.987 & $ 1024 \times 257 \times 512 $ & -     & 238.9 & -     & 16.9 & $\infty$ \\ 
      $10^7$ & 0     & $\infty$ & 0.5      & 0.975 & $ 1024 \times 257 \times 512 $ & -     & 245.7 & -     & 18.4 & $\infty$ \\[5pt] 
      $10^7$ & 500   & 10.0  & $\infty$ & 1     & $ 1024 \times 257 \times 512 $ &  71.0 & 250.0 & 40.3 & 13.9 & 9.743 \\ 
      $10^7$ & 500   & 10.0  & 4        & 0.997 & $ 1024 \times 257 \times 512 $ &  72.1 & 243.7 & 41.8 & 14.2 & 9.461 \\ 
      $10^7$ & 500   & 10.0  & 2        & 0.994 & $ 1024 \times 257 \times 512 $ &  73.2 & 239.9 & 43.2 & 14.4 & 9.213 \\ 
      $10^7$ & 500   & 10.0  & 1        & 0.987 & $ 1024 \times 257 \times 512 $ &  77.5 & 226.9 & 48.8 & 15.1 & 8.122 \\ 
      $10^7$ & 500   & 10.0  & 0.5      & 0.975 & $ 1024 \times 257 \times 512 $ &  83.9 & 235.7 & 58.4 & 17.2 & 7.275 \\[5pt] 
      $10^7$ & 885   & 3.19  & $\infty$ & 1     & $ 1024 \times 257 \times 512 $ &  97.7 & 263.4 & 24.4 & 12.6 & 3.373 \\
      $10^7$ & 885   & 3.19  & 4        & 0.997 & $ 1024 \times 257 \times 512 $ &  99.3 & 262.5 & 25.2 & 12.9 & 3.300 \\
      $10^7$ & 885   & 3.19  & 2        & 0.994 & $ 1024 \times 257 \times 512 $ & 101.0 & 252.5 & 26.2 & 13.1 & 3.174 \\
      $10^7$ & 885   & 3.19  & 1        & 0.987 & $ 1024 \times 257 \times 512 $ & 105.5 & 240.0 & 28.9 & 13.6 & 2.903 \\
      $10^7$ & 885   & 3.19  & 0        & 0.975 & $ 1024 \times 257 \times 512 $ & 111.7 & 236.1 & 33.0 & 15.2 & 2.729 \\[5pt]
      $10^7$ & 1581  & 1.0 & $\infty$ & 1     & $ 1024 \times 257 \times 512 $ & 134.2 & 304.1 & 14.4 & 11.9 & 1.231 \\ 
      \textcolor{rev}{$10^7$} & 1581  & 1 & 8 &  0.998 & $ 1024 \times 257 \times 512 $ & 135.2 & 300.6 & 14.7 & 12.0 & 1.210\\
      $10^7$ & 1581  & 1.0 & 4        & 0.997 & $ 1024 \times 257 \times 512 $ & 136.7 & 300.2 & 15.0 & 12.1 & 1.184 \\ 
      $10^7$ & 1581  & 1.0 & 2        & 0.994 & $ 1024 \times 257 \times 512 $ & 138.9 & 285.6 & 15.6 & 12.2 & 1.140 \\ 
      $10^7$ & 1581  & 1.0 & 1        & 0.987 & $ 1024 \times 257 \times 512 $ & 143.1 & 278.5 & 16.7 & 12.6 & 1.072 \\ 
      $10^7$ & 1581  & 1.0 & 0.5      & 0.975 & $ 1024 \times 257 \times 512 $ & 152.5 & 275.2 & 19.3 & 13.9 & 0.979 \\[5pt] 
    \end{tabular}
    \end{center}
\end{table}

\begin{table}
    \begin{center}
    \begin{tabular}[b]{cccccccccccccccc}
    $\mathrm{Ra}$ & $\mathrm{Re}_{b}$ & $\mathrm{Ri}_{b}$ & $S/\delta$ & $\delta_\textit{eff}/\delta$& {$N_x\times N_y\times N_z$} 
         &  $Re_\tau$ & $Re_k$ & $C_f (\cdot 10^{-3})$ & $\mathit{Nu}$ & $-\delta_\textit{eff}/L$  \\[3pt] 
       \hline
       $10^7$ & 2800  & 0.32 & $\infty$ & 1     & $ 1024\times 257\times 512 $ & 190.4 & 353.2 &  9.3 & 12.3 & 0.446 \\ 
      $10^7$ & 2800  & 0.32 & 4        & 0.997 & $ 1024\times 257\times 512 $ & 193.9 & 350.6 &  9.6 & 12.7 & 0.434 \\ 
      $10^7$ & 2800  & 0.32 & 2        & 0.994 & $ 1024\times 257\times 512 $ & 198.1 & 352.8 & 10.1 & 13.1 & 0.422 \\ 
      $10^7$ & 2800  & 0.32 & 1        & 0.987 & $ 1024\times 257\times 512 $ & 204.9 & 348.8 & 10.9 & 13.8 & 0.401 \\ 
      $10^7$ & 2800  & 0.32 & 0.5      & 0.975 & $ 1024\times 257\times 512 $ & 216.2 & 349.8 & 12.4 & 15.1 & 0.373 \\[5pt] 
      $10^7$ & 5000  & 0.1 & $\infty$ & 1     & $ 1024 \times 257 \times 512 $ & 304.5 & 499.1 & 7.4 & 17.1 & 0.152 \\ 
      $10^7$ & 5000  & 0.1 & 4        & 0.997 & $ 1024 \times 257 \times 512 $ & 306.4 & 497.0 & 7.5 & 17.3 & 0.150 \\ 
      $10^7$ & 5000  & 0.1 & 2        & 0.994 & $ 1024 \times 257 \times 512 $ & 311.4 & 501.7 & 7.8 & 17.7 & 0.146 \\ 
      $10^7$ & 5000  & 0.1 & 1        & 0.987 & $ 1024 \times 257 \times 512 $ & 316.9 & 499.9 & 8.2 & 18.2 & 0.143 \\ 
      $10^7$ & 5000  & 0.1 & 1        & 0.975 & $ 1024 \times 257 \times 512 $ & 328.9 & 504.0 & 8.9 & 19.3 & 0.135 \\[5pt] 
      $10^7$ & 10000 & 0.025  & $\infty$ & 1     & $ 1536 \times 513 \times 1024 $ & 552.2 & 847.6 & 6.1 & 28.4 & 0.042 \\ 
      $10^7$ & 10000 & 0.025  & 4        & 0.997 & $ 1536 \times 513 \times 1024 $ & 555.0 & 822.7 & 6.2 & 28.1 & 0.041 \\ 
      $10^7$ & 10000 & 0.025  & 2        & 0.994 & $ 1536 \times 513 \times 1024 $ & 558.3 & 841.3 & 6.3 & 28.4 & 0.041 \\ 
      $10^7$ & 10000 & 0.025  & 1        & 0.987 & $ 1536 \times 513 \times 1024 $ & 570.5 & 825.5 & 6.6 & 28.1 & 0.038 \\ 
      $10^7$ & 10000 & 0.025  & 0.5      & 0.975 & $ 1536 \times 513 \times 1024 $ & 585.2 & 831.9 & 7.0 & 27.1 & 0.034 \\[5pt] 
      %
      %
      $10^8$ & 2800  & 3.19 & $\infty$ & 1     & $2048 \times 513\times 1024$ & 246.1 & 749.3 & 15.5 &  25.6 & 4.297 \\ 
      $10^8$ & 2800  & 3.19 & 4        & 0.997 & $2048 \times 513\times 1024$ & 249.5 & 791.9 & 15.9 &  26.9 & 4.327 \\ 
      $10^8$ & 2800  & 3.19 & 2        & 0.994 & $2048 \times 513\times 1024$ & 259.3 & 718.1 & 17.3 &  27.6 & 3.964 \\ 
      $10^8$ & 2800  & 3.19 & 1        & 0.987 & $2048 \times 513\times 1024$ & 269.3 & 695.1 & 18.7 &  29.4 & 3.761 \\ 
      $10^8$ & 2800  & 3.19 & 0.5      & 0.975 & $2048 \times 513\times 1024$ & 295.1 & 710.6 & 22.8 &  33.7 & 3.277 \\ 
      \hline
    \end{tabular}
    \caption{List of simulation configurations with flow parameteres and resulting global flow properties.}
    \label{tbl:simulation_configuration}
    \end{center}
\end{table}
\section{Results}
\subsection{Global flow properties}

The results of the global flow properties for the different simulations are presented in table \ref{tbl:simulation_configuration}, where the configurations 
are arranged according to the parameter \textcolor{rev02}{triad} ($Ra$, $Re_b$, $S/\delta$).  
The smooth wall configurations are indicated by $S=\infty$ and the configurations with streamwise-aligned ridges are listed with decreasing $S$.
Since the forced and mixed convection cases are run at constant flow rate (CFR), the presence of the Gaussian ridges will increase the drag, 
which translates into an increase of $C_f$ and $Re_\tau$ compared to the respective case with smooth wall conditions.
As can be seen for all considered cases, the steady decrease of $S$ leads to a monotonic increase of $C_f$ and $Re_\tau$ compared to the smooth wall case. 
For forced convection the increase of $C_f$ is up to $26\%$, while the largest increase is found for $Ri_b = 10$ with $45\%$.
\textcolor{rev03}{
Due to the changing friction drag, the ridge height in wall units $h_g^+$ changes for all cases as well, and ranges  between $7 \le h_g^+ \le 60$ for the current configurations.
}
In case of pure forced convection, turbulent secondary motions induced by  streamwise-aligned ridges are known to increase the global friction as well as the heat transfer of the flow compared to smooth wall conditions \citep{stroh_secondary_2020}.
This behaviour is also observed for the present forced convection case with streamwise-aligned Gaussian ridges, where for the densest ridge spacing $S=0.5\delta$ the heat transfer increases $16\%$ compared to the smooth wall case. 
In a similar range with $17\%$ is the increase for the natural convection case, and the largest increase is found for $Ri_b=3.2$ with $32\%$.

\begin{figure}
    \centering
    \includegraphics[trim=3pt 0 0 0]{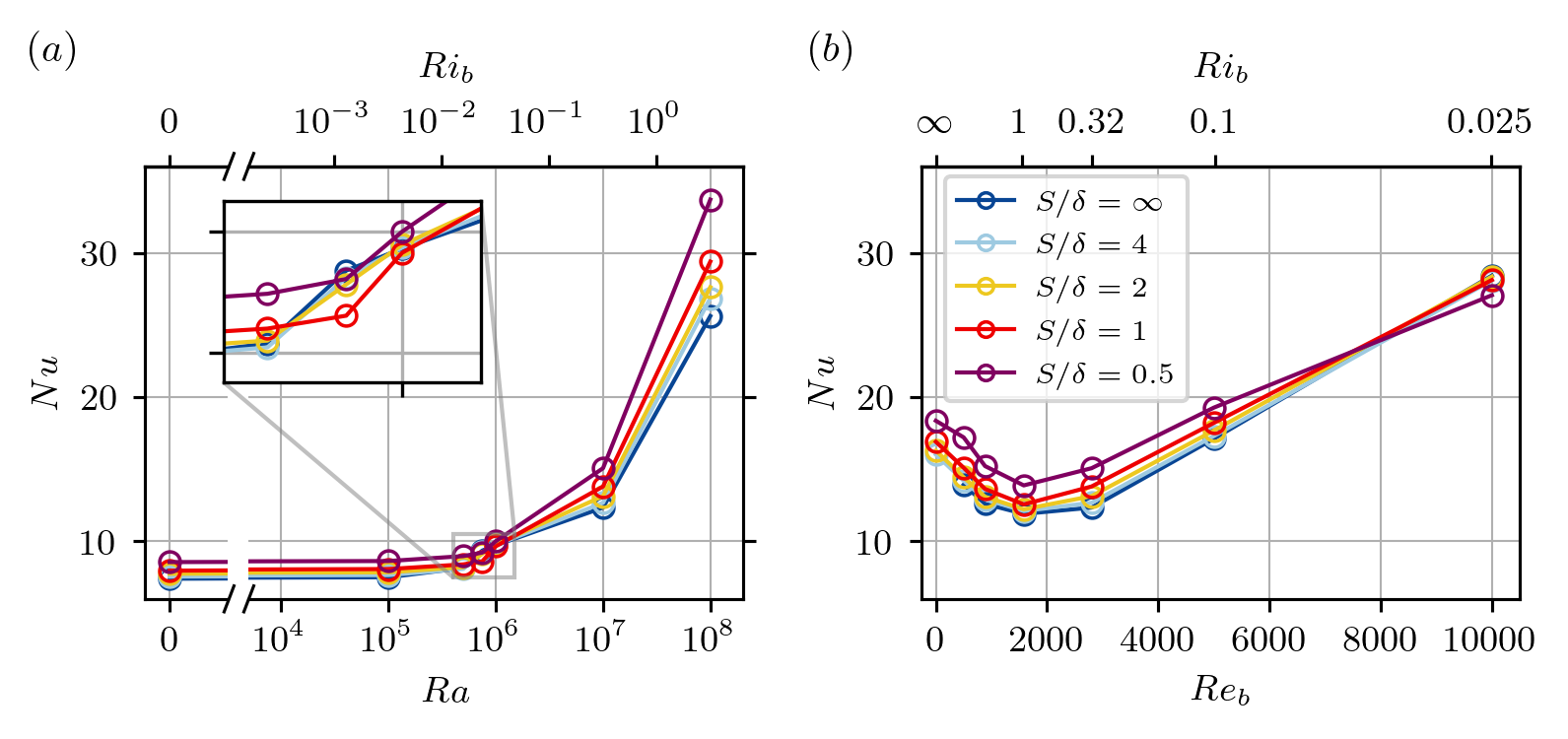}
    \caption{Nusselt number $\mathit{Nu}$ as a function of Rayleigh number $Ra$ in (a) and bulk Reynolds number $Re_b$ in (b) for different ridge spacings $S$.
    In (a) the bulk Reynolds number $Re_b = 2800$ and in (b) the Rayleigh number $Ra=10^7$ is kept constant.}
    \label{fig:Cf_Nu_Re_Ra}
\end{figure}

While the skin friction drag increases with decreasing $S$ and increasing wetted surface area, this behaviour is not found for the heat transfer for all present cases. 
This is illustrated for the two parameter sweeps \textcolor{rev03}{at constant $Re_b = 2800$ in figure \ref{fig:Cf_Nu_Re_Ra} $(a)$ and at constant $Ra=10^7$ in figure \ref{fig:Cf_Nu_Re_Ra} $(b)$}. 
\textcolor{rev03}{
The forced convection case $Ra=0$ and weak convective case $Ra=10^5$ in figure \ref{fig:Cf_Nu_Re_Ra} $(a)$ show the successive increase of $Nu$ with decreasing ridge spacing $S$. 
An increase of $Ra$ or $Ri_b$ introduces an additional buoyant contribution to the vertical mixing, resulting in larger heat transfer and for large Rayleigh numbers ($Ra>10^6$), which represent configurations where buoyancy is comparable to shear or even stronger, the monotonic increase of $Nu$ with decreasing $S$ is also found. 
However, for the particular cases $Ra=7.5\cdot 10^5$ ($Ri_b=0.024$) and $Ra=10^6$ ($Ri_b=0.032$)  heat transfer does not monotonically increase with decreasing $S$, which is visible in the inset of figure \ref{fig:Cf_Nu_Re_Ra} $(a)$.}


In figure \ref{fig:Cf_Nu_Re_Ra} $(b)$ the natural convection case is given by $Re_b = 0$ and the influence of buoyancy is successively reduced by increasing $Re_b$.
The minimum found for $Nu$ is associated with the break-up of the thermal plumes of the Rayleigh-Bénard case, when shear is added, and has been reported for unstable thermal stratification in Poiseuille and Couette flows \citep{scagliarini_heat_flux_2014,blass_flow_2020}.
The non-monotonic behaviour of $Nu$ with decreasing $S$ seen in figure \ref{fig:Cf_Nu_Re_Ra} $(a)$ also occurs for the largest $Re_b = 10000$ case ($Ri_b = 0.025$), which is in a similar bulk Richardson number range as the former cases.
Thus, all cases which depict a non-monotonic behaviour of $Nu$ with respect to $S$ fall within a range of bulk Richardson values $Ri_b = 0.016 - 0.032$, where shear effects are strong and buoyancy effects are weak.
It will be shown that this range of Richardson numbers marks the transition from forced convection structures to streamwise rolls and the ridge spacing $S$ affects and alters this transition.

\begin{figure}
    \centering
    \includegraphics[trim=3pt 0 0 0]{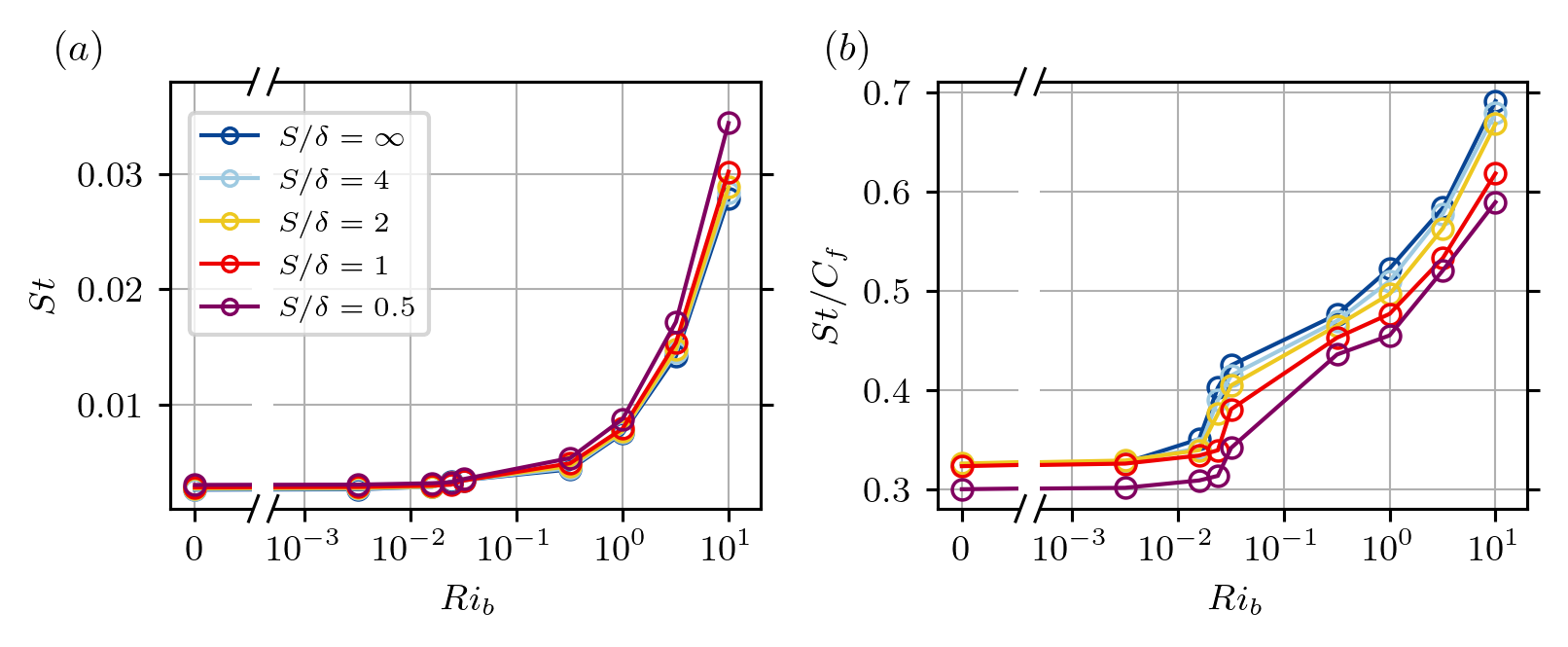}
    \caption{Stanton number $St$ in (a) and ratio of $St$ to $C_f$ in (b) as a function of bulk Richardson number $Ri_b$ for different ridge spacings $S$.
    The selected cases have values of the Reynolds number $Re_k$ in a similar range.}
    \label{fig:St_Cf_Ri}
\end{figure}

In case of mixed convection, turbulence is driven by two generation mechanisms, the production by shear and by buoyancy, and the exact turbulence level cannot be inferred a priori by $Ra$ and $Re_b$.
The same holds for the bulk Richardson number $Ri_b$. 
To rule out effects caused by the different turbulence levels or Reynolds number effects, \textcolor{rev}{which will be discussed in section \ref{sec:reynolds_effects}}, we separate those cases with \textcolor{rev02}{significantly} higher turbulence levels from the cases with comparable values.
\textcolor{rev}{
For the subsequent discussion and sections only cases which fall in a range of Reynolds number values $Re_k = 200 -355$ are considered. 
This selection includes only those cases of the two parameter sweeps for which $Ra\le 10^7$ and $Re_b \le 2800$. 
Instead of using the pairs of $Ra$ and $Re_b$, the bulk Richardson number $Ri_b$ is used in the following to characterize the relative importance of buoyancy and shear effects.
In analogy to $Ri=Ra/(4 Re_b^2 Pr)$, $Nu$ is replaced by the Stanton number $St=Nu/(Re_b Pr)$. The corresponding results are presented in figure \ref{fig:St_Cf_Ri} $(a)$ and reveal an increase of $St$ with $Ri_b$.}


\textcolor{rev}{
The relative increase between heat and momentum transfer is characterized by the ratio  $St/C_f$ which is shown in figure \ref{fig:St_Cf_Ri} $(b)$.
For each $S$ it can be seen that larger values of $Ri_b$ induce larger  $St/C_f$ values. 
This implies that buoyancy effects lead to a larger increase of heat transfer than of momentum transfer.
In addition, a consistent influence of the ridges can be observed in this representation. 
A decrease of $S$ results in less $St/C_f$ for all $Ri_b$.
This is also reflected in the stability parameter $-\delta_\textit{eff}/L$ in table \ref{tbl:simulation_configuration}, which is another quantity to compare the relative heat and momentum transfer. 
The most distinct property of figure \ref{fig:St_Cf_Ri} $(b)$ is the strong increase of $St/C_f$ in the range $Ri_b = 0.016-0.032$. This increase is delayed to larger $Ri_b$ with decreasing $S$ and can be linked to a reorganization of the turbulent flow structures as discussed in the following section. 
}

\subsection{Turbulent flow structures}
\label{sec:turbulent_structures}

\begin{figure}
    \centering
    \includegraphics[trim=3pt 0 0 0]{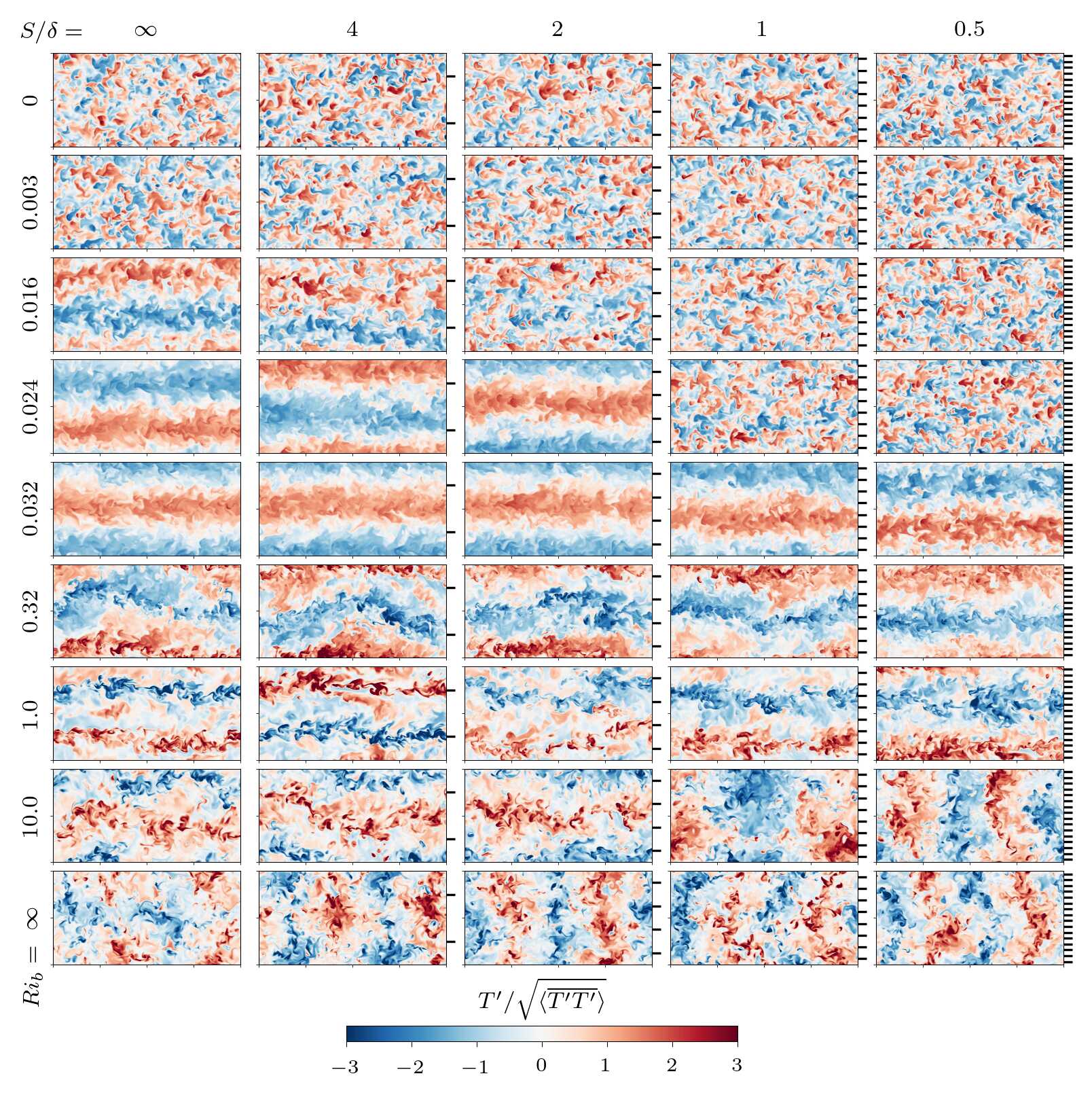}
    \caption{Instantaneous temperature fluctuation fields at the half-channel height position $y=\delta$ for varying Richardson number $Ri_b$ and different spanwise spacing $S$ of the Gaussian ridges.
    The spanwise position of the ridges is indicated by the black lines on the right outer frame of the figures.
    The horizontal sections show the full simulation domain of size $16\delta \times 8\delta$.
    }
    \label{fig:midplane_T_Ri_Reb}
\end{figure}

\begin{figure}
    \centering
    \includegraphics[trim=3pt 0 0 0]{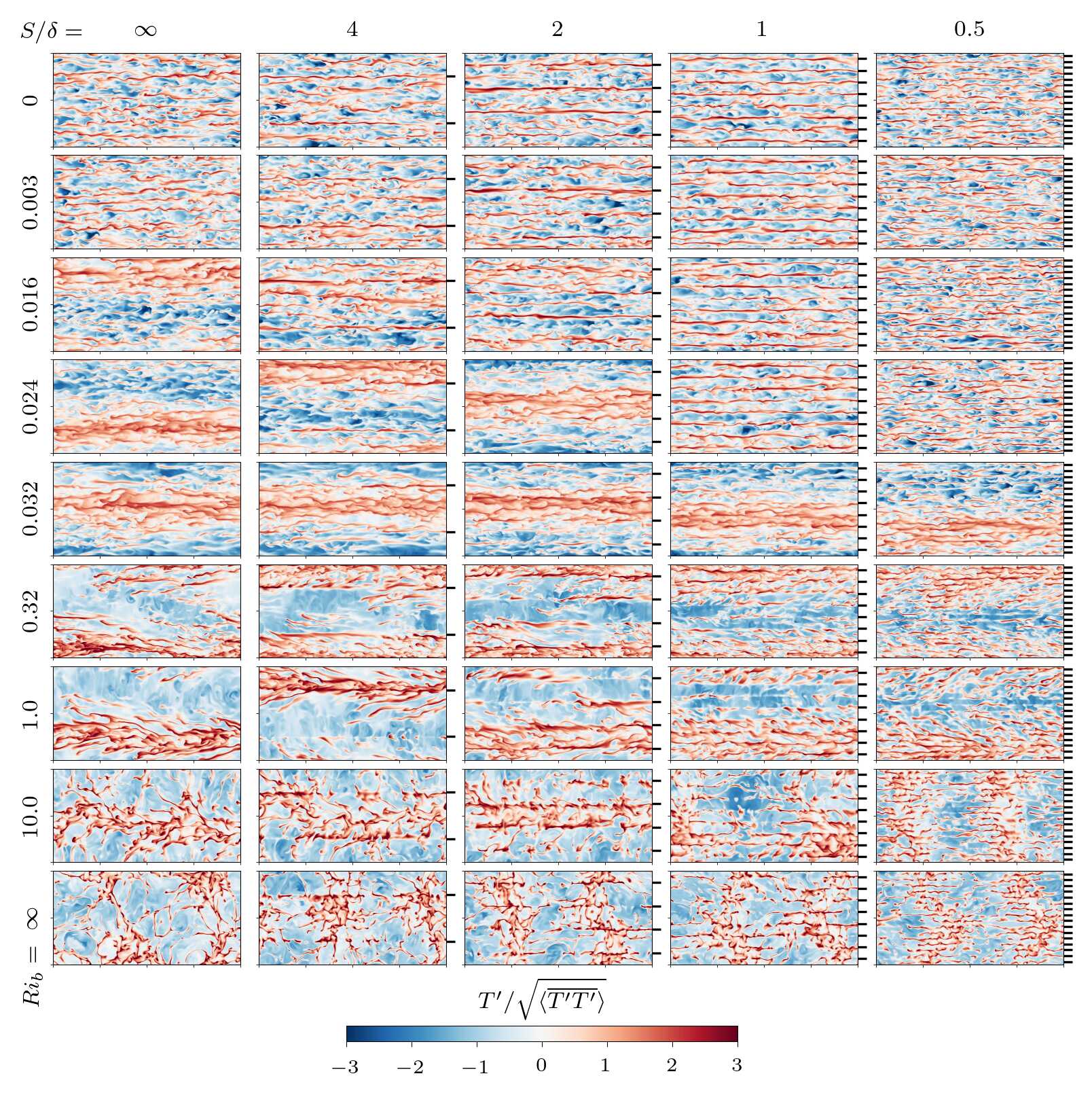}
    \caption{Instantaneous temperature fluctuation fields at $y=0.15\delta$ for varying Richardson number $Ri_b$ and different spanwise spacing $S$ of the Gaussian ridges.
    }
    \label{fig:yplane_bottom_T_Ri_Reb}
\end{figure}

\textcolor{rev03}{
The different flow organization of mixed convection flows observed over smooth wall conditions can be visualized by instantaneous velocity or temperature fluctuations in horizontal planes \citep{Salesky_nature_2017}.
The influence of heterogeneous surfaces on this flow organization is shown for various pairs of $Ri_b$ and $S$ for the
}
instantaneous temperature fluctuations in wall-parallel planes located at the half-channel height ($y=\delta$) in figure \ref{fig:midplane_T_Ri_Reb} and slightly above the top of the ridges ($y=0.15\delta$) in figure \ref{fig:yplane_bottom_T_Ri_Reb}.
\textcolor{rev02}{
Please note that  the discussion for the near-wall region refers to the bottom wall, unless stated otherwise.  
}
Both horizontal planes display the same instantaneous realization of the flow field and comprise cases that fall in a similar range of turbulent Reynolds number $Re_k$.
The $Ri_b$ increases from top to bottom, starting with the forced convection case $Ri_b = 0$ ($Re_b=2800$, $Ra=0$) at the top and the natural convection case $Ri_b = \infty$ ($Re_b=0$, $Ra=10^7$) at the bottom panel. 
The ridge spacing $S/\delta$ decreases from left to right, with the smooth-wall case at the outer left panel side. 

Considering the smooth wall cases $S=\infty$ first, the flow topology of the forced convection case has a spotty organization which is also the case for the mild convective case $Ri_b = 0.003$ ($Re_b=2800$, $Ra=10^5$) in figure \ref{fig:midplane_T_Ri_Reb}.
\textcolor{rev03}{The transition to streamwise rolls takes place for slightly larger buoyant forcing at case $Ri_b = 0.016$ ($Re_b=2800$, $Ra=5\cdot10^5$), but the rolls still display some patchiness.}
This transition also results in a change of the near-wall structures as can be seen in figure \ref{fig:yplane_bottom_T_Ri_Reb}, where strong elongated temperature fluctuations preferentially locate in regions of the streamwise roll updrafts, while less pronounced in the downdraft region.
These elongated temperature fluctuations coincide with the near-wall low-speed streaks (not shown here), since for neutral and moderately convective cases the temperature behaves like a passive scalar with strong correlation with the streamwise velocity \citep{khanna_ABL_1998}. 
The formation mechanism of streamwise rolls is strongly linked to localized buoyancy forces, which concentrate in low-speed streaks and thereby create linear updrafts \citep{khanna_ABL_1998}.
Multiple updrafts can merge to a strong buoyancy-enhanced streak, forming the updraft region of the streamwise roll in figure \ref{fig:midplane_T_Ri_Reb}.
This updraft is reaching the opposing wall and reduces or destroys the coherence of the low-speed streaks there.
At the same time, between these impingement region of the updrafts at the opposite wall, buoyancy enhanced low-speed streaks can form, which in turn generate a strong localized downdraft and in combination with the updrafts result in a large-scale streamwise roll motion in the cross-section.

In contrast to case $Ri_b = 0.016$, the streamwise roll of case $Ri_b = 0.024$ in figure \ref{fig:midplane_T_Ri_Reb} is more articulated in its structure\textcolor{rev01}{, which is associated with a sudden increase of $St/C_f$ in figure \ref{fig:St_Cf_Ri} $(b)$.}
The streamwise rolls persist up to $Ri_b = 1$ in figure \ref{fig:midplane_T_Ri_Reb} with a spanwise wavelength of approximately $8\delta$, such that the chosen domain size is able to accommodate a single pair of counter-rotating rolls as reported by \cite{Pirozzoli_mixed_2017}.
For the cases $Ri_b=0.32$ and $Ri_b=1.0$ in figure \ref{fig:midplane_T_Ri_Reb} the rolls show a strong waviness of the thermal up- and downdrafts, which is also seen in the near-wall region where the updraft region encompasses spanwise inclined near-wall streaks in figure \ref{fig:yplane_bottom_T_Ri_Reb}.
When buoyancy forces become more important, the streamwise roll is more disrupted and reduce its streamwise coherence, since thermal plumes become dynamically more important (see case $Ri_b = 10$ in figure \ref{fig:midplane_T_Ri_Reb}) \citep{Salesky_nature_2017}.
The increased influence of buoyancy also modifies the near-wall structures, where cell-like structures appear in the updraft region, which still depict some streamwise coherence, as seen in figure \ref{fig:yplane_bottom_T_Ri_Reb}.
For the Rayleigh-Bénard case $Ri_b=\infty$ ($Re_b=0$, $Ra=10^7$) in figure \ref{fig:midplane_T_Ri_Reb}, the streamwise roll is not present any more and the flow organizes in convective cells.
These structures have a preferential roll orientation in the $x$- and $z$-direction, also seen in the near-wall region in figure \ref{fig:yplane_bottom_T_Ri_Reb}, which is due to the rectangular domain size \citep{Pirozzoli_mixed_2017}

The introduction of the ridges displays no significant differences of the flow structures in the channel center for the forced convection case $Ri_b=0$ and weak buoyancy case $Ri_b=0.003$ compared to the smooth wall case.
However, the elongated high temperature fluctuations in the near-wall region are more coherent in the streamwise direction in case of ridge spacings $S=\delta, 2\delta, 4\delta$, for which they preferentially occur at the spanwise ridge position.
For mild buoyancy effects, these elongated temperature regions collapse with low-speed streaks (not shown here), forming low-momentum pathways directly above the ridges.
As such the preferential position of the low-speed streaks coincides with the mean upward motion of the secondary motions, which will be shown in section \ref{sec:mean_profiles}.
For the densest ridge spacing $S=0.5\delta$ these elongated structures still occur at the ridges, however they appear less coherent in the streamwise direction.

Significant effects of the ridges can be seen for the transition between forced convection structures and streamwise rolls in figure \ref{fig:midplane_T_Ri_Reb} for cases $Ri_b = 0.016$-$0.032$.
As shown before, the transition towards streamwise rolls for smooth wall conditions takes place at $Ri_b=0.016$ and this transition can be also observed for the coarsest ridge spacing $S=4\delta$, which however is more interrupted by individual turbulent spots than the smooth wall case. 
This is also reflected in a change of the near-wall structures, where for case $S=4\delta$ the elongated high temperature fluctuations still favour the updraft region, but in contrast to the smooth wall case, also occur inside the downdraft region at the ridge position (figure \ref{fig:yplane_bottom_T_Ri_Reb}).
For denser ridge spacing $S<4\delta$ the preferred concentration of low-speed streaks is not observed anymore.
Therefore the streamwise roll is not visible in the channel core and the flow structures resemble those seen for the forced convection and weak buoyancy cases $Ri_b = 0.003$.
For slightly larger bulk Richardson number $Ri_b = 0.024$ the streamwise roll is now clearly visible for the two coarsest ridge spacings $S=4\delta$ and $S=2\delta$ in figure \ref{fig:midplane_T_Ri_Reb}, and strong enough to reorganize the near-wall structures seen in figure \ref{fig:yplane_bottom_T_Ri_Reb}, while for the denser ridge spacings the streamwise roll is not present. 
Eventually, the streamwise roll is observed for all ridge spacings $S$ at $Ri_b = 0.032$, while for the denser ridge spacings still some spot-like structures overlap with the rolls.
\textcolor{rev01}{Considering figure \ref{fig:St_Cf_Ri} $(b)$ the delayed transition between forced convection structures and streamwise rolls with decreasing $S$ can be related to the increased drag introduced by the ridges.
Denser ridge spacings introduce more drag and shear in the near-wall region, and in consequence larger buoyancy forces are required to form the streamwise rolls which in turn induce an increase in heat transfer.}

The streamwise roll, present for the intermediate Richardson number $Ri_b=0.32$ and $Ri_b=1.0$ cases (figure \ref{fig:midplane_T_Ri_Reb}), displays no significant influence of the ridges. 
This is likewise the case for the near-wall region in figure \ref{fig:yplane_bottom_T_Ri_Reb}, where it can be seen that the formation of high temperature fluctuations occur inside the updraft region of the roll.

\textcolor{rev03}{At higher convective conditions for $Ri_b = 10.0$} the streamwise roll is present for the smooth wall case and the two coarsest ridge spacing $S =2\delta, 4\delta$, while it completely disappears for denser ridge spacings $S \le \delta$.
For $S = \delta$ the streamwise roll is replaced by convection cells, resembling the one found for the Rayleigh-Bénard case with a spacing of $S=4\delta$.
For the densest ridge spacing $S=0.5\delta$ rolls with a preferential orientation in the spanwise direction occur, which has similarities to the densest ridge spacing $S=0.5\delta$ of the Rayleigh-Bénard case.
This transition from roll to cell structures is also reflected in a transition of the near-wall structures in figure \ref{fig:yplane_bottom_T_Ri_Reb}.
The roll-to-cell transition is also observed for the lower bulk Richardson case $Ri_b = 3.2$ (higher $Re_k$) for the same ridge spacings $S$ (not shown).
This result is remarkable, since the transition between roll to cell structures over homogeneous wall conditions in atmospheric boundary layer is expected to begin at larger values of the stability parameter $-z_i/L \approx 26$ \citep{Salesky_nature_2017}, while the stability parameter for the cases $Ri_b = 3.2$ and $Ri_b = 10.0$ are ranging between $-\delta_\textit{eff}/L = 3.4-9.7$.
This illustrates that streamwise-aligned ridges reduce significantly the range of $Ri_b$ or $-\delta_\textit{eff}/L$ in which streamwise rolls appear.
This suggest, that heterogeneous rough surfaces can trigger the roll-to-cell transition at smaller buoyancy forces.

As can be seen for the Rayleigh-Bénard case $Ri_b = \infty$ in figure \ref{fig:midplane_T_Ri_Reb} decreasing the ridge spacings $S$ results in an increasingly preferential orientation of the convective cell towards the spanwise direction $z$.
The rolls with orientation in the $x$-direction experience increasing lateral drag as $S$ decreases, and for $S\le 2\delta$ these rolls can eventually no longer emerge and only rolls in the $z$-direction whose circulation is along the ridge direction occur. 
This observation will be discussed further  in section \ref{sec:turbulent_structures}.
\textcolor{rev03}{The increase of $Nu$ for smaller $S$ is also reflected by intensified thermal up- and downdrafts in the channel center plane.}
\textcolor{rev}{We note that additional simulations for $S=\infty$ and $S=0.5\delta$ in a wider domain ($L_x = L_z = 16\delta$) do not indicate a domain size dependence of the obtained results. }
\subsection{Mean properties}
\label{sec:mean_profiles}

\begin{figure}
    \centering
    \includegraphics[trim=3pt 0 0 0,scale=0.75]{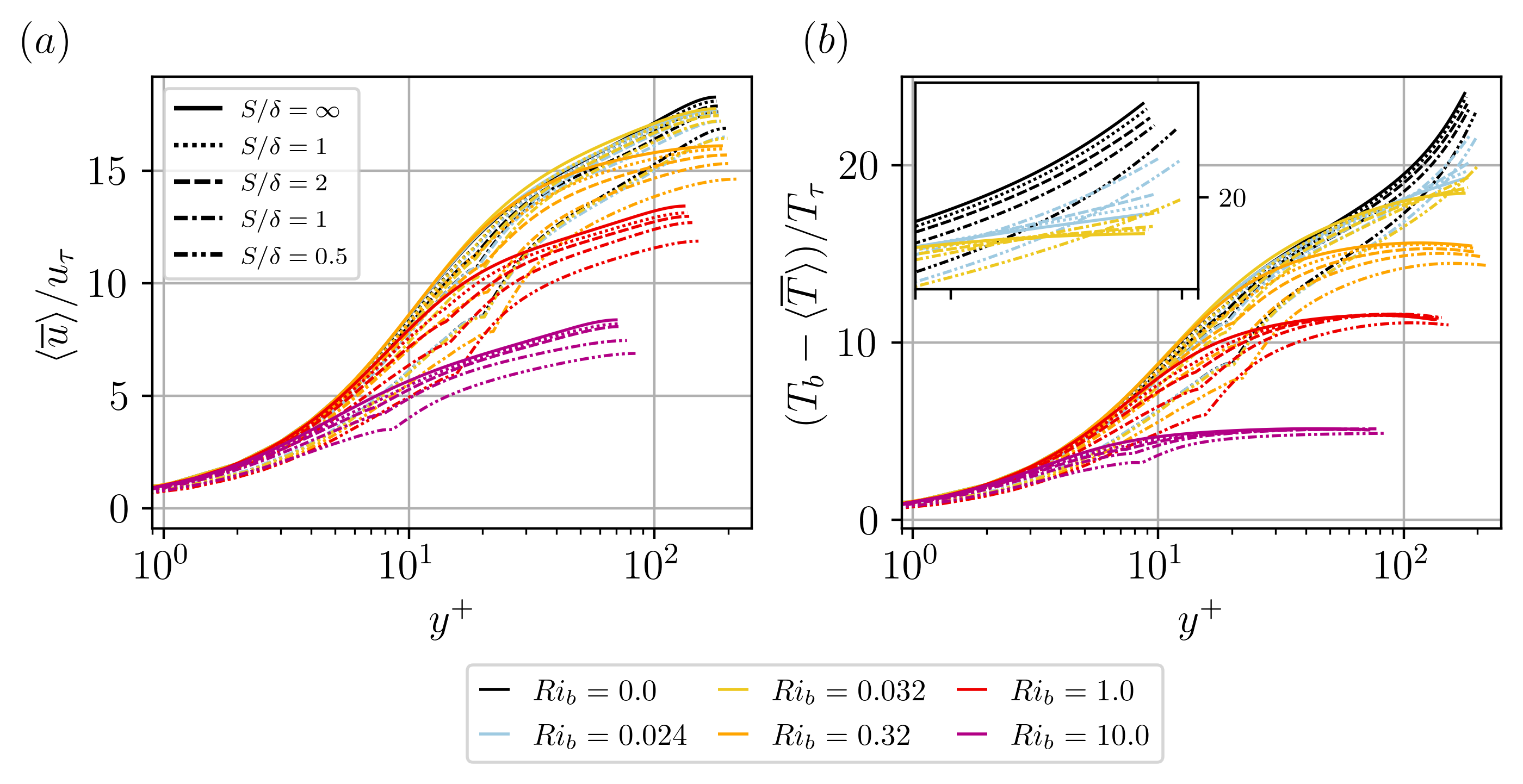}
    \caption{Effect of Richardson number $Ri_b$ and $S$ on wall-normal profiles of streamwise mean velocity and mean temperature for different ridge spacings $S$ scaled in wall units. 
    }
    \label{fig:mean_profile_um_Tm_S_Ri_sweep}
\end{figure}
\textcolor{rev03}{The effect of ridge spacing $S$ and the relative strength of shear and buoyancy effects on the time and horizontally averaged mean streamwise velocity and temperature profiles are shown in figure \ref{fig:mean_profile_um_Tm_S_Ri_sweep}.
The mean temperature is represented as the difference from the bottom wall temperature $T_b$ and scaled by the friction temperature $T_\tau = Q/\rho c_{\hspace{-1.5pt}p} u_\tau$. 
Starting from forced convection, the trend of the mean profiles to become flatter with increasing $Ri_b$ is consistent with the results of \cite{Pirozzoli_mixed_2017}.
The logarithmic region of $\langle \overline{u} \rangle$ found for weak convective condition starts to deviate at $Ri_b = 0.032$.
The reduction of the spanwise ridge spacing $S$ leads to a decrease of the mean streamwise velocity and temperature profile, which is in agreement with increased surface drag (see table \ref{tab:Grid_study} and figure \ref{fig:St_Cf_Ri} $(b)$).
However, temperature profiles within the transition range from forced convection structures to streamwise rolls show deviations from this behaviour close to the channel core, which is highlighted by the inset of figure \ref{fig:mean_profile_um_Tm_S_Ri_sweep} $(b)$. 
In this, it can be seen for $Ri_b=0.024$ and $Ri_b = 0.032$ that the temperature takes larger values with decreasing $S$ in the channel center and at the same time the slope of the temperature increases. 
This indicates that the thermal mixing of the flow is increasingly weakened by the ridges. 
The transition from streamwise rolls to spot-like structures for case $Ri_b = 0.024$ can be also inferred from the similar slope of $S=\delta$ and $S=0.5\delta$ to the one of the forced convection cases.
Interestingly, for case $Ri_b =0.032$, where all $S$ feature streamwise rolls, the slope of the temperature profile for $S=0.5\delta$ also resembles the one of the forced convection cases, which indicates that the transition point is already close.
}
As can be seen, the influence and effects of the ridge spacing $S$ is of the same order as a change of the bulk Richardson number $Ri_b$.

\begin{figure}
    \centering
    \includegraphics[trim=3pt 0 0 0]{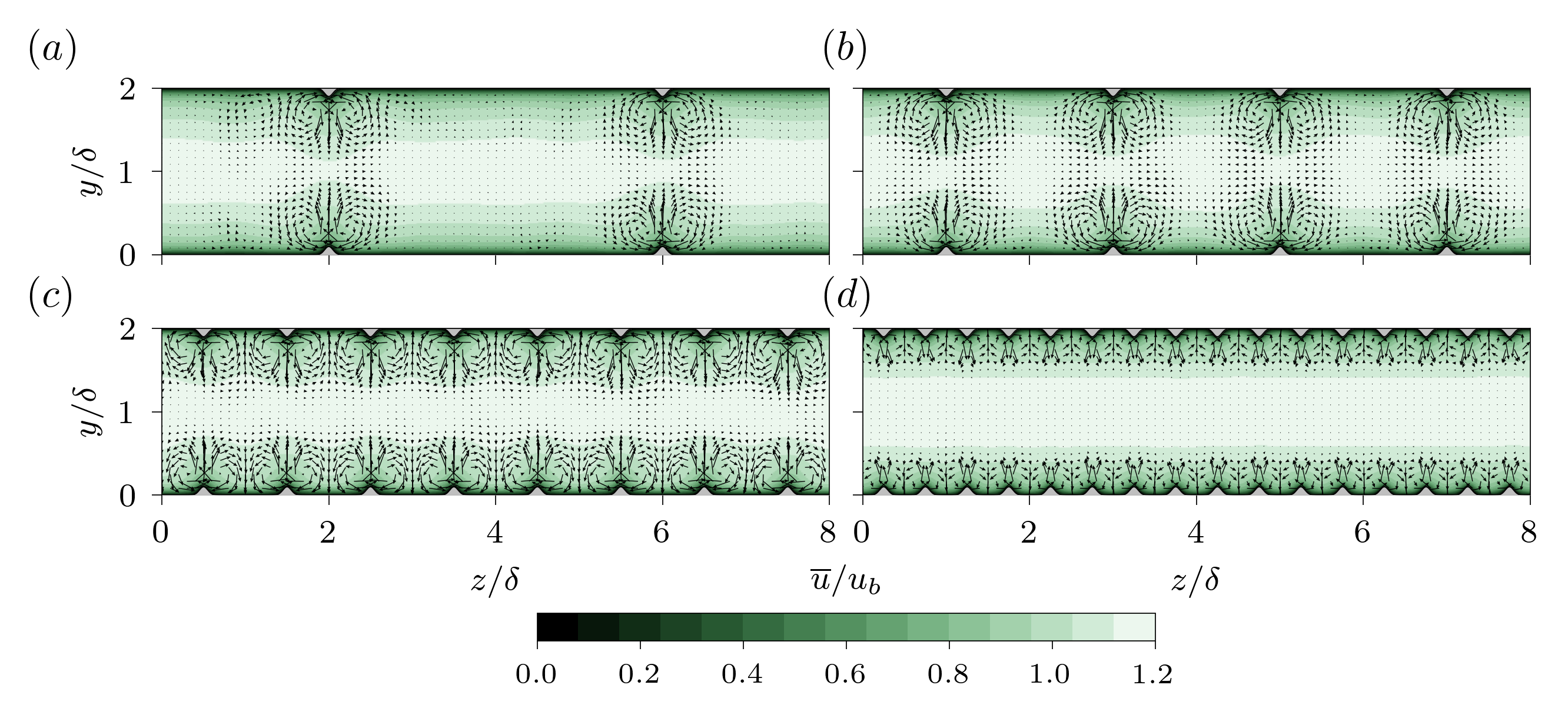}
    \caption{Effect of spanwise spacing on mean streamwise velocity for forced convection case $Ri_b = 0$ ($Re_b = 2800$, $Ra=0$).
    The spanwise spacing of the Gaussian ridges ranges from \textcolor{rev01}{$S = 4\delta$ $(a)$, $S = 2\delta$ $(b)$, $S = \delta$ $(c)$ and $S = 0.5\delta$ $(d)$}. Arrows indicate cross-sectional velocity components and are scaled by bulk velocity. 
    }
    \label{fig:um_xplane_sec_motion}
\end{figure}

The occurrence of secondary motions over streamwise-aligned ridges in forced convection flows is observed in time and streamwise averaged velocity fields in the channel cross section, which is shown for case $Ri_b=0$ in figure \ref{fig:um_xplane_sec_motion}.
\textcolor{rev}{The relative strength of the coherent motion among the different ridge spacings can be directly compared, due to the same scaling of the cross-sectional velocity components in bulk units.}
The smooth channel flow exhibits no coherent motion in the cross section (not shown here), while for streamwise-aligned ridges the secondary motions appear in the mean flow field as counter-rotating vortices at each ridge in figure \ref{fig:um_xplane_sec_motion}.
These vortices introduce an upward motion above each ridge and a downward motion is located to the side of each ridge. 
As can be seen the secondary motion induces a bulging of the mean streamwise velocity above the ridge, transporting low momentum into the bulk region, and for large spacings $S=4\delta$ and $S=2\delta$ reaching almost the half-channel height.
The spacing between the ridges of case $S=4\delta$ in figure \ref{fig:um_xplane_sec_motion} ($a$) is large enough, so that a homogeneous region unaffected by the secondary motion can form between the ridges. 
Decreasing the spanwise spacing $S$, the secondary motions fill almost the entire channel domain for case $S=2\delta$ (figure \ref{fig:um_xplane_sec_motion} ($b$) and case $S=\delta$ (figure \ref{fig:um_xplane_sec_motion} ($c$).
However, for case $S=\delta$ the wall-normal extent of the secondary motions is slightly reduced compare to case $S=2\delta$, which indicates that the secondary motions of adjacent ridges affect each other at a spacing of $S=\delta$.
Further decrease of the ridge spacing to $S=0.5\delta$ (figure \ref{fig:um_xplane_sec_motion} ($d$)) shows a significant reduction of the spatial extent of the secondary motion in the wall-normal direction. 

\begin{figure}
    \centering
    \includegraphics[trim=5pt 0 0 0]{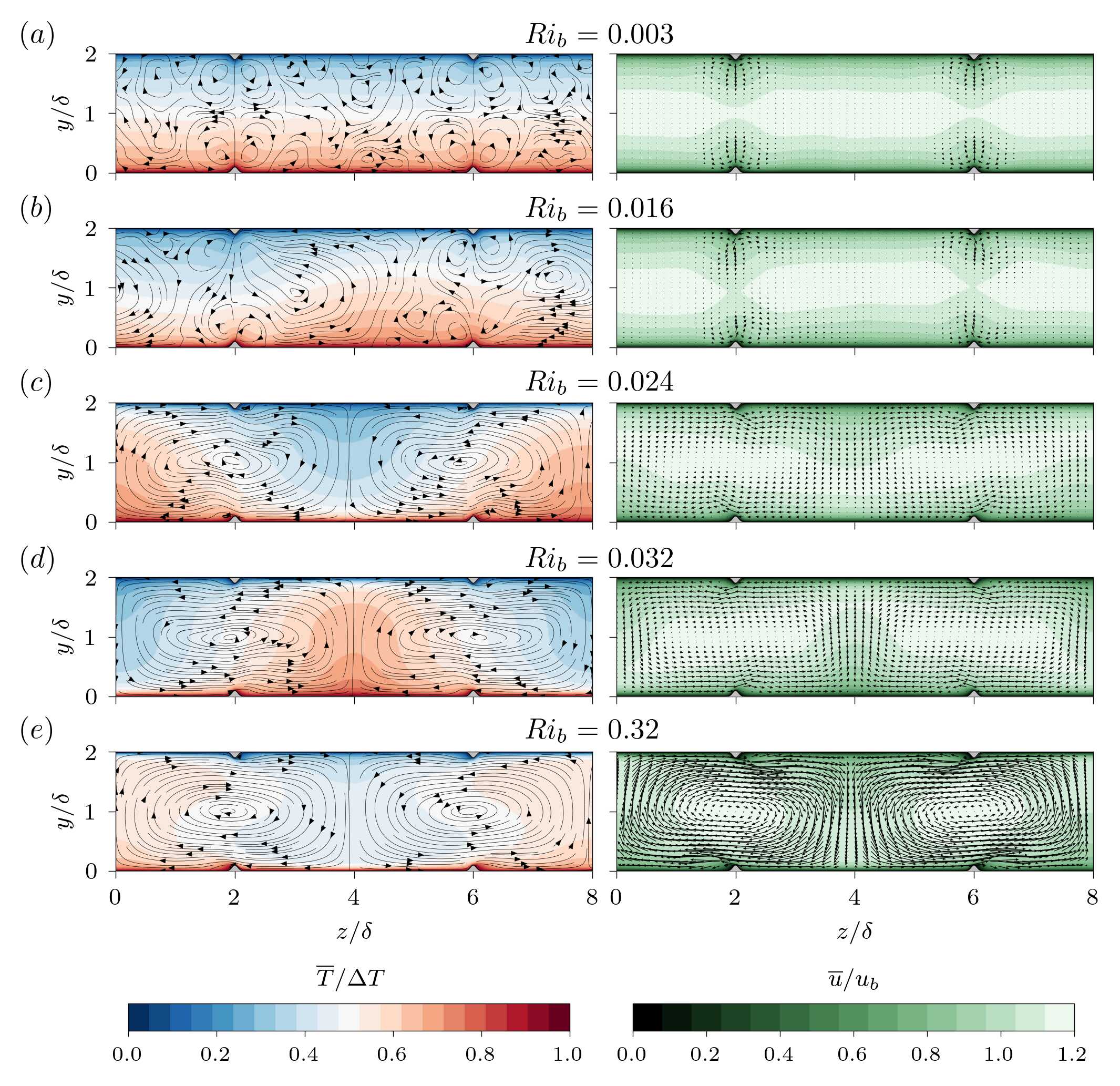}
    \caption{Effect of buoyancy on streamwise mean velocity and temperature for constant $Re_b=2800$ and $S=4\delta$ for different Richardson numbers.
    Arrows indicate cross-sectional velocity components and are scaled by bulk velocity.
    }
    \label{fig:mean_uT_xplane}
\end{figure}

The investigation of the horizontal fields of the instantaneous temperature have shown that the transition of the flow topology for smooth wall conditions is affected by the introduction of the streamwise-aligned ridges. 
This reorganisation is also reflected in the mean streamwise velocity and temperature field in the cross-section, which is shown in figure \ref{fig:mean_uT_xplane} for the transition from forced convection structures to streamwise rolls at a ridge spacing of \textcolor{rev02}{$S=4\delta$}.
The strength of the cross-sectional velocity components are represented by the arrows in the mean streamwise velocity field, while their flow topology is visualized by streamlines in the mean temperature field.
The weak buoyancy case $Ri_b = 0.003$ is shown in figure \ref{fig:mean_uT_xplane} (a) and similar as the forced convection case in figure \ref{fig:um_xplane_sec_motion} secondary motions are present in the mean velocity field which also leads to a bulging of the mean temperature at the ridges. 
\begin{figure}
    \centering
    \includegraphics[trim=5pt 0 0 0]{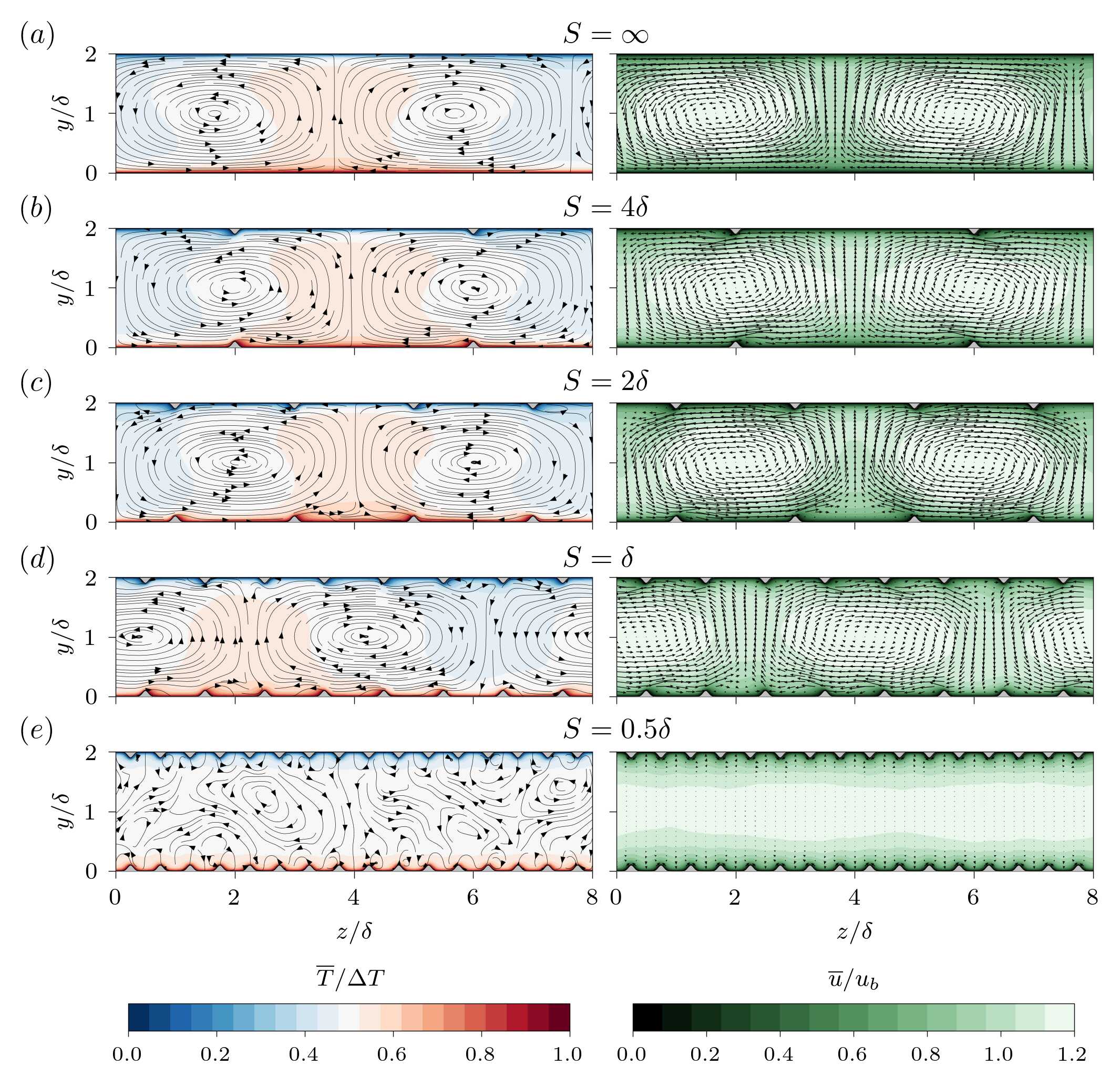}
    \caption{Effect of ridge spacing $S$ on streamwise mean velocity and temperature for constant $Ri_b = 10$.
    }
    \label{fig:mean_uT_xplane_Ri10}
\end{figure}

The transition towards streamwise rolls has been seen to occur for smooth wall conditions at $Ri_b=0.016$, while for $S=4\delta$ a slight tendency towards rolls was present. 
Figure \ref{fig:mean_uT_xplane} $(b)$ shows that secondary motions can still occur at the ridges, though two diminished roll structures emerge that extend to the opposite wall. 
This in turn replaces the local bulging of the mean temperature at the ridge by a significant wider bulging of the mean temperature.
As $Ri_b$ increases to $Ri_b = 0.024$ the secondary motions are replaced by streamwise roll.
The upward and downward motion of the convective rolls is located in the valley between two Gaussian ridges and each roll has a spanwise extent of four half-channel heights.
As can be seen in the streamwise mean velocity field, the rolls induce stronger cross-sectional velocity in the entire channel, which introduces a recirculation zone at the leeward side of the Gaussian ridges.
The cross-sectional velocities of the streamwise roll further intensify as $Ri_b$ increases, as can be seen for case $Ri_b = 0.32$ in figure \ref{fig:mean_uT_xplane} $(e)$.

Figure \ref{fig:mean_uT_xplane_Ri10} shows the effect of $S$ on the mean temperature and mean streamwise velocity for case $Ri_b =10$, which features the transition between streamwise rolls and convective cells.
The bulging of $\overline{T}$ and $\overline{u}$ due to the streamwise roll is found for smooth wall conditions and $S\ge\delta$, while this is not found for the densest ridge spacing $S=0.5\delta$.
For the latter case this reflects the transition from streamwise rolls to convective cells with preferential orientation in $z$-direction found in the instantaneous temperature fields in figure \ref{fig:midplane_T_Ri_Reb} and \ref{fig:yplane_bottom_T_Ri_Reb}, which results in the disappearance of the cross-sectional motion in figure \ref{fig:mean_uT_xplane_Ri10} $(e)$.
The up- and downdrafts of the streamwise rolls for $S=4\delta$ and $S=2\delta$ are located in the valleys between adjacent ridges.
For the former case the lateral movement of the roll encounters the ridge in the middle between up- and downdrafts, where large spanwise velocities of the roll occur. 
The ridges close to the up- and downdrafts for case $S=2\delta$ support the wall-normal motion of the roll by the upward deflection at the ridges, which results in stronger bulging of $\overline{u}$ at the up- and downdraft region compared to $S=4\delta$.
As can be seen for case of $S=\delta$ in figure \ref{fig:mean_uT_xplane_Ri10} $(d)$ the strength of the cross-sectional motion is reduced compared to the coarser $S$ cases, since the roll experience more lateral drag by crossing the ridges due to decreasing $S$.

\subsection{Turbulent properties}
\label{sec:turbulent_structures}

The mean velocity and temperature fields presented in the previous section have shown that secondary motions and streamwise rolls manifest as large-scale coherent motion in the cross-sectional plane.
\textcolor{rev}{The energetics of these structures is further analyzed by applying the decomposition procedures of equations \ref{equ:decomp01} \& \ref{equ:decomp02} to the turbulent kinetic energy $k= 0.5 \cdot \overline{u'_i u'_i}$.
This separates the turbulent kinetic energy $k$ into its coherent contribution $\tilde{k}$ and random contribution $k''$ given by $k = \tilde{k} + k''$.}
In order to extract the influence of the cross-sectional motion, the coherent turbulent kinetic energy $\tilde{k}$ is decomposed into its cross-sectional $\tilde{k}_c = 0.5 \cdot (\tilde{v}\tilde{v} + \tilde{w}\tilde{w})$ and streamwise part $\tilde{k}_s = 0.5 \cdot \tilde{u}\tilde{u}$.
Since the global mean velocity components $\langle \overline{v} \rangle$ and $\langle \overline{w} \rangle$ are zero, the coherent components $\tilde{v}$ and $\tilde{w}$ represent the mean velocity motion in the cross-sectional plane (seen for instance in figure \ref{fig:mean_uT_xplane}).
The coherence of the large-scale motion is quantified by \textcolor{rev}{$K_c$, which is} the volume average of the coherent turbulent kinetic energy of the cross-sectional components $\tilde{k}_c$ \textcolor{rev}{obtained by the integration formula of equation \ref{equ:global_average}.}
While $K_c$ is a good measure of coherence for the majority of cases considered here, it will be shown later that there are two cases for which $K_c$ is not a useful measure. 
The first case applies when the coherent motion involves strong temporal dynamics leading to a reduction of the coherent velocities $\tilde{v}$ and $\tilde{w}$ by long time averages. 
The second case concerns any coherent motion in the $x$-$y$ plane, so that this coherence is masked in the random velocity variance contribution.  
\begin{figure}
    \centering
    \includegraphics[trim=5pt 0 0 0]{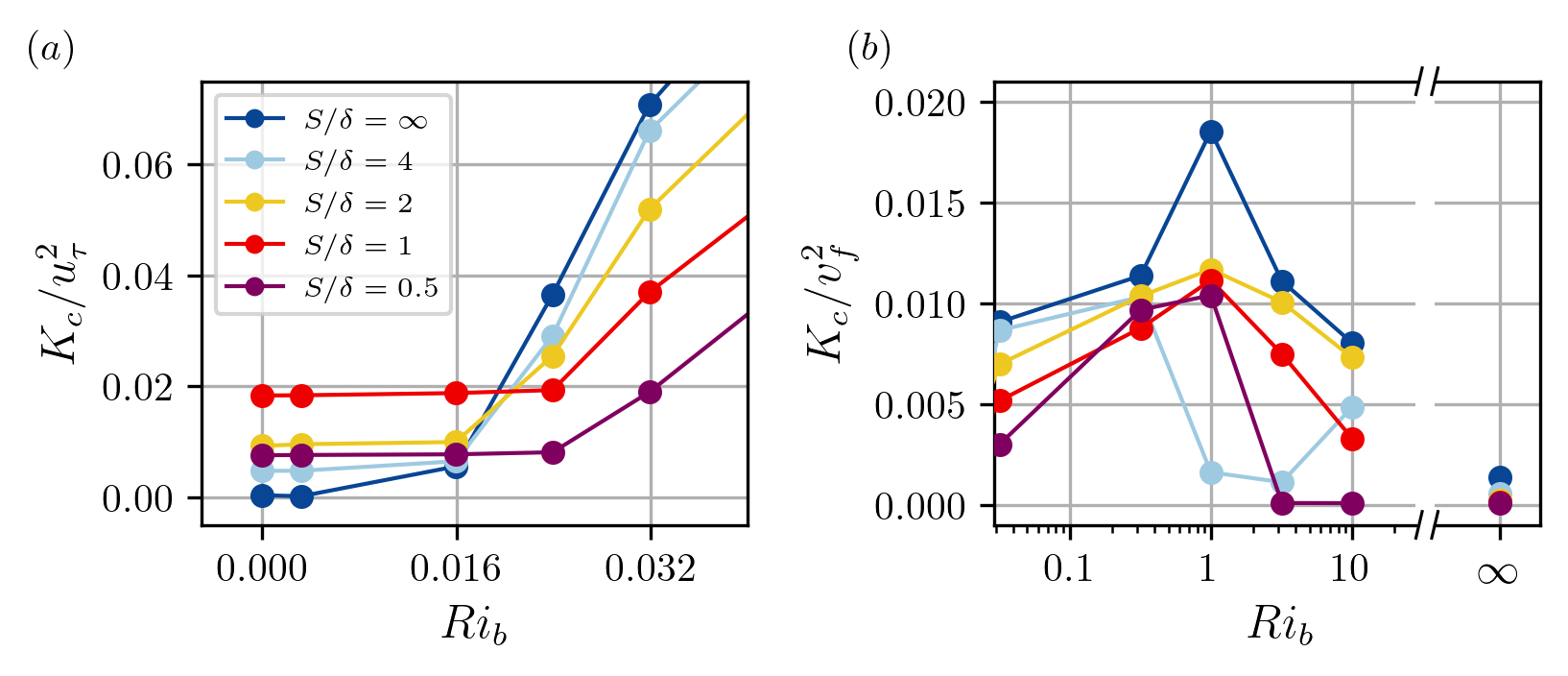}
    \caption{Volume-averaged coherent turbulent kinetic energy of the cross-sectional components for low $Ri_b$ cases scaled in wall-units in $(a)$ and for large $Ri_b$ cases scaled in free-fall units in $(b)$.}
    \label{fig:tke_coherent_global}
\end{figure}

The influence of the transition between forced convection structures and streamwise rolls on $K_c$ is illustrated in figure \ref{fig:tke_coherent_global} $(a)$.
As can be seen the forced convection case $Ri_b = 0$ and weakly convective case $Ri_b = 0.003$ for smooth wall conditions display no coherent energy in the cross-sectional components due to the missing presence of coherent motion. 
However, the introduction of the Gaussian ridges and the appearance of secondary motions results in a coherent kinetic energy contribution $K_c$ with the highest value for a ridge spacing of $S=\delta$, consistent with recent studies \citep{medjnoun_effects_2020,Wangsawijaya_effect_2020}.
At $Ri_b = 0.016$ the streamwise rolls emerge for the smooth wall case, which is reflected by an increase of $K_c$. 
This increase induced by the streamwise rolls is eventually also present for ridge spacings $S=4\delta$ and $S=2\delta$ for case $Ri_b = 0.024$.
As shown in the previous sections, the two densest ridge spacing cases display secondary motions at higher $Ri_b$, which leads to an unchanged and constant value of $K_c$ up to $Ri_b = 0.024$ for $S=\delta$ and $S=0.5\delta$.
This behavior is consistent with the observation of the delayed increase in $Nu$ with increasing $Ra$ for these two cases in figure \ref{fig:Cf_Nu_Re_Ra} $(a)$, illustrating the importance of the flow structures on the scaling of global quantities.
Due to the presence of streamwise rolls for all cases at $Ri_b = 0.032$, $K_c$ also increases for all $S$, whose values almost double compared to $Ri_b = 0.024$.

The change of coherence due to the transition between streamwise rolls and convective cells with increasing $Ri_b$ is shown in figure \ref{fig:tke_coherent_global} $(b)$.
For this range of values $Ri_b$, $K_c$ is scaled in free-fall units, which ease the comparison with the natural convection case $Ri_b = \infty$.
For the smooth wall condition the coherence increases up to $Ri_b = 1$ and subsequently decreases to the natural convection case. 
This maximum of $K_c$ occurs for a value of $-\delta_\textit{eff}/L = 1.23$, which is consistent with recent findings in ABL, for which the maximum coherence of streamwise rolls are found at $-z_i/L = 1.08$ \citep{jayaraman_transition_2021}.
For the rough wall cases the coherence decreases monotonically with decreasing $S$ only for $Ri_b =0.032$ and $Ri_b=\infty$, while this behaviour is not found for values in between. 
Comparing to the smooth wall cases the introduction of coarsely spaced ridges $S=4\delta$ yield a large
drop of $K_c$ for $Ri_b = 1$ and $Ri_b =3.2$. 
The reason for this reduction is the aforementioned temporal variability of the streamwise rolls, which causes the up- and downdrafts to slowly move in the spanwise direction over a long period of time instead of being fixed, thereby reducing averaged values of $\tilde{v}$ and $\tilde{w}$ and thus $K_c$. This will be discussed in more detail in the following section \ref{sec:variability_rolls}.

For $Ri_b = 10$ the value of $K_c$ for $S=4\delta$ is below $S=2\delta$, indicating that the coherence of the streamwise roll is more affected by the coarser ridge spacing. 
In this case the reduction of $K_c$ is not related to time variability of the streamwise rolls, but to 
the relative position of the ridges to the up- and downdrafts. 
As can be seen in figure \ref{fig:mean_uT_xplane_Ri10} the up- and downdrafts for $Ri_b = 10$ occur between the ridges, which for $S=4\delta$ results in the roll encountering a ridge in the middle of its lateral motion, which causes stronger lateral drag and thereby weaken the roll motion.  
For $S=2\delta$, the ridges do not interfere the roll motion at their strongest lateral velocity. 
Instead, the adjacent ridges at the up- and downdrafts support the upward motion of the roll by its wall-normal deflection at the ridges. 
Even though denser ridge spacings contribute to more drag, the support of the deflections compensate a part of the losses in $K_c$ for $S=2\delta$, while this does not occur for $S=4\delta$.

For $Ri_b = 3.2$ and $Ri_b = 10$ the value of $K_c$ vanishes for the densest ridge spacing $S=0.5\delta$, which reflects that streamwise rolls are not present for these cases as can be seen in the instantaneous temperature fields in figure \ref{fig:midplane_T_Ri_Reb}.
Also for the natural convection case the value of $K_c$ approaches zero for $S\le2\delta$, even though the instantaneous temperature fields in figure \ref{fig:midplane_T_Ri_Reb} suggest an increase of the coherence in the $x$-direction due to rolls aligned in the $z$-direction. 
This reflects the property of $K_c$ that only coherent motion in the $z$-direction can be detected, while any coherence in the $x$-direction is masked.
\textcolor{rev02}{Consequently, for the current ridge cases, a reduction of $S$ leads to a weakening of the coherence in $z$-direction, which is equivalent to a weakening of rolls with orientation in $x$-direction. 
Note, that for a further reduction of $S$ down to the limit of $S\rightarrow 0$ the surface will approach a smooth wall again with a reduced cross-sectional area.
Since $Re_b$ and $Ra$ are kept constant while varying $S$ we expect the flow for $S\rightarrow 0$ to be similar to the present smooth wall case $S=\infty$.}

\begin{figure}
    \centering
    \includegraphics[trim=5pt 0 0 0,scale=0.75]{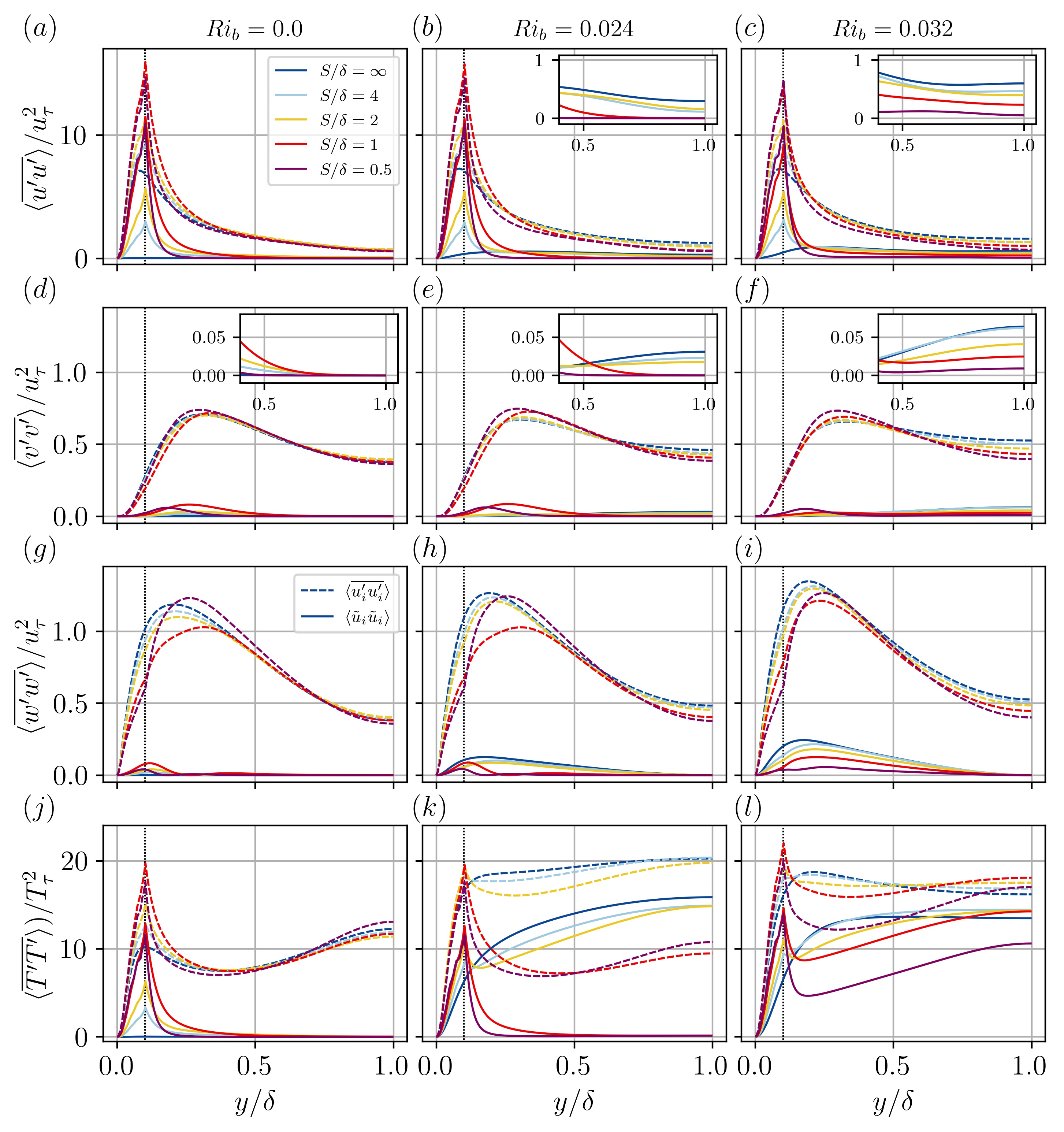}
    \caption{Velocity and temperature variances scaled in inner units for low bulk Richardson number $Ri_b$ cases at transition from forced convection structures to streamwise rolls.
    Black vertical dotted line indicates the height of the Gaussian ridges.}
    \label{fig:variance_S_sweep_low_Ri}
\end{figure}
\vspace{10pt}

The strength of the different large-scale coherent motion is associated with different wall-normal regions, where coherence is dominant.
This is shown for the transition between the forced convection structures and streamwise rolls for the  horizontally averaged velocity stresses and temperature variance in figure \ref{fig:variance_S_sweep_low_Ri} for three different $Ri_b$ cases.
The dashed line represents the Reynolds-stresses (e.g. $\langle\overline{u'u'}\rangle$), while the solid line indicates the coherent stress (e.g. $\langle\tilde{u}\tilde{u}\rangle$).
The difference of both contributions results in the random stress, e.g. $ \langle \overline{u'' u''} \rangle = \langle\overline{u' u'} \rangle - \langle\tilde{u}\tilde{u}\rangle$. 
For the forced convection case $Ri_b = 0$ and rough wall condition the coherent streamwise stress $\langle\tilde{u}\tilde{u}\rangle$ is concentrated close to the wall, with increasing peak values with decreasing $S$ down to $S=\delta$. 
The densest ridge spacing $S=0.5\delta$ has a similar peak value as $S=\delta$, however extending less into the bulk region, consistent with the reduced spatial extent of the mean secondary motions for this ridge spacing seen in figure \ref{fig:um_xplane_sec_motion}.
This is also reflected by the wall-normal location of the peak values of $\langle\tilde{v}\tilde{v}\rangle$ and $\langle\tilde{w}\tilde{w}\rangle$, which is located closer to the wall for $S=0.5\delta$.
The coherent temperature variance displays a similar trend as $\langle\tilde{u}\tilde{u}\rangle$ with respect to $S$, since temperature is a passive scalar for this case resulting in a strong correlation between the streamwise velocity and temperature.

As discussed before, streamwise rolls are present for case $Ri_b = 0.024$ and $S=\infty, 4\delta, 2\delta$, which can be seen most clearly by increased values of $\langle\tilde{T}\tilde{T}\rangle$ and $\langle \overline{T' T'} \rangle$ within the entire bulk region in figure \ref{fig:variance_S_sweep_low_Ri} $(k)$.
For these cases the coherent temperature variance $\langle\tilde{T}\tilde{T}\rangle$ contributes for a large fraction of the temperature variance $\langle \overline{T' T'} \rangle$, which reflects the strong bulging of the mean temperature seen in figure \ref{fig:mean_uT_xplane} $(c)$.
The induced coherence by streamwise rolls is also seen in the coherent velocity stresses but less pronounced.
Among them this is most noticeable for the spanwise coherent velocity stress $\langle\tilde{w}\tilde{w}\rangle$ in figure \ref{fig:variance_S_sweep_low_Ri} $(h)$, where stronger spanwise coherent stresses are observed with respect to the forced convection cases (figure \ref{fig:variance_S_sweep_low_Ri} $(g)$).
A slight increase of $\langle\tilde{u}\tilde{u}\rangle$ and $\langle\tilde{v}\tilde{v}\rangle$ can be also found for these cases in the bulk region, which is illustrated by the figure insets in figure \ref{fig:variance_S_sweep_low_Ri} $(b)$ and $(e)$.
Although the coherent velocity stresses of the streamwise rolls are for this case rather weak, this motion is sufficient to cause a strong imprint in the coherent temperature variance.
For the two densest ridge spacings $S=\delta$ and $S=0.5\delta$ where secondary motions occur, the velocity stresses and temperature variances remain similar to the forced convection cases at $Ri_b = 0$.
This further supports the fact that the increased drag and vertical mixing due to the ridges for these two cases is strong enough to inhibit the formation of streamwise rolls. 

 \begin{figure}
    \centering
    \includegraphics[trim=5pt 0 0 0,scale=0.75]{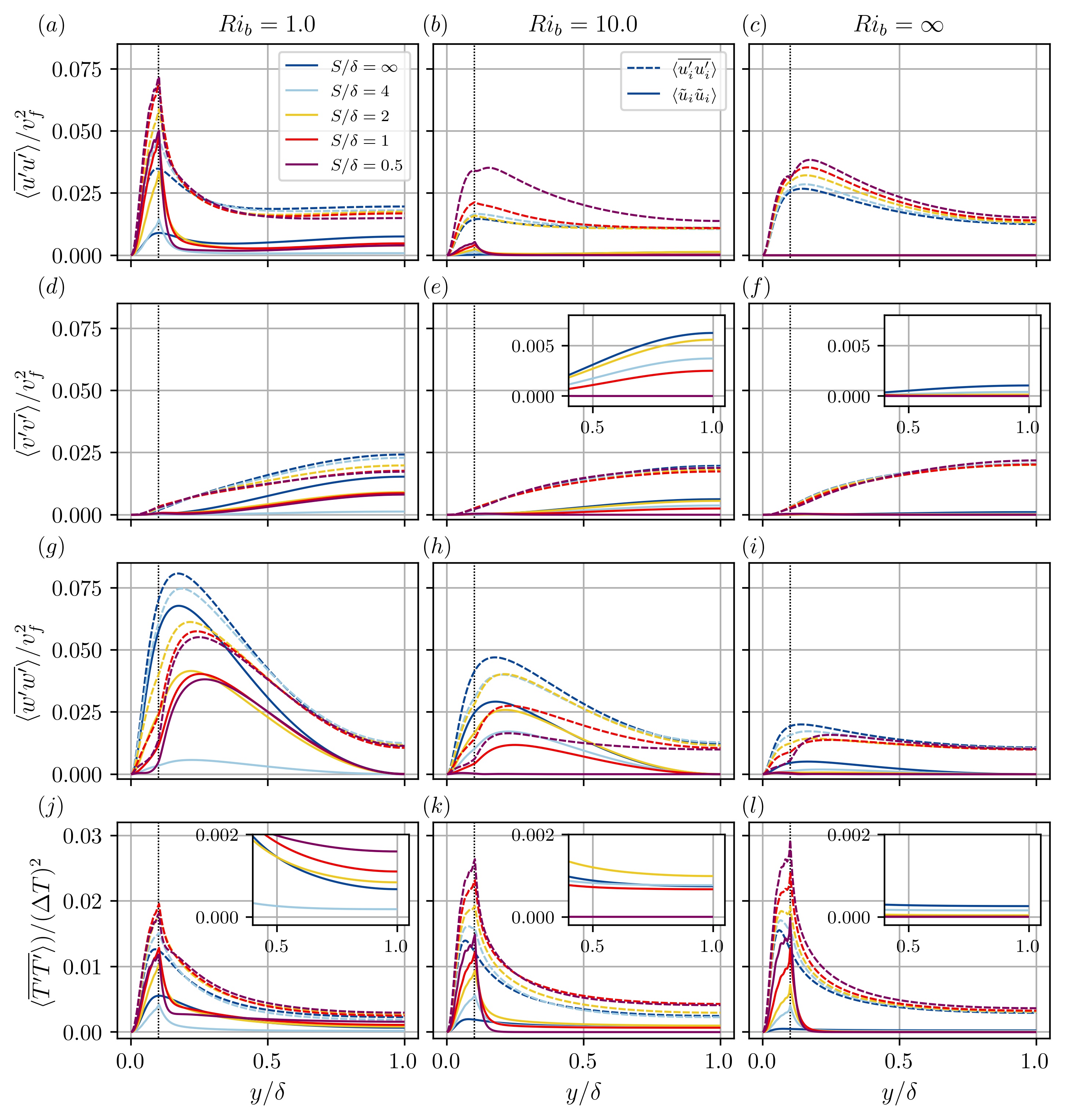}
    \caption{Velocity and temperature variances scaled in free fall units for high bulk Richardson number $Ri_b$ cases at transition from streamwise rolls to natural convection.}
    \label{fig:variance_S_sweep_high_Ri}
\end{figure}

For case $Ri_b = 0.032$, when all ridge configurations exhibit streamwise rolls, the two densest ridge spacings $S$ now also show a significant increase in the coherent temperature variance $\langle\tilde{T}\tilde{T}\rangle$, although their values are lower than for the streamwise roll cases of $Ri_b = 0.024$.
Likewise, the increase of coherent velocity stresses, which initiated at $Ri_b=0.024$, continues, which is clearly seen by  $\langle\tilde{w}\tilde{w}\rangle$ in figure \ref{fig:variance_S_sweep_low_Ri} $(i)$. 
As can be seen the successive reduction of the ridge spacing $S$ results in a decrease of the coherent velocity stresses, indicating that the streamwise rolls are damped by the presence of the ridges. 
For $S=0.5\delta$ only a mild increase of $\langle\tilde{u}\tilde{u}\rangle$ and $\langle\tilde{v}\tilde{v}\rangle$ is found in the bulk region (figure \ref{fig:variance_S_sweep_low_Ri} $(c,f)$), and the near-wall peak of $\langle \tilde{v}\tilde{v}\rangle$, introduced by the secondary motions, is still visible.
The persistence of stronger wall-normal coherent motions near the wall for $S=0.5\delta$, similar to the forced convection cases, is consistent with the streamwise rolls to appear more spot-like as seen in figure \ref{fig:midplane_T_Ri_Reb}.

Figure \ref{fig:variance_S_sweep_high_Ri} shows the velocity stresses and temperature variance for the transition from streamwise rolls to natural convection.
For case $Ri_b = 1$ the streamwise roll displays the strongest coherent cross-sectional motion, which is illustrated by larger values of $\langle\tilde{w}\tilde{w}\rangle$ compared to the $\langle\tilde{u}\tilde{u}\rangle$.
At the same time the wall-normal Reynolds stresses have comparable magnitude to the streamwise Reynolds stress in the channel center region.
The decrease of coherent velocity stresses with decreasing $S$ is found for the three densest ridge spacings, while case $S=4\delta$ display significantly lower values due to the time variability of the streamwise rolls.  
At the same time, the coherent temperature variance in the bulk region increases with decreasing $S$ (inset figure \ref{fig:variance_S_sweep_high_Ri} $(j)$) and $\langle \overline{T' T'} \rangle$ exhibits larger values in the near-wall region compared to the smooth wall case.

As discussed in relation to figure \ref{fig:tke_coherent_global} the coherence of streamwise rolls is reduced when increasing $Ri_b$ beyond values of $Ri_b \approx 1$, which is reflected by a reduction of the coherent velocity stresses for $Ri_b = 10$.
For $Ri_b = 10$ and dense ridge spacings $S=\delta$ and $S=0.5\delta$ the instantaneous temperature fields in figure \ref{fig:midplane_T_Ri_Reb} have shown a transition from the streamwise rolls to convection cells, which is reflected here by significant lower coherent stresses $\langle \tilde{v}\tilde{v}\rangle$ and $\langle \tilde{w}\tilde{w}\rangle$ compared to the smooth wall case.
The increased peak value of $\langle \overline{T'T'} \rangle$ with decreasing $S$ indicates that ridges are more efficient in mixing temperature close to the wall. 
For the densest ridge spacing $S=0.5\delta$ the convection cells are oriented along the spanwise direction, which is in agreement with the observation of zero coherent spanwise stresses $\langle \tilde{w}\tilde{w}\rangle$ in figure \ref{fig:variance_S_sweep_high_Ri} $(h)$.

The observation that the coherence of convective cells oriented along the $x$-direction is significantly reduced for $S\le2\delta$ is also reflected by the coherent velocity stresses and coherent temperature variance. 
The coherent contribution of $\langle \tilde{w}\tilde{w}\rangle$ for $S=4\delta$ is significantly reduced compared to smooth wall conditions, as can be seen in figure \ref{fig:variance_S_sweep_high_Ri} $(i)$. 
The only increase of coherence can be found for $\langle \tilde{T}\tilde{T}\rangle$ in the near-wall region with decreasing $S$ in figure \ref{fig:variance_S_sweep_high_Ri} $(l)$, while it vanishes in the bulk region for dense spacings.
The preferred orientation of the convective cells in the spanwise direction, as seen in figure \ref{fig:midplane_T_Ri_Reb}, needs to result in larger streamwise motion, which is reflected in the steady increase of $\langle \overline{u'u'}\rangle$ with $S$ (figure \ref{fig:variance_S_sweep_high_Ri} $(c)$).
This is also accompanied by a steady increase of $\langle \overline{T'T'}\rangle$ with decreasing $S$, which suggest that the aligned ridges induce stronger thermal plumes. 
Note, that in a square domain with smooth wall condition the streamwise and spanwise stresses have the same distribution due to the directional invariance of the cells \citep{pandey_turbulent_2018}. 
However, the smooth wall case already displays slightly larger values for $\langle \overline{u' u'} \rangle$ than $\langle \overline{w' w'} \rangle$ and this indicates that the convection cell are slightly more oriented in the spanwise direction before introducing the aligned ridges.

\subsection{Variability of streamwise rolls}
\label{sec:variability_rolls}
\begin{figure}
    \centering
    \includegraphics[trim=6pt 0 0 0]{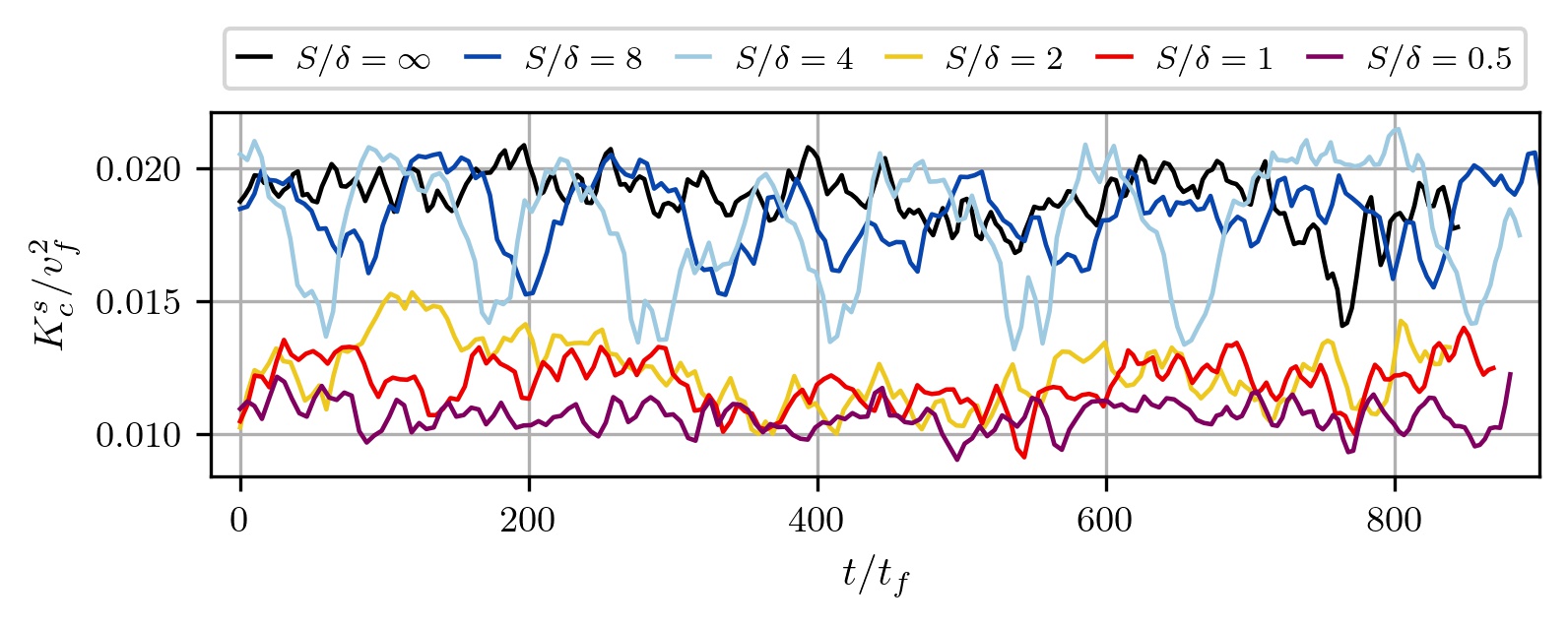}
    \caption{Short-time averaged coherent kinetic energy $K^s_c$ of case $Ri_b = 1$ and different ridge spacings over time. The values of $K^s_c$ are averaged for time intervals $\Delta t_s \approx 3.4t_f$.}
    \label{fig:time_coherent_kinetic_energy}
\end{figure}

In the previous section it was found that the coherence of the streamwise rolls for coarse ridge spacings $S=4\delta$ drops significantly for case $Ri_b = 1$ and $Ri_b = 3.2$ compared to the denser values of $S$, indicating weaker cross-sectional motion of the roll.
However, the instantaneous temperature visualizations in figure \ref{fig:midplane_T_Ri_Reb} and \ref{fig:yplane_bottom_T_Ri_Reb} for this specific case do not indicate weaker streamwise rolls, suggesting that a time-varying behaviour of the streamwise rolls might be present. 
For this purpose, the volume-averaged coherent turbulent kinetic energy of the cross-sectional components $K_c$, which is based on the average of the entire time series, is now averaged for shorter time-windows.
The short-time averaged coherent turbulent kinetic energy $K_c^s$ (superscript $s$ indicates the short-time average) is computed over a time range of $\Delta t_s \approx 3.4t_f$.
Note, that the value of the short-time average is the shortest available data for the present simulations. 
The time evolution of $K_c^s$ for consecutive short-time intervals is shown for case $Ri_b = 1$  in figure \ref{fig:time_coherent_kinetic_energy}. 
As can be seen, all cases feature relatively slow dynamics and for $S=\infty$ and $S\le2\delta$ the time variation vary mildly around their full time-averaged value $K_c$ in figure \ref{fig:tke_coherent_global} $(b)$.
For $S=4\delta$ the time variation is more pronounced and the dynamics show clearly visible periodic reduction of $K_c^s$ with a period of $\mathcal{O}(t) \approx 100\, t_f$, which corresponds to $\mathcal{O}(t) \approx 200\, t_b$. 
\textcolor{rev}{For this analysis an additional simulation with $S=8\delta$ is performed and it shows similar time variation compared to $S=4\delta$ however with smaller amplitude.}
These observed dynamics are \textcolor{rev02}{significantly} slower than observations in ABL where the dynamic of the flow reaches a statistically quasi-steady state in roughly $6\, t_f$ \citep{moeng_comparison_1994}.
The time mean value of $K_c^s$ is significantly larger than the value $K_c$, which suggest, that the streamwise rolls of $S=4\delta$ \textcolor{rev}{and $S=8\delta$} feature some time variability, which is masked by considering quantities based on the average of the entire time series. 
This time-varying behaviour is also found for $Ri_b =3.2$ and $S=4\delta$, which also displayed a significant reduction of $K_c$ in figure \ref{fig:tke_coherent_global}.

\begin{figure}
    \centering
    \includegraphics[trim=6pt 0 0 0]{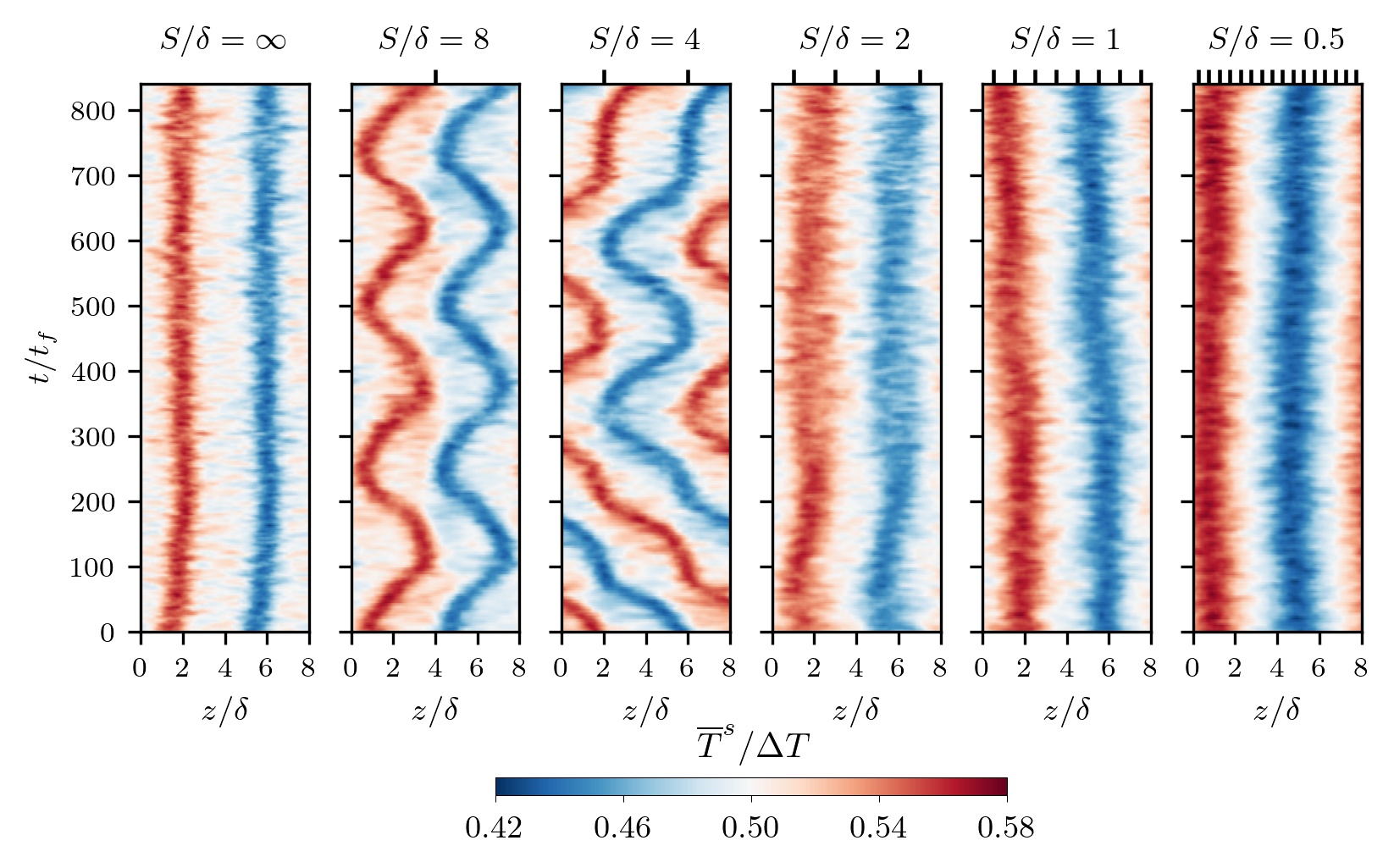}
    \caption{Streamwise and short-time averaged temperature $\overline{T}^s$ over time and spanwise position at the wall-normal channel center location $y=\delta$ for cases $Ri_b = 1$. The spanwise position of the ridges is indicated by the black lines at the top figure frame. 
}
    \label{fig:roll_position_time}
\end{figure}

The variability of the streamwise rolls is illustrated by the time evolution of the short-time and streamwise averaged temperature $\overline{T}^s$ at the wall-normal channel center $y=\delta$ along the spanwise direction $z$ in figure \ref{fig:roll_position_time}.
The spanwise position of the thermal up- and downdrafts of the streamwise roll are represented by the higher and lower temperature values, respectively. 
While the spanwise location of the up- and downdrafts remain at the same position for $S=\infty$ and $S\le2\delta$, the spanwise location of the up- and downdrafts of case $S=4\delta$ \textcolor{rev}{and $S=8\delta$} are strongly varying in time.
The up- and downdrafts \textcolor{rev}{for these two cases} exhibit strong lateral movement.
\textcolor{rev}{While this movement is almost periodic for $S=8\delta$ and remains between the ridges, the up- and downdrafts of $S=4\delta$ are able to cross} the ridges at some time instances, e.g. $t/t_f \approx 100$, while they are not able to cross them at other time instances, e.g. $t/t_f \approx 350$.
The large values of $K_c^s$ of $S=4\delta$ \textcolor{rev}{and $S=8\delta$} in figure \ref{fig:time_coherent_kinetic_energy} corresponds to occasions, when the up- and downdrafts are located \textcolor{rev}{close or directly} at the ridges, e.g. $t = 700-800t_f$, while small values of $K_c^s$ corresponds to locations of the up- and downdrafts in between the ridges. 
This increase of $K_c^s$ can be interpreted by the formation mechanism of streamwise rolls proposed by \cite{khanna_ABL_1998}, which relates them to the organization of localized buoyancy forces within near-wall streaks.
When the up- and downdrafts are located at ridges, the ridges support the formation of strong localized buoyancy forces, leading to strong local up- and downdrafts. 
Due to the symmetric arrangement of the ridges at both walls, the up- and downdrafts impinge at another ridge on the opposite wall, which is supposed to counteract the impinging roll by the formation of localized buoyancy forces with opposing direction.
Thus, for \textcolor{rev}{$S=4\delta$} the ridges support the formation of strong up- and downdrafts, however they cannot remain at the spanwise location due to the counteraction of the opposing ridge, leading to the lateral evasion. 
\textcolor{rev}{For case $S=8\delta$ this enhancement at the ridges occurs only for the up- or the downdraft and not simultaneously as for $S=4\delta$, which might be an explanation for the observation that up- and downdrafts are not able to cross the ridges.}

For case $S=2\delta$ a short time interval $t \approx 120t_f$ with large values of $K_c^s$ is present, which corresponds to a time interval in which the up- and downdrafts are located above ridges.
However, for most of the time the up- and downdrafts remain in between adjacent ridges as has been shown in figure \ref{fig:mean_uT_xplane_Ri10} and only a slight meandering within this range is observed. 
The examination of time series of cases $Ri_b = 0.32$ and $Ri_b = 10$ do not reveal this strong lateral movement of the up- and downdraft location, while for $Ri_b = 3.2$ and $S=4\delta$ a similar lateral movement is found. 
\textcolor{rev}{The results suggest that the dynamics of the streamwise rolls is very sensitive to ridge spacings in the order of the spanwise rolls' width as seen for $S=4\delta$ and $S=8\delta$.}
For denser ridge spacings $S\le 2\delta$ several adjacent ridges contribute by localized buoyancy forces to the formation of the up- and downdrafts which might be strong enough to inhibit disturbances by the opposing ridges and thereby prevent lateral movement of the streamwise rolls.
\textcolor{rev}{
Future investigations with a staggered ridge arrangement or ridges placed only at the bottom or top wall might further shed light on the influence of ridges on the roll formation.
For instance, the comparison of a symmetric and staggered arrangement of streamwise-aligned ridges in forced convection flows \citep{stroh_secondary_2020} has shown that a staggered arrangement promotes the coherence of the large-scale secondary motion, which might also be valid for rolls and may lead to a fixation of the rolls for $S=4\delta$.
}

\subsection{Reynolds number effects}
\label{sec:reynolds_effects}

\begin{figure}
    \centering
    \includegraphics[trim=3pt 0 0 0]{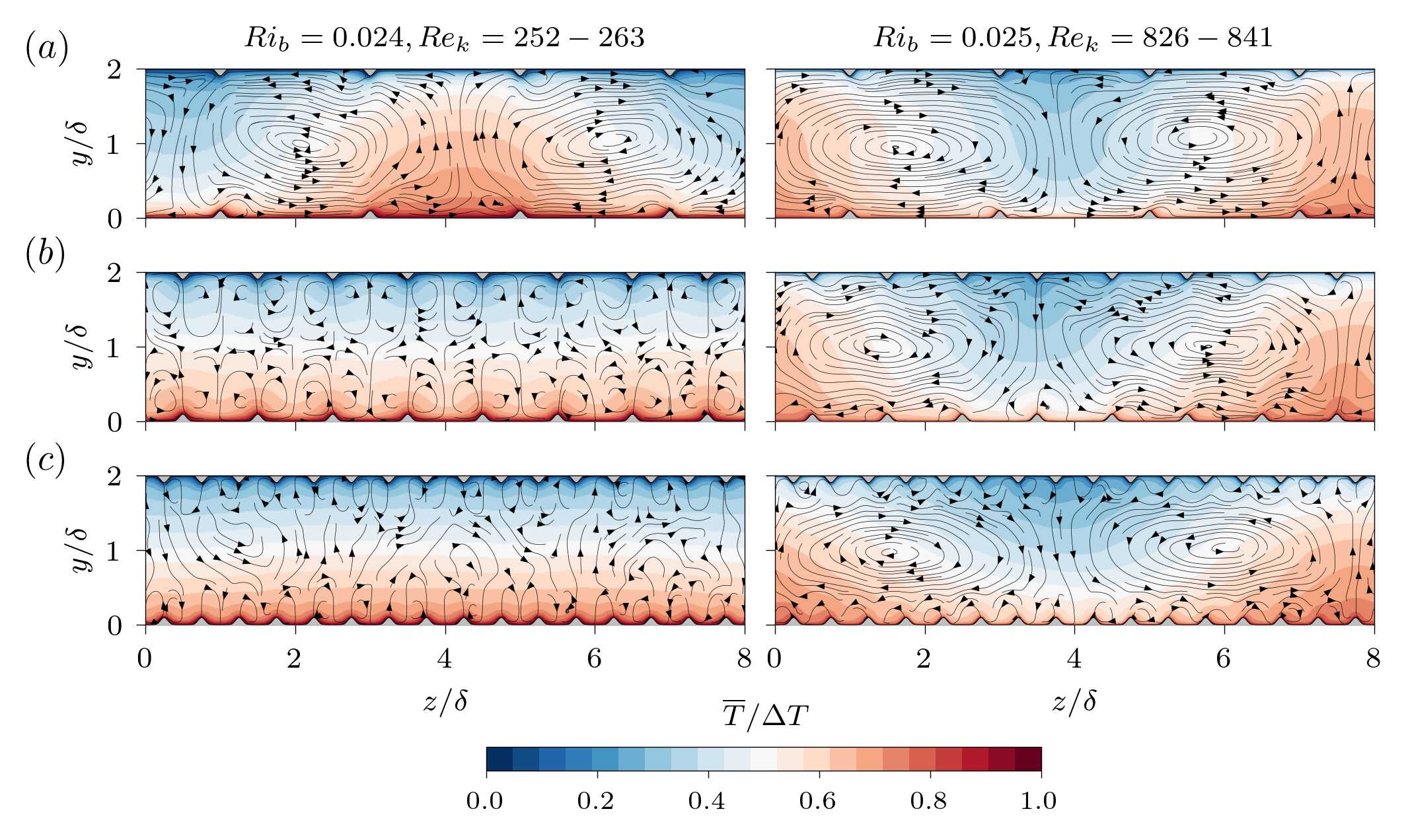}
    \caption{Effect of turbulent Reynolds number $Re_k$ an spanwise ridge spacing $S$ on mean temperature for case $Ri_b=0.024, Re_k = 252$-$263$ ($Ra=7.5\cdot10^5$, $Re_b=2800$) on the left side and case $Ri_b=0.025, Re_k = 826$-$841$ ($Ra=10^7$, $Re_b=10000$) on the right side. 
    The spanwise spacing of the Gaussian ridges ranges from $S = 2\delta$ $(a)$, $S = \delta$ $(b)$ and $S=0.5\delta$ $(c)$. 
    }
    \label{fig:mean_T_xplane_S_sweep_compare_Ri_low}
\end{figure}

In the previous section the flow organization of the mixed convection flows was considered in terms of varying bulk Richardson number $Ri_b$ and ridge spacing $S$, while the Reynolds number $Re_k$ was approximately constant.
\textcolor{rev03}{
The effect of $Re_k$ on the transition between forced convection structures and streamwise rolls is shown in figure \ref{fig:mean_T_xplane_S_sweep_compare_Ri_low}, which presents the cross-sectional mean temperature and flow topology for  case $Ri_b=0.024$ with values of $Re_k = 252$-$263$ and case $Ri_b=0.025$ with threefold larger values of $Re_k = 826$-$841$.
}
\textcolor{rev03}{As can be seen figure \ref{fig:mean_T_xplane_S_sweep_compare_Ri_low} $(a)$ both cases feature a streamwise roll down to a ridge spacing of $S=2\delta$ (not shown for $S=\infty$ and $S=4\delta$).}
The comparison of the temperature fields between the low and high $Re_k$ cases depicts that the thermal boundary layer is reduced for higher $Re_k$ due to the more efficient mixing of the flow in the near wall region.
As shown in the previous sections, the streamwise rolls are replaced by secondary motions for the lower $Re_k$ cases with $Ri_b = 0.024$ and $S\le\delta$, \textcolor{rev03}{while for the larger $Re_k$ cases the streamwise roll remains for these ridge spacings.
However, the streamwise roll appears more distorted and affected by the ridges as can be seen for $S=\delta$ and $S=0.5\delta$.} 
In addition, secondary motion in form of one pair of counter-rotating vortices emerge at one ridge at the bottom wall lying in the downdraft region of the roll. 
This illustrates, that the ridges on the opposing wall of the up- and downdraft region are able to \textcolor{rev03}{form coherent structures that} counteract the large-scale roll formation.
For the densest ridge spacing $S=0.5\delta$ in figure \ref{fig:mean_T_xplane_S_sweep_compare_Ri_low} $(c)$ the streamwise roll for the large $Re_k$ case is now confined to a smaller wall-normal region in the bulk flow. 
\textcolor{rev03}{This is associated with} the recirculation zones at the leeward side of the ridges, which are connected between the up- and downdraft regions, and thereby forming a roughness sublayer which inhibits the attachment of the lateral movement of the streamwise roll at the wall. 
These results suggest, that the transition range between forced convection structures and streamwise rolls with heterogeneous rough surfaces is not solely determined by the pair of $Ri_b$ and $S$, but also by the value of the Reynolds number $Re_k$. 
\textcolor{rev03}{Due to the increased turbulent mixing for larger values of $Re_k$ the streamwise rolls can counteract the additional shear by the ridges, such that the transition between forced convection structures and streamwise rolls is shifted towards smaller values of $Ri_b$.}

\begin{figure}
    \centering
    \includegraphics[trim=3pt 0 0 0]{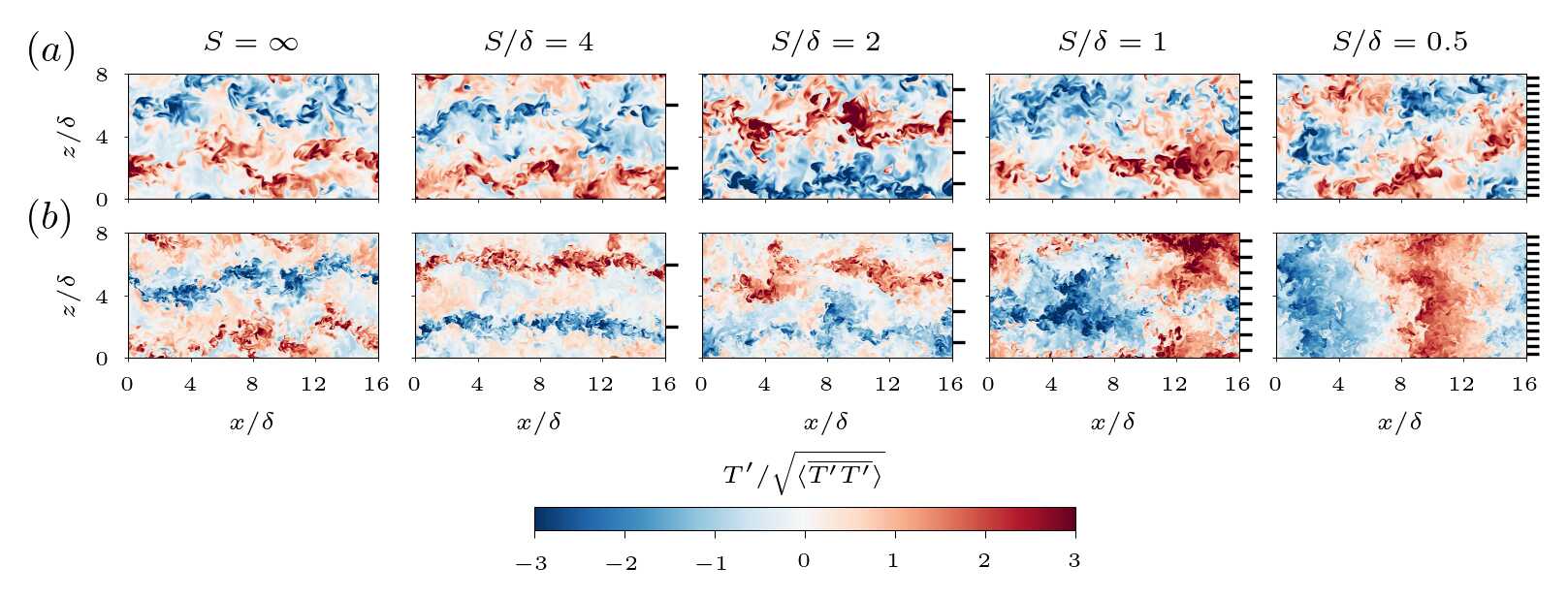}
    \caption{Instantaneous temperature fluctuation fields at the half-channel height position $y=\delta$ for cases $Ri_b = 3.2$ with $Re_k = 236$-$263$ ($Ra=10^7$, $Re_b=885$) in $(a)$ and $Re_k = 695$-$792$ ($Ra=10^8$, $Re_b=2800$) in $(b)$ for different spanwise ridge spacing $S$.}
    \label{fig:comparison_T_instant_3.2}
\end{figure}

\textcolor{rev03}{The influence of $Re_k$ on the roll-to-cell transition is illustrated for two cases with $Ri_b = 3.2$ and different $Re_k$ by the temperature fluctuation in the horizontal mid-plane in figure \ref{fig:comparison_T_instant_3.2}.}
As can be seen both cases exhibit streamwise rolls for smooth wall conditions and $S\ge2\delta$, while differences in the flow structures start to appear at $S=\delta$.
At this ridge spacing the lower $Re_k$ case still shows streamwise rolls (figure \ref{fig:comparison_T_instant_3.2} $(a)$), while for the larger $Re_k$ case the streamwise rolls are disturbed by strong thermal plumes spanning almost the entire spanwise domain.
For the lower $Re_k$ case with $S=0.5\delta$ no coherent streamwise rolls can be observed and similar to $S=\delta$ for the larger $Re_k$ case thermal plumes emerge, which indicates the beginning of the transition to convective cells.
This shows that the increasing $Re_k$ has a comparable effect on the flow organization as the reduction of the ridge spacing $S$. 
While smaller values of $S$ increase the friction of the flow and weaken the lateral motion of the streamwise rolls, the higher $Re_k$ increases the thermal mixing near the wall, and both effects promote the formation of thermal plumes.
The loss of coherence of the streamwise rolls for case $S=0.5\delta$ in figure \ref{fig:comparison_T_instant_3.2} ($a$) is also supported by a vanishing value of $K_c$ in figure \ref{fig:tke_coherent_global} $(b)$.
Also the inspection of the time series of $K_c^s$ reveals only a very weak contribution, which is an order magnitude lower than for case $S=\delta$. 
The observation that only convective cells with preferential orientation in $z$-direction occur for the natural convection cases $S\le2\delta$ and for $Ri_b=10$ with $S=0.5\delta$ is also seen for the large $Re_k$ case with $S=0.5\delta$.
The influence of $Re_k$ on the transition between streamwise rolls and convective cells is such that higher values of $Re_k$ initiate this transition to convective cells at smaller $Ri_b$, thus reducing the range of streamwise rolls.

\section{Discussion and conclusion}

The present study shows that heterogeneous \textcolor{rev01}{surfaces} in form of streamwise-aligned Gaussian ridges have a significant influence on the flow organization of mixed convection flows.
The appearance of streamwise rolls is considerably reduced \textcolor{rev}{for dense ridge spacings $S$, which is related to the increased drag introduced by the ridges.}
\textcolor{rev}{Therefore, the formation of the rolls requires larger buoyancy forces, such that} the transition from forced convection structures to streamwise rolls is delayed by the ridges towards higher $Ri_b$ values than expected for smooth-wall conditions.
\textcolor{rev}{
Specifically, this transition occurs for the smooth channel at $Ri_b = 0.016$, while for large ridge spacings of $S\ge 2\delta$ this transition occurs first at $Ri_b=0.024$ and for denser ridge spacings $S\le\delta$ at  $Ri_b = 0.032$.    
}

\textcolor{rev}{The strongest influence of the heterogeneous surface on the flow organization} occurs between the roll-to-cell transition range, where a change of the \textcolor{rev01}{surface} properties has a comparable effect as a change of $Ri_b$ for homogeneous wall conditions. 
This behaviour is observed by the inspection of instantaneous and mean cross-sectional velocity and temperature fields.
In the range of $Ri_b =3.2 - 10$, where streamwise rolls are present for smooth-wall conditions, dense ridge spacings already trigger the transition from roll to cell structures. 
This is surprising, since this range of bulk Richardson number, which corresponds to a range of stability parameter $-\delta_\textit{eff}/L = 3.4 - 9.7$, is below the range where commonly cell structures are observed in the ABL \textcolor{rev01}{\citep{Salesky_nature_2017}}. 
The results show that the increased lateral drag introduced by the densely-spaced ridges, diminish the coherence of the streamwise rolls, and eventually lead to the transition to convective cells \textcolor{rev}{at smaller $Ri_b$}. 
\textcolor{rev}{In addition to the earlier roll-to-cell transition the ridges also affect the orientation of the convection cells for denser ridge spacings.}
\textcolor{rev02}{
While the convective cells have no preferential orientation for the smooth-wall natural convection case, they increasingly prefer to orient perpendicular to the ridges with decreasing $S$. 
}
\textcolor{rev}{This is also explained by the additional drag, which is experienced by ridge-aligned convective cells, such that the lateral near-wall motion of these cells is increasingly disturbed for smaller $S$. 
This will eventually lead to their breakdown and the flow prefers to stream only along the ridges resulting in the occurrence of spanwise coherent convective cells.}

\textcolor{rev02}{
For the moderate values of Reynolds numbers that can be afforded for the present simulations, we find that an increase in $Re$ favors the transition from forced convection structures to streamwise rolls at smaller $Ri_b$, which is associated with the increased thermal vertical mixing at larger $Re$.
At the roll-to-cell transition range, an increase of $Re$ promotes the transition towards convective cells, such that convective cells appear for larger $S$ if $Re$ is increased.
}

\textcolor{rev}{One particular observation is that the dynamics of streamwise rolls is very sensitive to ridge spacings in the order of the rolls' width, which is found for $Ri_b = 1$ and $Ri_b = 3.2$.}
For \textcolor{rev}{the specific ridge spacing $S=4\delta$} the up- and downdraft regions move over the entire channel slowly in time, with time periods of roughly 100 free-fall time units or 200 time bulk units, 
\textcolor{rev}{which is in contrast to} denser ridge spacings and smooth-wall conditions, where the spanwise location of the rolls is fixed. 
Due to this variation of streamwise rolls in the former case, some statistical features of the rolls are masked by long time integration.
This is seen for example for the strength of the roll's coherence, which almost vanish for long time intervals. 
Inspection of consecutive short-time averages reveal, that the strength of the roll's coherence depends on the spanwise location of the up- and downdraft region.
The coherence is reduced if the up- and downdraft regions occur in the valley of adjacent ridges, and is increased if they occur \textcolor{rev}{in the vicinity of} the ridges. 
In the former case the rolls experience stronger lateral drag due to their horizontal movement above the ridges, while in the latter case the ridges support the formation of localized buoyancy forces at the ridges, which in turn strengthens the up- and downdraft region. 
Although the ridges reinforce the rolls, they do not reside there permanently.
This is likely due to the symmetric arrangement of the ridges at both walls, since the up- and downdrafts impinge on an opposing ridge, which disturbs the roll formation.
While the formation mechanism of streamwise rolls is still not clear and under debate  \citep{etling_roll_1993,Salesky_nature_2017}, 
the present observations indicate that \textcolor{rev}{the formation and} the dynamics of streamwise rolls are very sensitive to 
\textcolor{rev01}{ heterogeneous surfaces}.
\backsection[Acknowledgements]{
This work was performed on the computational resources of HOREKA
and used the storage facility LSDF funded by the Ministry of Science,
Research and the Arts Baden-W\"urttemberg, and Deutsche Forschungsgemeinschaft (DFG) within the framework programme bwHPC.
}

\backsection[Funding]{
KS and BF acknowledge funding through DFG project number 423710075.
JPM acknowledges support
through project PID2019-105162RB-I00 funded by MCIN/AEI/10.13039/501100011033 
}

\backsection[Declaration of interests]{The authors report no conflict of interest.}


\backsection[Author ORCID]{\\[2pt]
K. Schäfer,  \href{https://orcid.org/0000-0002-1704-8233}{https://orcid.org/0000-0002-1704-8233}; \\[2pt]
B. Frohnapfel,  \href{https://orcid.org/0000-0002-0594-7178}{https://orcid.org/0000-0002-0594-7178}; \\[2pt]
J. P. Mellado, \href{https://orcid.org/0000-0001-7506-6539}{https://orcid.org/0000-0001-7506-6539} 
}

\backsection[Author contributions]{
KS designed the computational framework, carried out the numerical simulations, performed the data post-processing and statistical analysis with supervision by BF and JPM. All authors contributed to the conceptualization of the study as well as to the discussion and interpretation of the data. KS wrote the original draft of the paper with review and editing support from BF and JPM.
Funding was acquired by BF.}


\appendix
\section{Validation of code implementation }
\label{app:validation_code}

\begin{figure}
    \centering
    \includegraphics[trim=3pt 0 0 0]{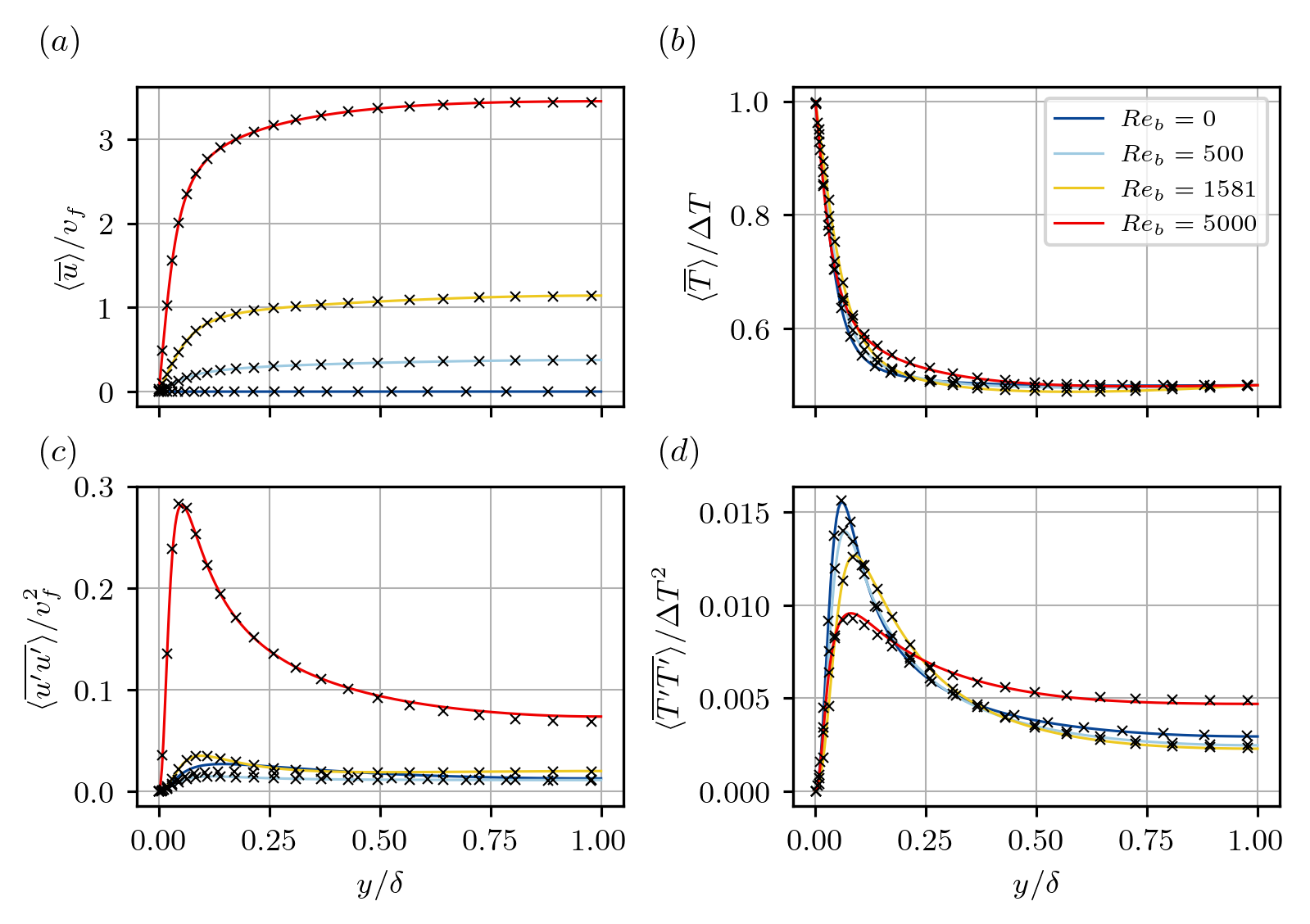}

    \caption{Mean profiles for validation study for $Ra=10^7$ and different bulk Reynolds numbers.
            The marks indicate the reference data of \cite{Pirozzoli_mixed_2017}, for clarity every fifth data point is shown.}
    \label{fig:validation_study}
\end{figure}

The implementation of the active scalar in Xcompact3d is validated against the Rayleigh-B\'enard and mixed convection cases of \cite{Pirozzoli_mixed_2017} at $Ra = 10^6$ and $10^7$. 
For a direct comparison the same grid resolution is used as in \cite{Pirozzoli_mixed_2017}, which is given in Table \ref{tab:validation_study}. 
The mean difference in skin friction coefficient and Nusselt number with respect to the reference data is indicated by $\varepsilon_{C_f}$ and $\varepsilon_{\mathit{Nu}}$.
While for the skin friction coefficient the two low Reynolds number cases at $Re=10^6$ show deviations up to $3.7\%$, this is reduced below $1.7\%$ for the higher Rayleigh number cases.
The Nusselt number is in very good agreement for both chosen Rayleigh numbers and stays below $0.8\%$ for all 
simulation cases. 
The mean velocity and mean temperature profiles, as well as the variances $\overline{u'u'}$- and $\overline{T'T'}$-profiles, are shown in Figure \ref{fig:validation_study} and the comparison 
to the reference data shows very good agreement between the considered flow cases. 

\begin{table}
    \begin{center}
    \begin{tabular}[b]{cccccccccccccccc}
    $\mathrm{Ra}$ & $\mathrm{Re}_{b}$ & $\mathrm{Ri}_{b}$ & $S/\delta$ & {$N_x\times N_y\times N_z$} 
         & & $C_f$ & $C_{f,\textit{ref}}$ & $\varepsilon_{C_f}$ & & $Nu$ & $Nu_{\textit{ref}}$ & $\varepsilon_{Nu}$ \\[3pt]
       \hline
      $10^6$ & 0     & $\infty$ & $\infty$ & $512 \times 193\times 256$  & & -       & -       & -        & & 8.257  &  8.288  & $0.38\%$    \\
      $10^6$ & 158.1 & 10       & $\infty$ & $512 \times 193\times 256$  & & 0.0719  & 0.0745  & $3.49\%$ & & 7.284  &  7.318  & $0.46\%$ \\ 
      $10^6$ & 500   & 1        & $\infty$ & $512 \times 193\times 256$  & & 0.0267  & 0.0277  & $3.69\%$ & & 6.312  &  6.356  & $0.70\%$ \\
      $10^6$ & 1581  & 0.1      & $\infty$ & $512 \times 193\times 256$  & & 0.0100  & 0.0102  & $1.79\%$ & & 6.798  &  6.780  & $0.26\%$ \\
      $10^6$ & 5000  & 0.01     & $\infty$ & $1024\times 257\times 512$  & & 0.00712 & 0.00715 & $0.39\%$ & & 12.360 & 12.419  & $0.48\%$ 
      \\[5pt]
      $10^7$ & 0     & $\infty$ & $\infty$ & $1024 \times 257\times 512$ & & -       & -       & -        & & 15.687 & 15.799  & $0.71\%$   \\
      $10^7$ & 500   & 10       & $\infty$ & $1024 \times 257\times 512$ & & 0.0403  & 0.0403  & $0.06\%$ & & 13.921 & 14.000  & $0.56\%$ \\ 
      $10^7$ & 1581  & 1        & $\infty$ & $1024 \times 257\times 512$ & & 0.0144  & 0.0146  & $1.28\%$ & & 11.911 & 11.880  & $0.26\%$ \\ 
      $10^7$ & 5000  & 0.1      & $\infty$ & $1024 \times 257\times 512$ & & 0.00742 & 0.00754 & $1.63\%$ & & 17.112 & 17.250  & $0.80\%$\\ 
      \hline
    \end{tabular}
    \caption{Simulation parameters and global flow properties of validation study for Rayleigh-B\'enard and Mixed Convection at $Ra = 10^6$ and $Ra=10^7$.
    The skin friction coefficient and Nusselt number of \cite{Pirozzoli_mixed_2017} are given by $C_{f,\textit{ref}}$ and $Nu_{\textit{ref}}$.}
    \label{tab:validation_study}
    \end{center}
\end{table}

\section{Grid study with Gaussian ridges}
\label{app:grid_study}

The grid resolution requirements for the simulations with Gaussian ridges is studied for different flow configurations to show that the chosen grid resolution 
is sufficiently fine to capture the investigated flow physics. 
The grid refinement study is performed for three different flow configurations, namely pure forced convection, mixed convection and pure Rayleigh-B\'enard flow. 
The domain size for this study was reduced to $L_x \times L_y \times L_z = 8\delta \times 2\delta \times 4\delta$ to keep the grid study computationally affordable. 
In all cases the spanwise spacing of the Gaussian ridges is $S/\delta=1$, corresponding to four Gaussian ridges at each side wall.
The different grid resolutions of the simulation cases and the resulting global flow properties are given in Table \ref{tab:Grid_study}.
\\
For the pure forced convection case the mean variation in skin friction coefficient and Nusselt number from the coarsest to the finest grid simulation is within a range 
of $0.5\%$ and $0.7\%$, respectively.
The grid refinement does not reveal any significant changes in the mean velocity, temperature and covariance profiles between all considered cases (not shown here). 
In order to satifsfy the grid requirements proposed by \cite{Pirozzoli_mixed_2017} for pure forced convection flows and being conservative with the spanwise grid resolution for the representation of the Gaussian ridges,
 the grid $N_x \times N_y \times N_z = 256 \times 193 \times 192$ is chosen to be appropriate.  
This results for the large domain simulation ($L_x \times L_y \times L_z = 16\delta \times 2\delta \times 8\delta$) in a grid of $N_x \times N_y \times N_z = 512 \times 193 \times 384$ for pure convection flows with Gaussian ridges. 

\begin{table}
    \begin{center}
    \begin{tabular}[b]{cccccccccccccc}
    $\mathrm{Ra}$ & $\mathrm{Re}_{b}$ & $Ri_b$ & $S/\delta$ & {$N_x\times N_y\times N_z$} 
         & $C_f (\cdot 10^{-3})$  & $\mathit{Nu}$  & $\Delta t_{tot}//t_b$ & $\Delta t_{tot}//t_f$\\[3pt]
       \hline
      0 & 2800  & 0 & 1 & $256 \times 193\times 128$ &  8.834 & 7.991 & 4719 & - \\
      0 & 2800  & 0 & 1 & $192 \times 193\times 192$ &  8.766 & 7.957 & 4719 & -  \\ 
      0 & 2800  & 0 & 1 & $256 \times 193\times 192$ &  8.781 & 7.958 & 4719 & - \\
      0 & 2800  & 0 & 1 & $256 \times 193\times 256$ &  8.822 & 7.997 & 5271 & - \\
      0 & 2800  & 0 & 1 & $256 \times 257\times 192$ &  8.823 & 8.014 & 4719 & - \\[5pt]
      $10^7$ & 5000 & 0.1 & 1 & $512 \times 193\times 192$ &  8.265 & 18.098 & 5423 &  864 \\
      $10^7$ & 5000 & 0.1 & 1 & $512 \times 193\times 256$ &  8.201 & 18.170 & 4254 &  677 \\
      $10^7$ & 5000 & 0.1 & 1 & $512 \times 257\times 192$ &  8.341 & 18.216 & 4301 &  684 \\ 
      $10^7$ & 5000 & 0.1 & 1 & $512 \times 257\times 256$ &  8.194 & 18.081 & 4150 &  660 \\
      $10^7$ & 5000 & 0.1 & 1 & $512 \times 257\times 320$ &  8.193 & 18.163 & 4123 &  657 \\[5pt]
      $10^7$ & 0 & $\infty$ & 1 & $512 \times 193\times 192$ &  0 & 16.974 & - & 1171 \\
      $10^7$ & 0 & $\infty$ & 1 & $512 \times 193\times 256$ &  0 & 16.999 & - & 1026 \\
      $10^7$ & 0 & $\infty$ & 1 & $512 \times 257\times 192$ &  0 & 16.921 & - & 1038 \\ 
      $10^7$ & 0 & $\infty$ & 1 & $512 \times 257\times 256$ &  0 & 17.034 & - & 1135 \\
      $10^7$ & 0 & $\infty$ & 1 & $512 \times 257\times 320$ &  0 & 16.981 & - & 1052  \\[5pt]
      \hline
    \end{tabular}
    \caption{Grid refinement study for pure forced convection, mixed convection and pure Rayleigh-B\'enard flow with Gaussian ridges at each side wall ($S/\delta = 1$). 
    The domain size for the study is set to $L_x \times L_y \times L_z = 8\delta \times 2\delta \times 4\delta$.}
    \label{tab:Grid_study}
    \end{center}
\end{table}

The grid refinement study for the mixed convection case is performed at $Ra=10^7$, which requires a finer grid compared to the grid study of the pure forced convection case at $Re_b = 2800$ according to 
smooth wall cases (see Table \ref{app:validation_code}). 
Furthermore, we increase the bulk Reynolds number to $Re_b = 5000$ in order to make this grid study more demanding in terms of the requirements of the shear induced turbulence. 
The mean difference of all cases in $C_f$ and $\mathit{Nu}$ with respect to the finest grid case, lies below  $1.81\%$ for the skin friction coefficient and $0.45\%$ for Nusselt number. 
This demonstrates, that the grid resolution for the plane wall mixed convection cases is already sufficient for the additional numerical representation of Gaussian ridges by the immersed boundary method based on polynomial reconstruction. 
Similar results are obtained for the pure Rayleigh-B\'enard case at $Ra=10^7$, where the mean difference in $\mathit{Nu}$ with respect to the finest grid case, is below $0.35\%$ for all simulation cases.
Consequently, for the investigation of mixed and natural convection at $Ra=10^7$ a grid of $N_x \times N_y \times N_z = 1024 \times 257 \times 512$ for the large domain cases is chosen. 
For lower $Ra$ cases, the chosen grid resoltion of the pure forced convection study marks the lower bound to sufficiently represent the Gaussian ridges in these cases. 

\bibliographystyle{jfm}
\bibliography{lit.bib}


\end{document}